\documentclass[12pt,a4paper]{article}

\usepackage{ifthen} 
\newboolean{pdflatex}
\setboolean{pdflatex}{true} 

\newboolean{articletitles}
\setboolean{articletitles}{true} 

\newboolean{uprightparticles}
\setboolean{uprightparticles}{false} 

\newboolean{inbibliography}
\setboolean{inbibliography}{false} 

\def\paperauthors{LHCb collaboration} 
\def\paperasciititle{Study of Upsilon production in pPb  collisions at sqrt(s)=8.16 TeV} 
\def\papertitle{Study of $\PUpsilon$ production in $p$Pb collisions at \sqsnn~=~8.16~\tev} 
\def\paperkeywords{{High Energy Physics}, {LHCb}} 
\def\papercopyright{\the\year\ CERN for the benefit of the LHCb collaboration} 
\def\paperlicence{CC-BY-4.0 licence}
\def\paperlicenceurl{https://creativecommons.org/licenses/by/4.0/}

\usepackage{tabto}
\usepackage[frak=pxtx]{mathalfa}
\DeclareMathAlphabet{\mathbbmsl}{U}{bbm}{m}{sl}
\DeclareMathAlphabet{\mathfrakmsl}{U}{frak}{m}{sl}

\usepackage[top=1in, bottom=1.25in, left=1in, right=1in]{geometry}

\usepackage[utf8]{inputenc}
\usepackage{microtype}
\usepackage{lineno}  
\usepackage{xspace} 
\usepackage{caption} 

\usepackage{graphicx}  
\usepackage{color}
\usepackage{colortbl}
\graphicspath{{./figs/}} 

\usepackage{booktabs}
\usepackage{multirow}
\usepackage{rotating} 
\usepackage{placeins} 
\usepackage{xfrac} %

\usepackage{amsmath} 
\usepackage{amssymb}
\usepackage{amsfonts}
\usepackage{upgreek} 
\usepackage{siunitx}

\newcommand*\patchAmsMathEnvironmentForLineno[1]{%
\expandafter\let\csname old#1\expandafter\endcsname\csname #1\endcsname
\expandafter\let\csname oldend#1\expandafter\endcsname\csname
end#1\endcsname
 \renewenvironment{#1}%
   {\linenomath\csname old#1\endcsname}%
   {\csname oldend#1\endcsname\endlinenomath}%
}
\newcommand*\patchBothAmsMathEnvironmentsForLineno[1]{%
  \patchAmsMathEnvironmentForLineno{#1}%
  \patchAmsMathEnvironmentForLineno{#1*}%
}
\AtBeginDocument{%
\patchBothAmsMathEnvironmentsForLineno{equation}%
\patchBothAmsMathEnvironmentsForLineno{align}%
\patchBothAmsMathEnvironmentsForLineno{flalign}%
\patchBothAmsMathEnvironmentsForLineno{alignat}%
\patchBothAmsMathEnvironmentsForLineno{gather}%
\patchBothAmsMathEnvironmentsForLineno{multline}%
\patchBothAmsMathEnvironmentsForLineno{eqnarray}%
}

\usepackage{hyperxmp}

\usepackage[pdftex,
            pdfauthor={\paperauthors},
            pdftitle={\paperasciititle},
            pdfkeywords={\paperkeywords},
            pdfcopyright={Copyright (C) \papercopyright},
            pdflicenseurl={\paperlicenceurl}]{hyperref}

\usepackage[all]{hypcap} 

\usepackage[colorinlistoftodos,textsize=scriptsize]{todonotes}

\usepackage{xspace} 
\usepackage{upgreek}

\def\lhcb {\mbox{LHCb}\xspace}

\def\MagUp {\mbox{\em Mag\kern -0.05em Up}\xspace}

\ifthenelse{\boolean{uprightparticles}}%
{

 \def\Pmu         {\ensuremath{\upmu}\xspace}

 \def\Ppsi        {\ensuremath{\uppsi}\xspace}

 \def\PDelta      {\ensuremath{\Delta}\xspace}                 
 \def\PXi      {\ensuremath{\Xi}\xspace}                 
 \def\PLambda      {\ensuremath{\Lambda}\xspace}                 
 \def\PSigma      {\ensuremath{\Sigma}\xspace}                 
 \def\POmega      {\ensuremath{\Omega}\xspace}                 
 \def\PUpsilon      {\ensuremath{\Upsilon}\xspace}                 
 
 \def\PB      {\ensuremath{\mathrm{B}}\xspace}                 
                  
 \def\PD      {\ensuremath{\mathrm{D}}\xspace}

 \def\PJ      {\ensuremath{\mathrm{J}}\xspace}                 
 \def\PK      {\ensuremath{\mathrm{K}}\xspace}

 \def\Pb      {\ensuremath{\mathrm{b}}\xspace}                 
 \def\Pc      {\ensuremath{\mathrm{c}}\xspace}

 \def\Pi      {\ensuremath{\mathrm{i}}\xspace}

}
{

 \def\Pmu         {\ensuremath{\mu}\xspace}

 \def\Ppsi        {\ensuremath{\psi}\xspace}                 
                  
 \mathchardef\PDelta="7101
 \mathchardef\PXi="7104
 \mathchardef\PLambda="7103
 \mathchardef\PSigma="7106
 \mathchardef\POmega="710A
 \mathchardef\PUpsilon="7107
                  
 \def\PB      {\ensuremath{B}\xspace}                 
                  
 \def\PD      {\ensuremath{D}\xspace}

 \def\PJ      {\ensuremath{J}\xspace}                 
 \def\PK      {\ensuremath{K}\xspace}

 \def\Pb      {\ensuremath{b}\xspace}                 
 \def\Pc      {\ensuremath{c}\xspace}

 \def\Pi      {\ensuremath{i}\xspace}

}

\makeatletter
\ifcase \@ptsize \relax
  \newcommand{\miniscule}{\@setfontsize\miniscule{4}{5}}
\or
  \newcommand{\miniscule}{\@setfontsize\miniscule{5}{6}}
\or
  \newcommand{\miniscule}{\@setfontsize\miniscule{5}{6}}
\fi
\makeatother

\DeclareRobustCommand{\optbar}[1]{\shortstack{{\miniscule (\rule[.5ex]{1.25em}{.18mm})}
  \\ [-.7ex] $#1$}}

\def\mup        {{\ensuremath{\Pmu^+}}\xspace}
\def\mun        {{\ensuremath{\Pmu^-}}\xspace} 
\def\mumu       {{\ensuremath{\Pmu^+\Pmu^-}}\xspace}

\def\cquark    {{\ensuremath{\Pc}}\xspace}

\def\bquark    {{\ensuremath{\Pb}}\xspace}

  \def\Kbar    {{\kern 0.2em\overline{\kern -0.2em \PK}{}}\xspace}

\def\KorKbar    {\kern 0.18em\optbar{\kern -0.18em K}{}\xspace}

  \def\Dbar    {{\kern 0.2em\overline{\kern -0.2em \PD}{}}\xspace}

\def\DorDbar    {\kern 0.18em\optbar{\kern -0.18em D}{}\xspace}

\def\Bbar    {{\ensuremath{\kern 0.18em\overline{\kern -0.18em \PB}{}}}\xspace}

\def\BorBbar    {\kern 0.18em\optbar{\kern -0.18em B}{}\xspace}

\def\jpsi     {{\ensuremath{{\PJ\mskip -3mu/\mskip -2mu\Ppsi\mskip 2mu}}}\xspace}

  \def\Y#1S{\ensuremath{\PUpsilon{(#1S)}}\xspace}
\def\OneS  {{\Y1S}}
\def\TwoS  {{\Y2S}}
\def\ThreeS{{\Y3S}}

\def\Lbar        {{\ensuremath{\kern 0.1em\overline{\kern -0.1em\PLambda}}}\xspace}
\def\LorLbar    {\kern 0.18em\optbar{\kern -0.18em \PLambda}{}\xspace}

\newcommand{\decay}[2]{\ensuremath{#1\!\to #2}\xspace}         

\def\to                 {\ensuremath{\rightarrow}\xspace}

\def\OneSTomm    {\decay{\OneS}{\mup\mun}}
\def\TwoSTomm    {\decay{\TwoS}{\mup\mun}}
\def\ThreeSTomm    {\decay{\ThreeS}{\mup\mun}}

\def\AT#1     {\ensuremath{A_{\mathrm{T}}^{#1}}\xspace}           

\def\C#1      {\ensuremath{\mathcal{C}_{#1}}\xspace}                       
\def\Cp#1     {\ensuremath{\mathcal{C}_{#1}^{'}}\xspace}                    
\def\Ceff#1   {\ensuremath{\mathcal{C}_{#1}^{\mathrm{(eff)}}}\xspace}        
\def\Cpeff#1  {\ensuremath{\mathcal{C}_{#1}^{'\mathrm{(eff)}}}\xspace}       
\def\Ope#1    {\ensuremath{\mathcal{O}_{#1}}\xspace}                       
\def\Opep#1   {\ensuremath{\mathcal{O}_{#1}^{'}}\xspace}                    

\newcommand{\tev}{\ifthenelse{\boolean{inbibliography}}{\ensuremath{~T\kern -0.05em eV}}{\ensuremath{\mathrm{\,Te\kern -0.1em V}}}\xspace}
\newcommand{\gev}{\ensuremath{\mathrm{\,Ge\kern -0.1em V}}\xspace}
\newcommand{\mev}{\ensuremath{\mathrm{\,Me\kern -0.1em V}}\xspace}
\newcommand{\kev}{\ensuremath{\mathrm{\,ke\kern -0.1em V}}\xspace}
\newcommand{\ev}{\ensuremath{\mathrm{\,e\kern -0.1em V}}\xspace}
\newcommand{\gevc}{\ensuremath{{\mathrm{\,Ge\kern -0.1em V\!/}c}}\xspace}
\newcommand{\mevc}{\ensuremath{{\mathrm{\,Me\kern -0.1em V\!/}c}}\xspace}
\newcommand{\gevcc}{\ensuremath{{\mathrm{\,Ge\kern -0.1em V\!/}c^2}}\xspace}
\newcommand{\gevgevcccc}{\ensuremath{{\mathrm{\,Ge\kern -0.1em V^2\!/}c^4}}\xspace}
\newcommand{\mevcc}{\ensuremath{{\mathrm{\,Me\kern -0.1em V\!/}c^2}}\xspace}

\def\mum  {\ensuremath{{\,\upmu\mathrm{m}}}\xspace}

\def\nb {\ensuremath{\mathrm{ \,nb}}\xspace}
\def\invnb {\ensuremath{\mbox{\,nb}^{-1}}\xspace}

\newcommand{\stat}{\ensuremath{\mathrm{\,(stat)}}\xspace}
\newcommand{\syst}{\ensuremath{\mathrm{\,(syst)}}\xspace}

\def\gsim{{~\raise.15em\hbox{$>$}\kern-.85em
          \lower.35em\hbox{$\sim$}~}\xspace}
\def\lsim{{~\raise.15em\hbox{$<$}\kern-.85em
          \lower.35em\hbox{$\sim$}~}\xspace}

\def\sqs   {\ensuremath{\protect\sqrt{s}}\xspace}
\def\sqsnn {\ensuremath{\protect\sqrt{s_{\scriptscriptstyle\rm NN}}}\xspace}

\def\pt         {\mbox{$p_{\mathrm{ T}}$}\xspace}

\def\geant      {\mbox{\textsc{Geant4}}\xspace}

\def\pythia     {\mbox{\textsc{Pythia}}\xspace}

\def\tell1  {TELL1\xspace}
\def\ukl1   {UKL1\xspace}

\renewcommand{\epsilon}{\varepsilon}
\renewcommand{\theta}{\vartheta}

\newcommand{\TeV}{\tev}
\newcommand{\GeVc}{\ensuremath{\gevc}}

\newcommand{\MeVcc}{\ensuremath{\mevcc}}
\def\nS {\Y nS}
\def\sqs  {\ensuremath{\protect\sqrt{s}}\xspace}

\usepackage{cite} 
\usepackage{mciteplus}
\usepackage[nameinlink,capitalise]{cleveref}

\usepackage{longtable} 

\begin{document}
\renewcommand{\thefootnote}{\fnsymbol{footnote}}
\setcounter{footnote}{1}


\begin{titlepage}
\pagenumbering{roman}

\vspace*{-1.5cm}
\centerline{\large EUROPEAN ORGANIZATION FOR NUCLEAR RESEARCH (CERN)}
\vspace*{1.5cm}
\noindent
\begin{tabular*}{\linewidth}{lc@{\extracolsep{\fill}}r@{\extracolsep{0pt}}}
\ifthenelse{\boolean{pdflatex}}
{\vspace*{-1.5cm}\mbox{\!\!\!\includegraphics[width=.14\textwidth]{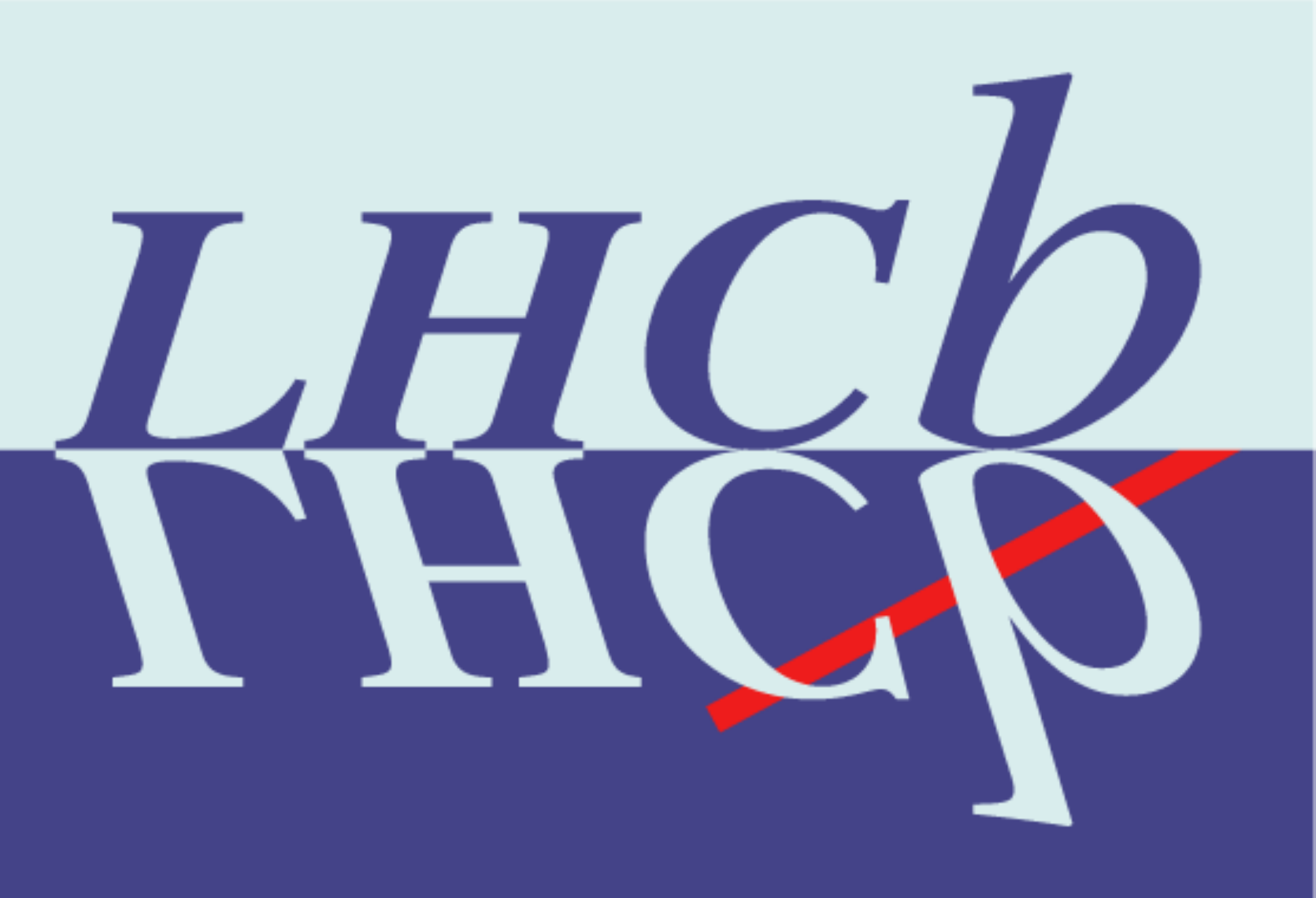}} & &}%
{\vspace*{-1.2cm}\mbox{\!\!\!\includegraphics[width=.12\textwidth]{lhcb-logo.eps}} & &}%
\\
 & & CERN-EP-2018-267 \\  
 & & LHCb-PAPER-2018-035 \\  
 & & \today \\ 
\end{tabular*}

\vspace*{3.5cm}

{\normalfont\bfseries\boldmath\huge
\begin{center}
  \papertitle 
\end{center}
}

\vspace*{2.0cm}

\begin{center}
\paperauthors\footnote{Authors are listed at the end of this paper.}
\end{center}

\vspace{\fill}

\begin{abstract}\noindent
The production of $\PUpsilon(nS)$ mesons ($n=1,2,3$) in $p$Pb and
  Pb$p$ collisions at a centre-of-mass energy per nucleon pair \sqsnn~=~8.16~\TeV\ is measured by the LHCb experiment, using a data sample corresponding to an integrated luminosity of 
  31.8~$\invnb$. 
  The \nS mesons are reconstructed through
  their decays into two opposite-sign muons. The measurements comprise  
  the differential production cross-sections 
  of the $\PUpsilon(1S)$ and $\PUpsilon(2S)$ states, 
  their forward-to-backward ratios and nuclear modification factors, 
   performed as a function of the transverse 
  momentum \pt and  rapidity in the nucleon-nucleon centre-of-mass frame $y^*$ of the \nS states, in the 
  kinematic range $\pt<25\gevc$
  and $1.5<y^*<4.0$ ($-5.0<y^*<-2.5$) for $p$Pb (Pb$p$) collisions. 
  In addition,  production cross-sections for \ThreeS\ are measured integrated over phase space and the production ratios
  between all three \nS states are determined. 
  The measurements are compared to theoretical predictions and suppressions for quarkonium in $p$Pb collisions are observed.   
\end{abstract}

\vspace*{2.0cm}

\begin{center}
  Submitted to JHEP
\end{center}

\vspace{\fill}

{\footnotesize 
\centerline{\copyright~\papercopyright. \href{\paperlicenceurl}{\paperlicence}.}}
\vspace*{2mm}

\end{titlepage}


\newpage
\setcounter{page}{2}
\mbox{~}
%
%
%
%

\cleardoublepage


\renewcommand{\thefootnote}{\arabic{footnote}}
\setcounter{footnote}{0}



\pagestyle{plain} 
\setcounter{page}{1}
\pagenumbering{arabic}


%

\section{Introduction}
\label{sec:Introduction}

Existing experimental results in collisions of ultra-relativistic heavy nuclei are
 consistent with the formation of  a deconfined state of hot partonic matter,  referred to as Quark-Gluon Plasma (QGP)~\cite{Adams:2005dq,Maciula:2013wg}.
One of the signatures of QGP is the suppression of heavy-quarkonia production in the collisions of heavy nuclei (AA collisions) with respect to $pp$ collisions,
 an effect that is enhanced for states with lower binding energies, such as the \ThreeS\ meson~\cite{Matsui:1986dk}.
However, the suppression of heavy-quarkonia production can also occur in the collisions of protons with heavy nuclei
($p$A collisions),  
where traditionally it was assumed that there was no QGP created.

In $p$A collisions, this suppression is caused by 
nuclear phenomena unrelated to deconfinement, commonly called cold nuclear matter (CNM) effects.
The CNM effects that are expected to affect quarkonia production are of two types, 
``initial-state" effects happening at a early stage of the collision, such as nuclear effects on parton densities~\cite{
deFlorian:2011fp,Owens:2012bv,Eskola:2009uj,Kovarik:2015cma} or coherent energy 
losses~\cite{Arleo:2014oha,Arleo:2013zua,Arleo:2012rs}, and ``final-state" effects, as 
quarkonia absorption by nucleons~\cite{GERSCHEL1988253}, expected to be negligible
at LHC energies~\cite{Albacete:2013ei,Adeluyi:2013tuu,Chirilli:2012jd,Chirilli:2012sk}.
Another final-state effect is the breaking of the 
$q\bar{q}$ pair caused by collisions with comoving particles with similar
rapidities (the so-called ``comovers" model~\cite{Ferreiro:2013pua,Ferreiro:2018wbd,Ferreiro:2014bia,
Du:2015wha,Ma:2017rsu}), 
whose density is determined from the particle 
multiplicity measured in that region of rapidity. 
This model could explain the relative suppression observed among the \nS states both in PbPb~\cite{Chatrchyan:2011pe} and in $p$A
collisions~\cite{Aaboud:2017cif}. The size of  nuclear effects can be quantified by measuring the nuclear modification factor $R_{p\mathrm{A}}$, which is defined as the ratio of the cross-section in $p$A collisions to that in $pp$ collisions scaled by the number of nucleons in the nucleus. In the absence of modifications, $R_{p\mathrm{A}}$ is  unity.

Previous measurements in AA collisions at RHIC~\cite{Ye:2017fwv}
and LHC~\cite{Khachatryan:2016xxp,Chatrchyan:2013nza,Aaboud:2017cif,Sirunyan:2017lzi,Abelev:2014oea}
{have revealed sizable nuclear modification factors for the \nS  states which increase with $n$.}
Using a data sample corresponding to an integrated luminosity of 
about 1.5~$\invnb$, the LHCb collaboration measured~\cite{Aaij:2014mza} the production of $\nS$ mesons in $p$Pb 
collisions at a per-nucleon centre-of-mass energy of $\sqsnn=5\TeV$. 
Moreover, the measurement of 
nuclear modification and forward-backward production ratios for \OneS, 
as well as  \nS\ to \OneS\ ratios were performed. 

In this paper, the production of $\nS$ mesons 
is studied in $p$Pb collisions using data collected at $\sqsnn=8.16\TeV$
with the LHCb detector, 
corresponding to a total integrated luminosity of 31.8$\invnb$.
 This dataset has been used already for the study of the production 
 of prompt $\jpsi$ and \jpsi\ coming from $b$-hadron 
decays (called nonprompt \jpsi\ in the following)~\cite{LHCb-PAPER-2017-014}.
The measurements presented here comprise
the differential production cross-sections 
of the $\PUpsilon(1S)$ and $\PUpsilon(2S)$ states, 
their forward-to-backward ratios and nuclear modification factors, and the production ratios
between all three \nS states.
In addition, the ratio of \OneS\ to nonprompt \jpsi\ cross-sections 
is determined as a function of proton-nucleon centre-of-mass rapidity, $y^*$,  integrated over the transverse momenta, \pt, of the mesons, a measurement 
that allows direct comparison of open heavy-flavour and quarkonia 
production in the environment of heavy-nuclei collisions.

\section{Detector description and data samples}
\label{sec:Detector}

The \lhcb detector~\cite{Alves:2008zz,LHCb-DP-2014-002} is a single-arm forward
spectrometer 
designed for the study of particles containing \bquark or \cquark
quarks. The detector includes a high-precision tracking system
consisting of a silicon-strip vertex detector surrounding the beam-beam
interaction region~\cite{LHCb-DP-2014-001}, a large-area silicon-strip detector located
upstream of a dipole magnet with a bending power of about
$4{\mathrm{\,Tm}}$, and three stations of silicon-strip detectors and straw
drift tubes~\cite{LHCb-DP-2013-003} placed downstream of the magnet.
The tracking system provides a measurement of the momentum of charged particles with
a relative uncertainty that varies from 0.5\% at low momentum to 1.0\% at 200\gevc.
The minimum distance of a track to a collision vertex, the impact parameter, 
is measured with a resolution of $(15+29/\pt)\mum$,
where \pt is in\,\gevc.
Different types of charged hadrons are distinguished using information
from two ring-imaging Cherenkov detectors~\cite{LHCb-DP-2012-003}. 
Photons, electrons and hadrons are identified by a calorimeter system consisting of
scintillating-pad and preshower detectors, an electromagnetic
calorimeter and a hadronic calorimeter. Muons are identified by a
system composed of alternating layers of iron and multiwire
proportional chambers~\cite{LHCb-DP-2012-002}.

The trigger~\cite{LHCb-DP-2012-004} consists of a
hardware stage, based on information from the calorimeter and muon
systems, followed by a software stage, in which all charged particles
with ${\pt>300\mevc}$ are reconstructed.  
The alignment and calibration of the detector is
performed in near real-time~\cite{LHCb-PROC-2015-011}.
 {This alignment is also used later in the off\-line reconstruction,}
 ensuring consistent and high-quality particle identification (PID)
information in the online and off\-line processing. 
The identical performance of the
online and off\-line reconstruction offers the opportunity 
to perform physics analyses directly
using candidates reconstructed in the 
trigger~\cite{LHCb-DP-2012-004, LHCb-DP-2016-001} 
as well as storing information about all reconstructed
particles in the event~\cite{LHCB-PAPER-2016-064}.
The storage of only the triggered candidates enables a
reduction of the event size by an order of magnitude.

For this analysis, at least one muon with $\pt > 500\mevc$ is required at the hardware trigger stage 
and at the software trigger stage, two
muon tracks with $\pt>300\mevc$ and a high-quality reconstructed decay vertex are required to form an
\nS candidate with invariant mass $m(\mu^+\mu^-)> 4.7$\gevcc.
In addition, a small fraction of events with a  large number of tracks in the vertex detector are rejected.

Simulation is used in the determination of efficiencies. 
{The $p$Pb collisions are simulated with  EPOS-LHC~\cite{EPOS}  and the \decay{\nS}{\mumu} decays with  \pythia 8.1~\cite{Sjostrand:2006za,*Sjostrand:2007gs} in $pp$ collisions
where the proton energy is equal to that in $p$Pb collisions.}
The interaction of the generated particles with the detector and its response
are implemented using the \geant
toolkit~\cite{Allison:2006ve, *Agostinelli:2002hh}, as described in
Ref.~\cite{LHCb-PROC-2011-006}.
The $\nS$ mesons are produced unpolarised, justified by the 
fact that the polarisation of $\nS$ mesons has been measured by LHCb
in $pp$ collisions at similar energies and found to be
small~\cite{Aaij:2017egv}.
Consistently with what was done in previous LHCb analyses~\cite{LHCb-PAPER-2017-014}, no systematic uncertainty is associated with this assumption.

The asymmetric layout of the LHCb experiment~\cite{Alves:2008zz}, 
which covers the pseudorapidity range  of $2<\eta< 5$, 
results in two configurations: in the 
{\em  forward} $p$Pb ({\em  backward} Pb$p$) configuration,
the proton (lead) beam travels from the VELO detector to the muon chambers,
taking advantage of the inversion of the proton and lead beams during the 
$p$Pb data-taking run. 
The energy of the proton beam is 6.5\tev, while that of the lead beam is
2.56\TeV per nucleon, resulting in a centre-of-mass energy of the proton-nucleon system
of 8.16\TeV.
Since the energy per nucleon in the proton beam is significantly larger than that in the
lead beam, the proton-nucleon centre-of-mass system has a rapidity in the laboratory
frame of $+0.465$ ($-0.465$) for {$p$Pb (Pb$p$)} collisions,
resulting in a shift of the range of the centre-of-mass rapidity $y^*$
in the proton nucleus case.
In this analysis, \nS\ mesons are measured in the kinematic range of  $\pt<25~\GeVc$, and  $1.5<y^*<4.0$ for $p$Pb forward and  $-5.0<y^*<-2.5$ for $p$Pb backward collisions.
This is the first measurement 
of \ThreeS\ production in $p$Pb collisions 
in this kinematic range.
The data samples correspond to an integrated luminosity of 
$12.5\pm 0.3\invnb$ in the forward configuration and $19.3\pm 0.5\invnb$ in the backward configuration. The luminosities are determined using  van der Meer scans~\cite{LHCb-PAPER-2014-047}, which were performed for both beam configurations.

\section{Definition of the observables}
\label{sec:observables}

The observables are measured in bins of \pt and $y^*$ of the \OneS\ and \TwoS\ mesons, where both \pt and $y^*$ are defined with respect to the direction of the proton beam in the centre-of-mass frame. For the \ThreeS\ meson, due to the limited signal yield, only integrated observables are measured.

The differential cross-section is measured in a fixed  bin size of 0.5 units for  $y^*$  and  variable  bin sizes for \pt in the 0--25\GeVc\ range.
The  $\PUpsilon(nS)$ meson double-differential production cross-section in the proton-lead collisions is defined as

\begin{equation}\label{eq:cross-section}    
  \frac{\text{d}^2\sigma}{\text{d} \pt\text{d} y^*}  
  = \frac{N(\nS\rightarrow \mu^+\mu^-)}                   
  {\mathcal{L} \times\epsilon^{\nS}_{\rm tot}\times \mathcal{B}^{\nS}_{\mu\mu} 
\times\Delta \pt\times\Delta y^*},
 \end{equation}   
 where $N(\nS\rightarrow \mu^+\mu^-)$ is the raw yield of the \nS\ decays
reconstructed in the given rapidity and
transverse momentum bin, $\epsilon^{\nS}_{\rm tot}$  is the total efficiency in that bin, including 
acceptance, $\mathcal{B}^{\nS}_{\mu\mu}$ is the branching fraction of the  \nS\ state to the $\mu^+\mu^-$ final state, and $\mathcal{L}$ is  the integrated luminosity of the   
data sample. The values of the branching fractions used in this
 measurement are $(2.48 \pm 0.05)\%$ for \OneSTomm,
 $(1.93 \pm 0.17)\%$ for \TwoSTomm,  and $(2.18 \pm 0.21)\%$ for 
\ThreeSTomm\cite{PDG2018}. 

The nuclear modification factor for ${}^{208}$Pb is defined for the $p$Pb and Pb$p$ configurations as

\begin{equation}                                                                                                                            
 R_{p{\rm Pb}} ( \pt,y^*) = \frac{1}{208} \frac{{\rm d}^2 \sigma_{p{\rm Pb}}( \pt,y^*)/{\rm d} \pt{\rm d}y^*}
 {{\rm d}^2\sigma_{pp}( \pt,y^*)/{\rm d} \pt{\rm d}y^*}, 
 \end{equation}
 where $\sigma_{pp}$ is the 
 reference cross-section from $pp$ collisions
{interpolated  to \sqs= 8.16\tev using the LHCb measurements at  \sqs=2.76, 7, 8, and 13~\tev.}

The forward-to-backward ratio is 
defined as
 \begin{equation}                                                                                                                            
 R_{\rm FB} (\pt,|y^*|) = \frac{{\rm d}^2 \sigma_{p{\rm Pb}}(\pt,+|y^*|)/{\rm d}\pt{\rm d}y^*}{{\rm d}^2 \sigma_{{\rm Pb}p}(\pt,-|y^*|)/{\rm d}\pt{\rm d}y^*},
 \end{equation} 
and is evaluated in the rapidity range of $2.5 < \vert y^* \vert < 4.0$, which
 is common to $p$Pb and Pb$p$ collisions.

 The ratio of excited \TwoS\ and \ThreeS\ states  to the \OneS\ ground state in proton-lead collisions is defined  as 

 \begin{equation}                                                                                                                            
  R(\nS) = \frac{\left[{{\rm d}^2\sigma}/{{\rm d}\pt dy^*}\right](\nS)}{\left[{{\rm d}^2\sigma}/{{\rm d}\pt dy^*}\right](\OneS)}.
  \end{equation}  
  In addition,  the ratio of \OneS\ to non-prompt $J/\psi$ cross-sections in proton-lead collisions is measured in the same way. 
  The double ratio 
   \begin{equation}  
  \mathfrak{R}^{\nS/\OneS}_{(p{\rm Pb}|{\rm Pb}p)/pp} =  \dfrac{R(\PUpsilon(nS))_{p{\rm Pb}|{\rm Pb}p}}{R(\PUpsilon(nS))_{pp}}
  \end{equation}  
   compares the ratio $R(\nS)$ in  $p$Pb or Pb$p$ collisions  to $R(\nS)$ in $pp$ collisions.

\section{Event selection}
\label{sec:triggerselections}

\begin{figure}[t]
   \begin{center}
     \includegraphics[width=0.49\linewidth]{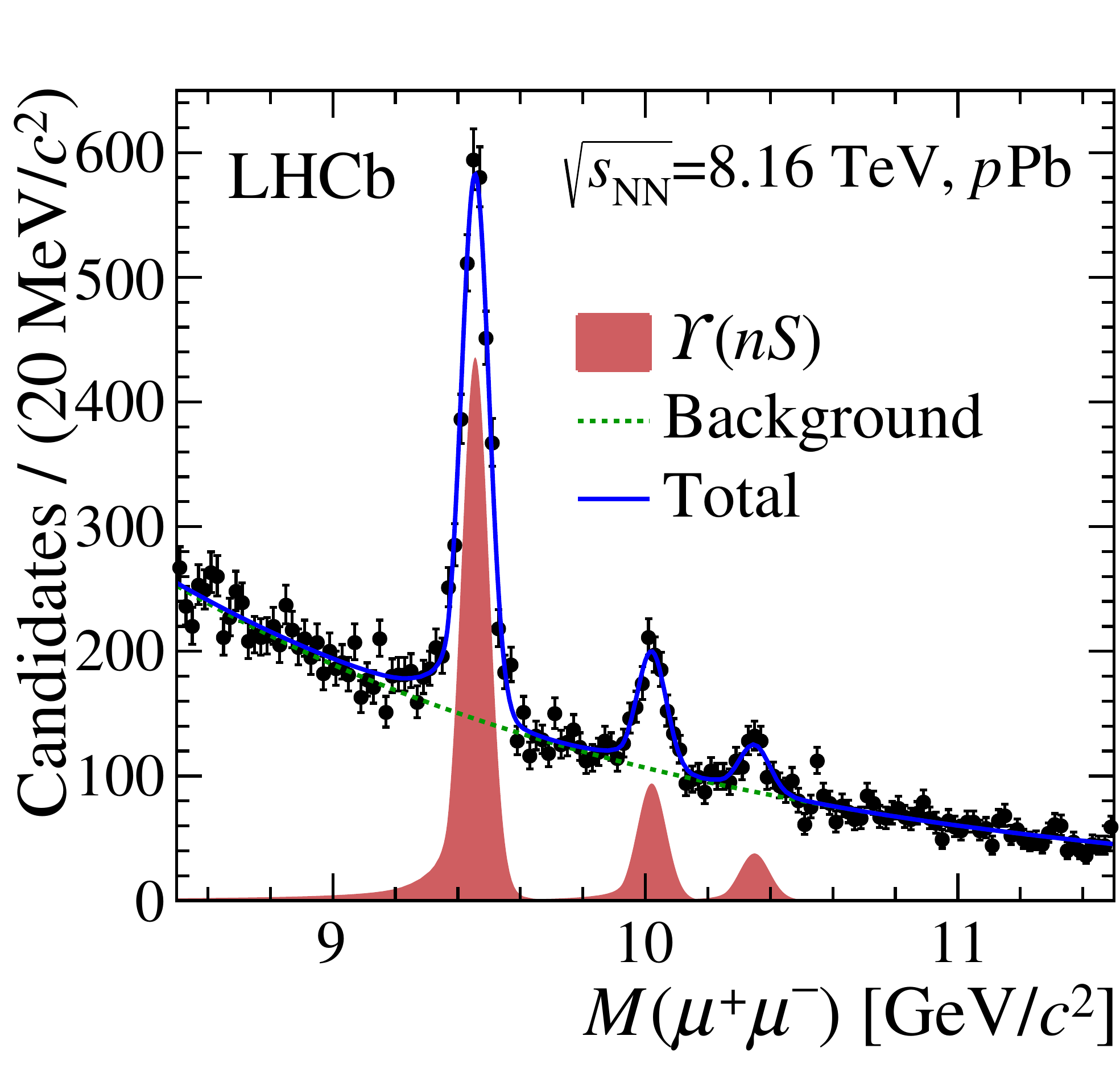}
     \includegraphics[width=0.49\linewidth]{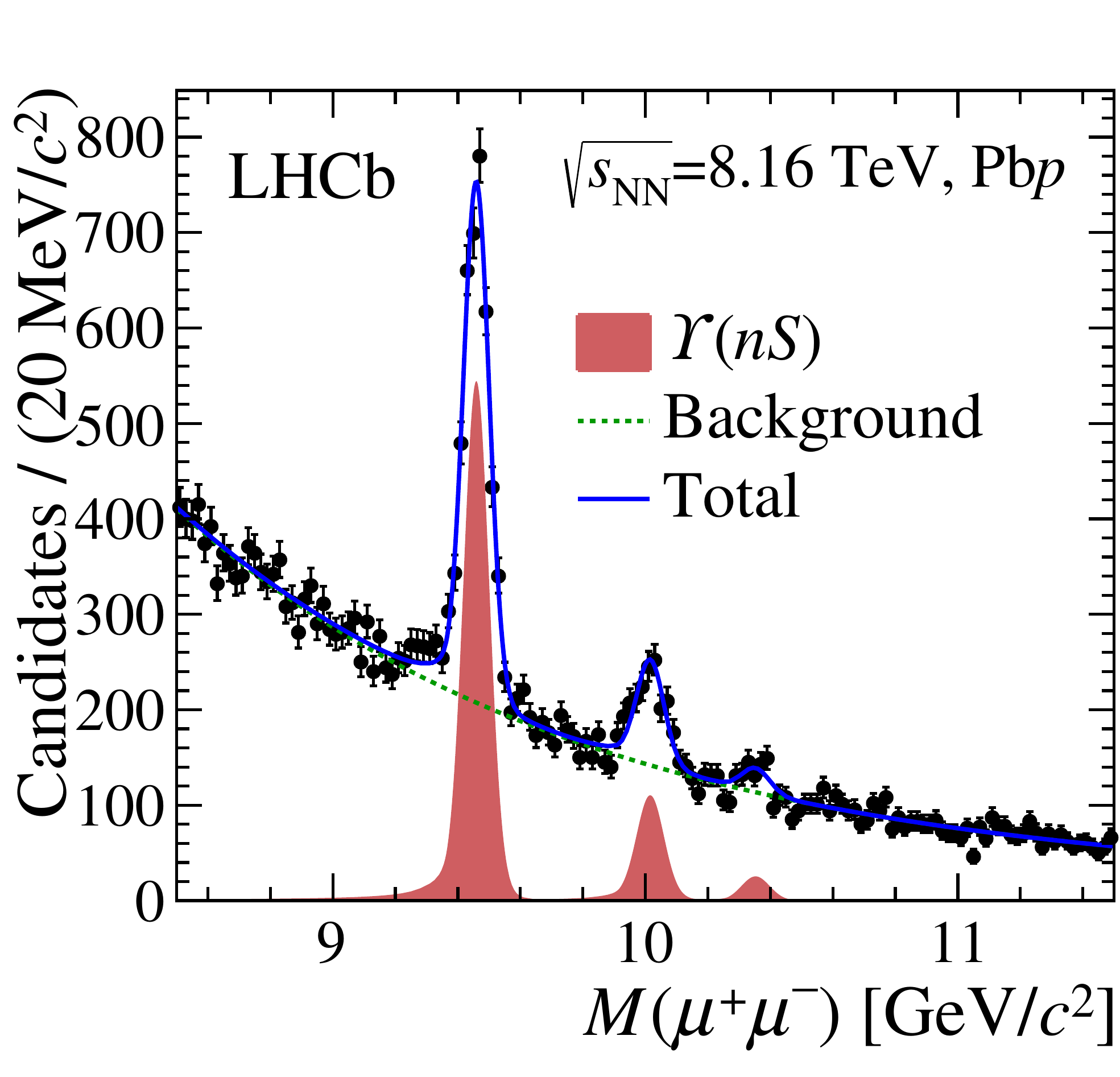}
   \end{center}
   \caption{
     Invariant-mass distribution of $\mu^+\mu^-$ pairs from the (left) $p$Pb and (right) Pb$p$ samples
     after the trigger and off\-line selections. 
}
   \label{fig:mass_pPb_Pbp}
 \end{figure}

The candidates reconstructed in the trigger are further filtered by means of an offline selection.
In the off\-line selection, 
muon tracks are required to have $\pt > 1\GeVc$, to be in the geometrical acceptance of the 
spectrometer ($2.0<\eta<5.0$), to satisfy PID requirements, 
and to have a good track-fit quality.
The dimuon invariant-mass distribution of off\-line-selected candidates is shown in Fig.~\ref{fig:mass_pPb_Pbp} 
for the 
$p$Pb and Pb$p$ samples. 
\label{sec:binning}

The dimuon invariant-mass distribution is fitted with  an exponential function for the background and three separate 
peaking functions, each consisting of the sum of two Crystal Ball functions~\cite{Skwarnicki:1986xj} for the \nS peaks. 
The shape parameters of the double Crystal Ball functions ($n$ and $\alpha$) are 
fixed to the values obtained in the simulation.
The yields of \OneS, \TwoS, \ThreeS\ 
mesons in the $p$Pb and Pb$p$ samples are summarised in Table~\ref{tab:yields}. 
The probability that the background can produce a fluctuation greater than or equal to the excess observed in data is calculated as the local $p$-value. For the exponential-background-only fits in the range of $\pm 100 \MeVcc$ around the expected \ThreeS\ mass peak, the local $p$-values are below $10^{-13}$ in $p$Pb sample and below $10^{-7}$ in Pb$p$ sample.

\begin{table}[bt]                  
   \caption{
   Yields of  $\OneS$, $\TwoS$, $\ThreeS$ mesons in $p$Pb and Pb$p$ samples as given by the fit. 
The uncertainties
are statistical only.
    }                                                                                                        
 \begin{center}\begin{tabular}{ccccc}                                                                                              \hline                                                                                                                   Samples & $\OneS$ &  $\TwoS$ & $\ThreeS$ & $\mathcal{L}$ \\
     \hline                    
 $p$Pb & $2705\pm 87$ & $584 \pm 49$ & $262\pm 44$ & 12.5~$\invnb$ \\
 Pb$p$ & $3072\pm 82$ & $679 \pm 54$ & $159 \pm 39$ & 19.3~$\invnb$ \\
      \hline                                                                                                             
   \end{tabular}\end{center}                                                                                                                 
 \label{tab:yields}                                                                                                                         
 \end{table}

\section{Efficiencies}
\label{sec:efficiency}

\def\effTot{\ensuremath{\epsilon_{\mathrm{tot}}}\xspace}
\def\effAcc{\ensuremath{\epsilon_{\mathrm{acc}}}\xspace}
\def\effReco{\ensuremath{\epsilon_{\mathrm{rec}}}\xspace}
\def\effTrk{\ensuremath{\epsilon_{\mathrm{trk}}}\xspace}
\def\effSel{\ensuremath{\epsilon_{\mathrm{sel}}}\xspace}
\def\effID{\ensuremath{\epsilon_{\mathrm{ID}}}\xspace}
\def\effTrigger{\ensuremath{\epsilon_{\mathrm{trigger}}}\xspace}

The signal yields are corrected bin-by-bin by the total efficiencies to obtain the cross-section measurements.
The total efficiency \effTot includes contributions from the geometrical acceptance, 
the tracking and trigger efficiencies, 
and the efficiency of the selection including the requirement on the PID of the muons.
All efficiencies are determined from simulation, apart from
  the tracking and  particle-identification efficiencies, where data are used
to correct the efficiencies obtained from the simulation. The same 
procedure is used for each of the three \nS states. 

The muon tracking efficiency is calculated using simulated $\nS$ events in $p$Pb
 and Pb$p$ collisions, and the efficiency in simulation is calibrated using efficiencies
estimated from \jpsi candidates selected in $p$Pb data using a tag-and-probe method similar to that 
adopted in the measurement of \jpsi\ production using the same
data set~\cite{LHCb-PAPER-2017-014}.
    
The PID efficiency for muons is measured using a statistically independent  sample of \jpsi\ decays  
in $p$Pb and $pp$ data.  
In regions where the number of 
\jpsi\ decays is small, 
the efficiency is determined using weighted data from  $pp$ collisions 
to reproduce the kinematics
and detector occupancies of $p$Pb collisions.

\begin{figure}[tb]
   \begin{center}
     \includegraphics[width=0.49\linewidth]{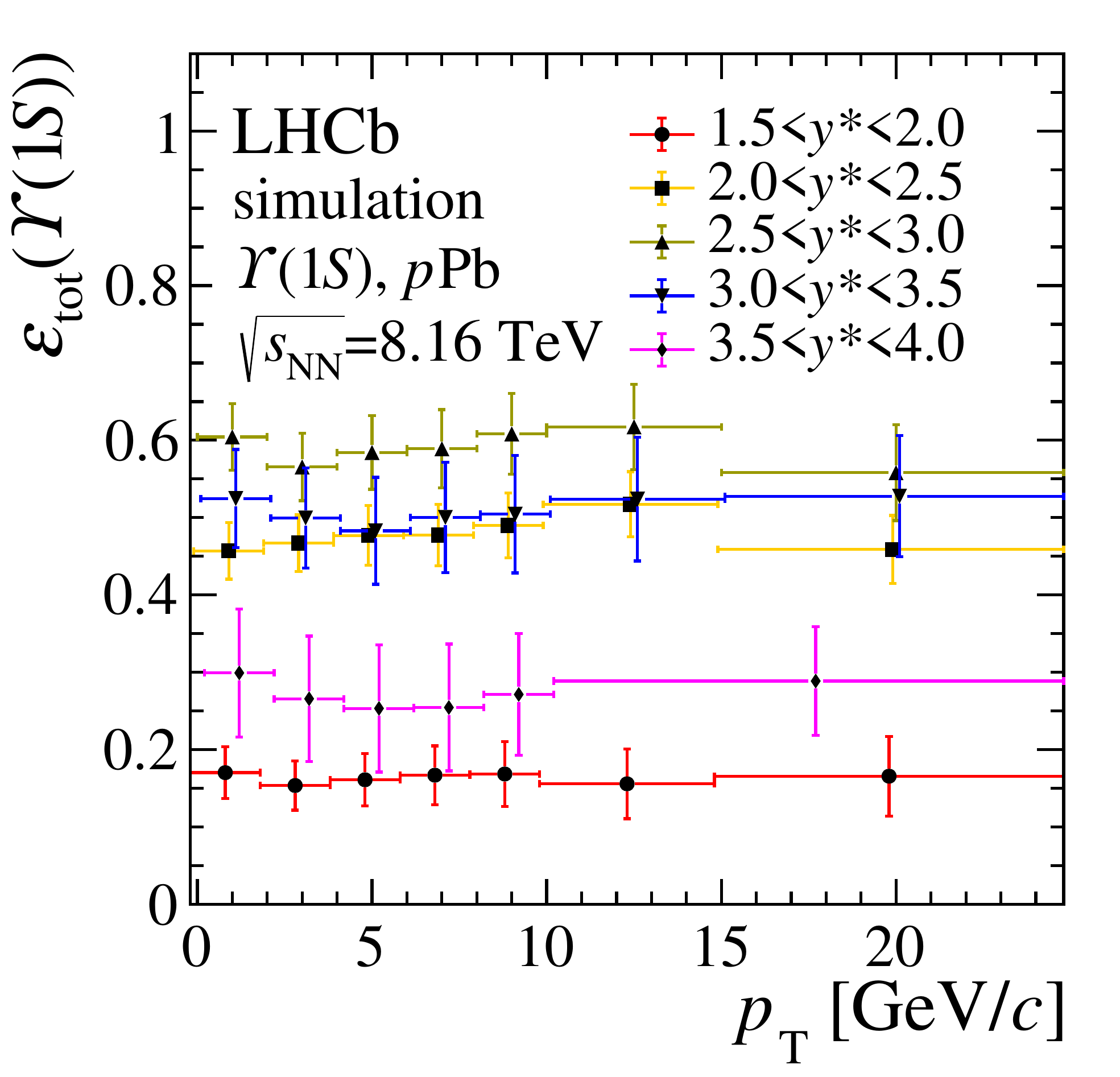}
     \includegraphics[width=0.49\linewidth]{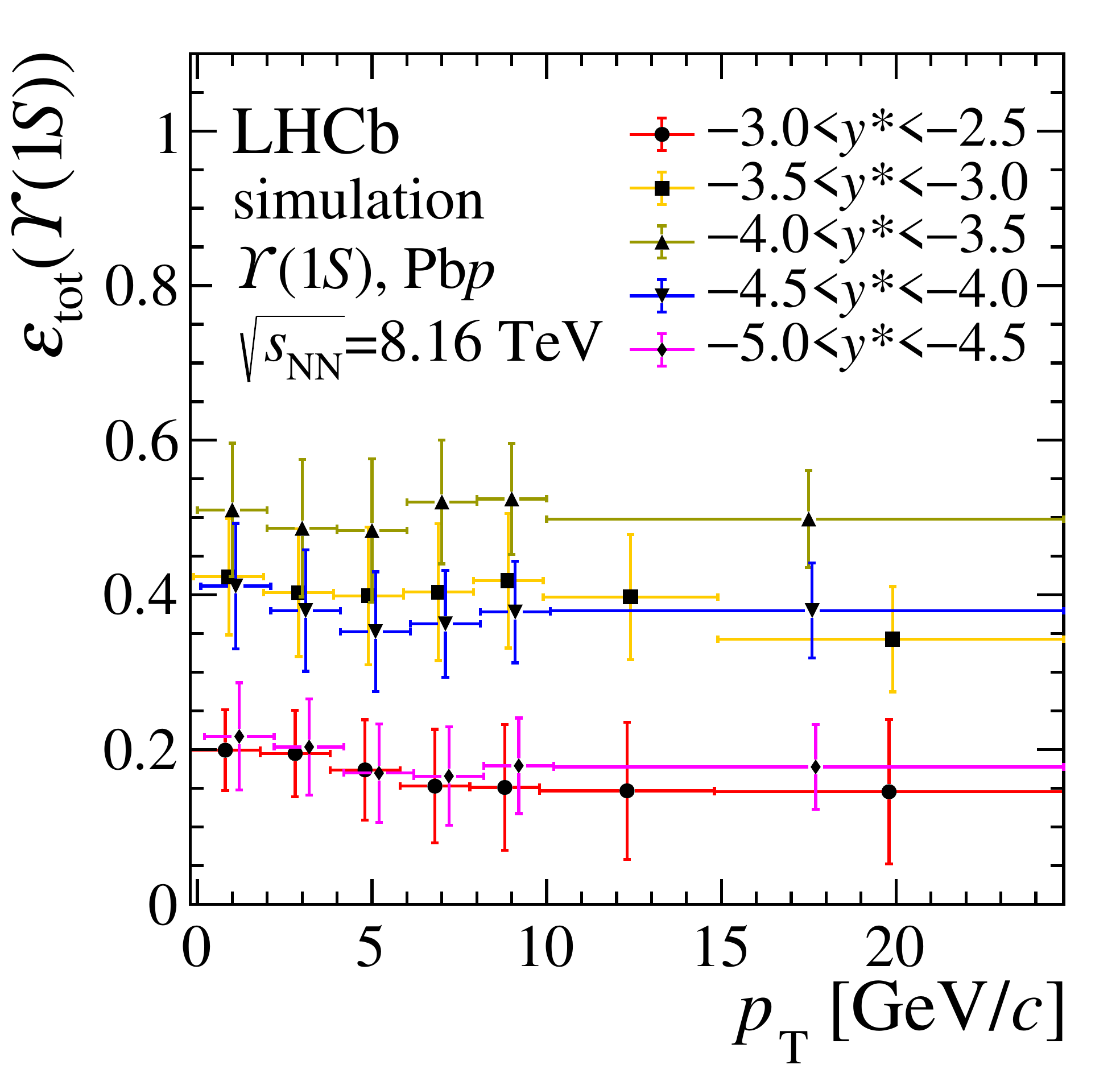}
   \end{center}
   \caption{Total efficiency $\effTot$ of the \OneS\ meson
 as a function of its \pt in different $y^*$ bins in (left) $p$Pb and (right) Pb$p$ collisions. The horizontal locations of the markers are roughly the centroids of the bins, with offsets from centre to aid in readability.}
   \label{fig:eff_total}
 \end{figure}

 The total efficiency for the \OneS\ state is shown in
  Fig.~\ref{fig:eff_total}. 
  The efficiencies for the \TwoS\ and \ThreeS\ states are similar. 
  The uncertainties shown are statistical, due to the limited size of the
simulated samples, and systematic, which will be discussed in the next section.
The difference in efficiencies as a function of rapidity is largely due to acceptance effects.

\section{Systematic uncertainties}

Table~\ref{tab:UncertaintySum} summarises the systematic uncertainties,  
which are different for each of the \nS\ states. 
The finite size of the simulation samples leads to an uncertainty on the
efficiency estimation, which is uncorrelated among bins and $\nS$ states and contributes to the uncertainties in acceptance, offline selection and trigger efficiencies. 
These uncertainties are 
small compared to the other systematic uncertainties and barely affect the overall systematic uncertainty. 
All other uncertainties are correlated among bins.

 \begin{table}[tb]   
   \caption{                               
     Systematic uncertainties (in percent) on the cross-section measurements. The ranges indicate the minimum and maximum values in different bins, among all \nS\ states. 
    }                                                   
 \begin{center}\begin{tabular}{lcc}  
     \hline                                                       
     Source & $p$Pb & Pb$p$\\          
     \hline   
     Signal detemination & 5.7\% & 5.7\%  \\
     Acceptance & 0.7\% -- 3.4\% &  0.5\% -- 3.5\%\\
     Reconstruction efficiency & 2.1\% -- 7.9\% & 2.5\% -- 8.1\%  \\
     Offline selection efficiency & 0.1\% -- 0.8\% & 0.1\% -- 1.4\% \\
     PID efficiency  &  1.1\% -- 4.4\% & 1.9\% -- 6.0\%  \\
     Trigger efficiency  & 2.0\% -- 2.8\% & 2.0\% -- 2.4\% \\  
     Luminosity & 2.6\% & 2.5\% \\
     Branching ratio & 2.0\% -- 9.6\% & 2.0\% -- 9.6\% \\
     \hline     
   \end{tabular}\end{center}   
 \label{tab:UncertaintySum}  
 \end{table}

The choice of the fit model for the mass distributions affects the signal yields.
 The uncertainty associated with the choice of the fit functions is estimated using
 different functions (single Crystal Ball functions for signal, and a second-order polynomial for background), and by 
 modifying the fit range for the signal fit
to account for the uncertainty due to the radiative tail.
The uncertainty due to the choice of the fit models 
is estimated to be 5.7\%.

The track reconstruction efficiency calibration has uncertainties from three sources: 
 the size of the calibration samples, the selection efficiency, and the signal yield determination of the calibration data sample. 
 Considering all these effects, the total uncertainty from the reconstruction 
of the tracks varies from 2.1\% to 7.9\% for the  $p$Pb sample and from 2.5\% to 8.1\% for the Pb$p$ sample.

The uncertainty on the offline selection efficiency is only due to the finite size of the simulation sample, varying from 0.1\% to
1.4\%.

The PID uncertainties are related to the limited size of the $pp$ and $p$Pb (Pb$p$)  calibration samples, and to the difference between the 
$pp$ and $p$Pb (Pb$p$) PID calibration samples. 
The latter effects lead to an uncertainty on the PID efficiency 
varying from 1.1\% to 3.9\% for the $p$Pb sample and from 1.9\% to 2.8\% for the 
Pb$p$ sample.
The total PID uncertainty including all effects
 varies from 1.1\% to 4.4\% for the $p$Pb sample and from 1.9\% to 6.0\% for the 
Pb$p$ sample. 

The trigger efficiency is obtained from simulation. The limited size of the simulated samples contributes to kinematic-bin-dependent uncertainties that vary between 0.2\% and 2.0\% for the $p$Pb sample and between 0.2\% and 1.2\% in the Pb$p$ sample. An additional uncertainty of 2.0\% is assigned based on a study of the trigger efficiency on a calibration data sample.

The relative uncertainty  on the $p$Pb luminosity determined by  the van der Meer scan is 2.6\% and that on the Pb$p$ luminosity is 2.5\%.

The uncertainties from the decay branching fractions of the \nS states contribute to the systematic uncertainty for values between 2.0 and 9.6\%\cite{PDG2018}.

\section{Results}
 
\begin{figure}[tb]
   \begin{center}
     \includegraphics[width=0.49\linewidth]{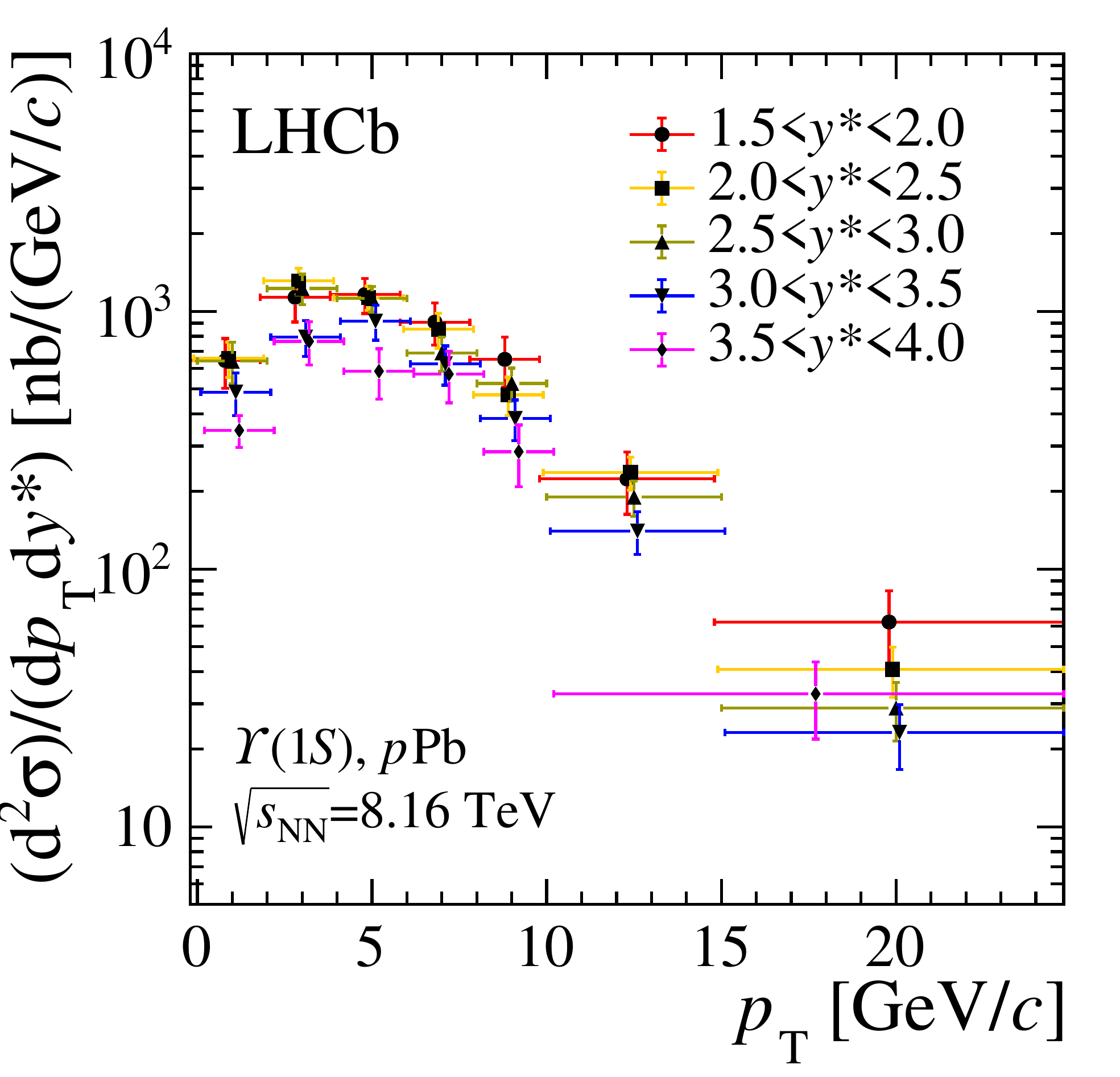}
     \includegraphics[width=0.49\linewidth]{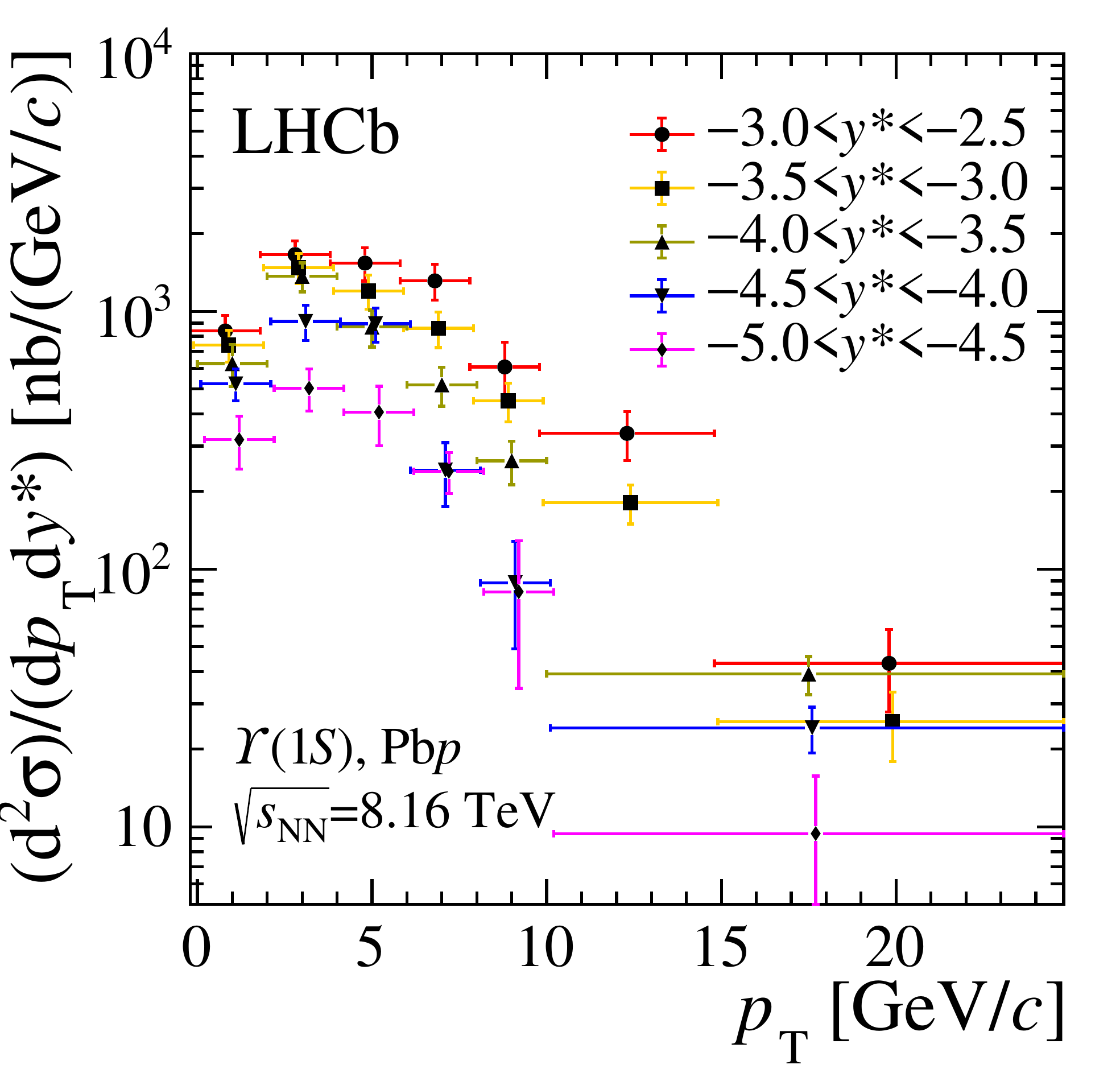}
     
   \end{center}
   \caption{
Double-differential cross-section for the  \OneS\ meson as a function of \pt for different 
values of  $y^*$ 
for the (left) forward $p$Pb and  (right) backward Pb$p$ samples.
The uncertainties are 
the sums in quadrature of the statistical and systematic components.
The horizontal locations of the markers are roughly the centroids of the bins, with offsets from centre to aid in readability.
    }
       \label{fig:CS_pt_ystr_1S}
 \end{figure}

\begin{figure}[tb]
   \begin{center}
     \includegraphics[width=0.49\linewidth]{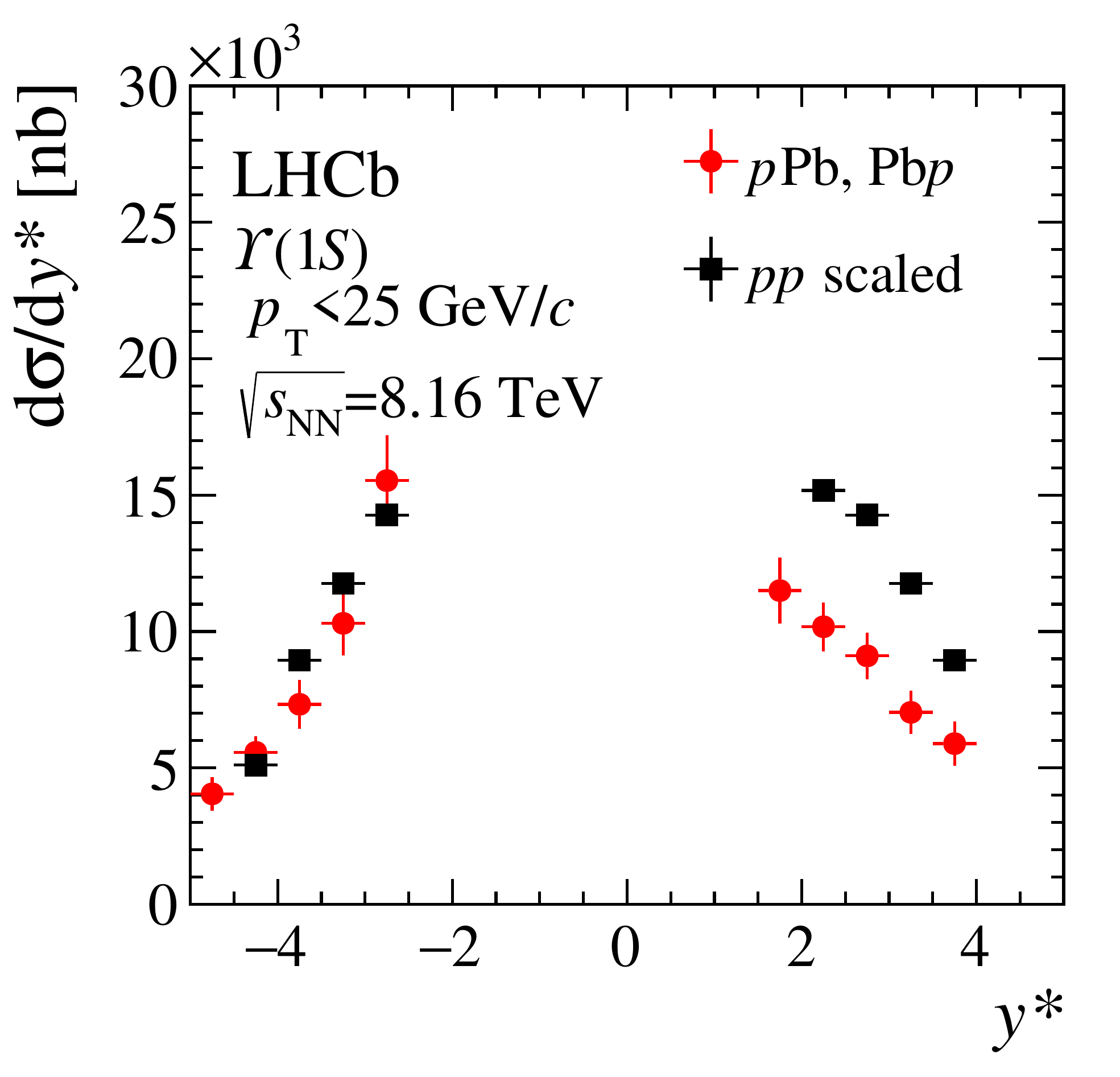}
     \includegraphics[width=0.49\linewidth]{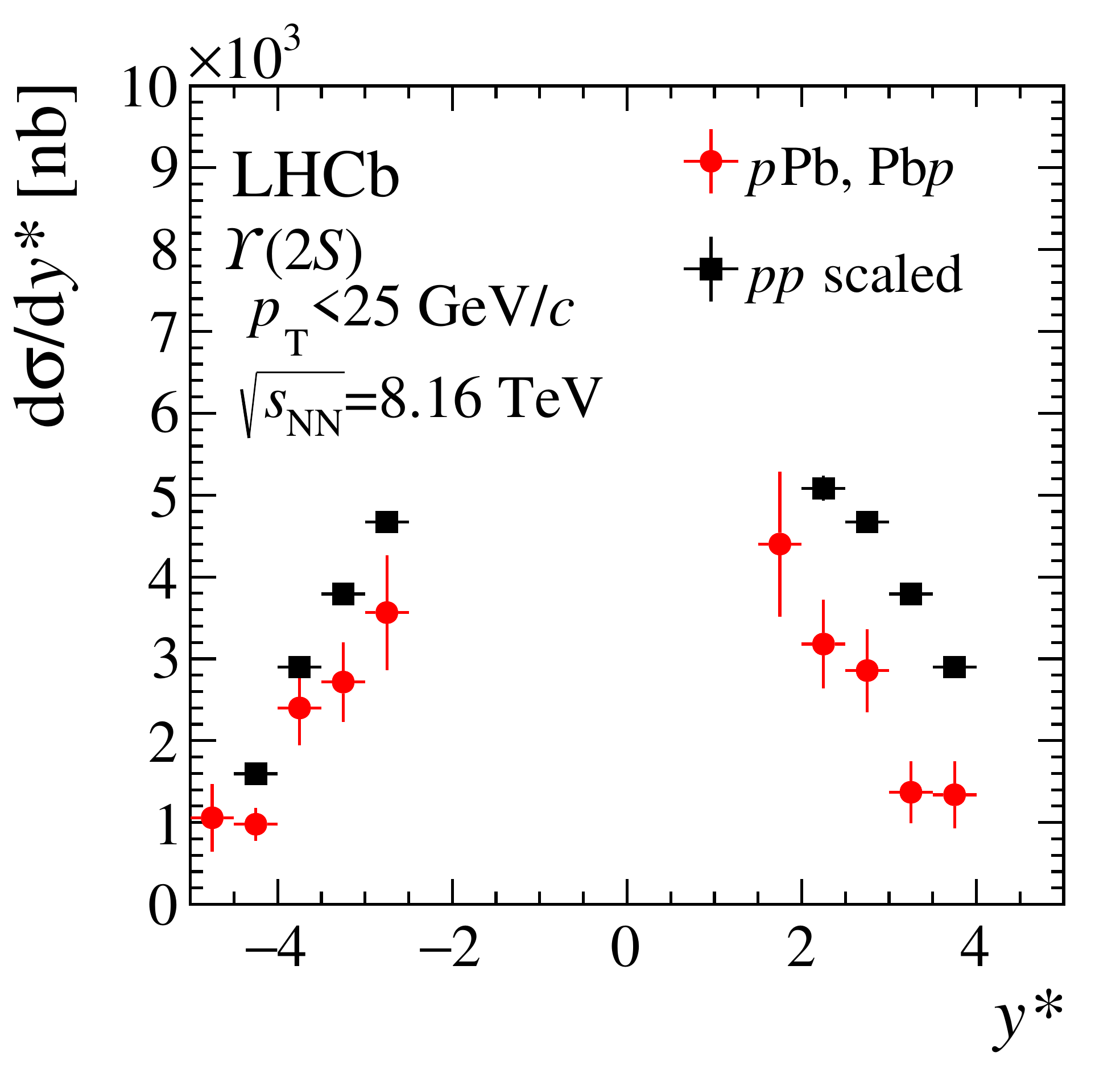}
     
   \end{center}
   \caption{
Cross-section of (left) \OneS\ and (right) \TwoS\ production as a function of $y^*$ integrated over \pt 
for the backward (negative $y^*$) and forward (positive $y^*$) samples, compared to 
the cross-section measured in $pp$,  interpolated to $\sqsnn=8.16\TeV$.
In this and subsequent figures, the uncertainties shown are the sums in quadrature of the statistical and systematic components.
    }
       \label{fig:CS_ystr_1S}
 \end{figure}

\begin{figure}[tb]
   \begin{center}
     \includegraphics[width=0.49\linewidth]{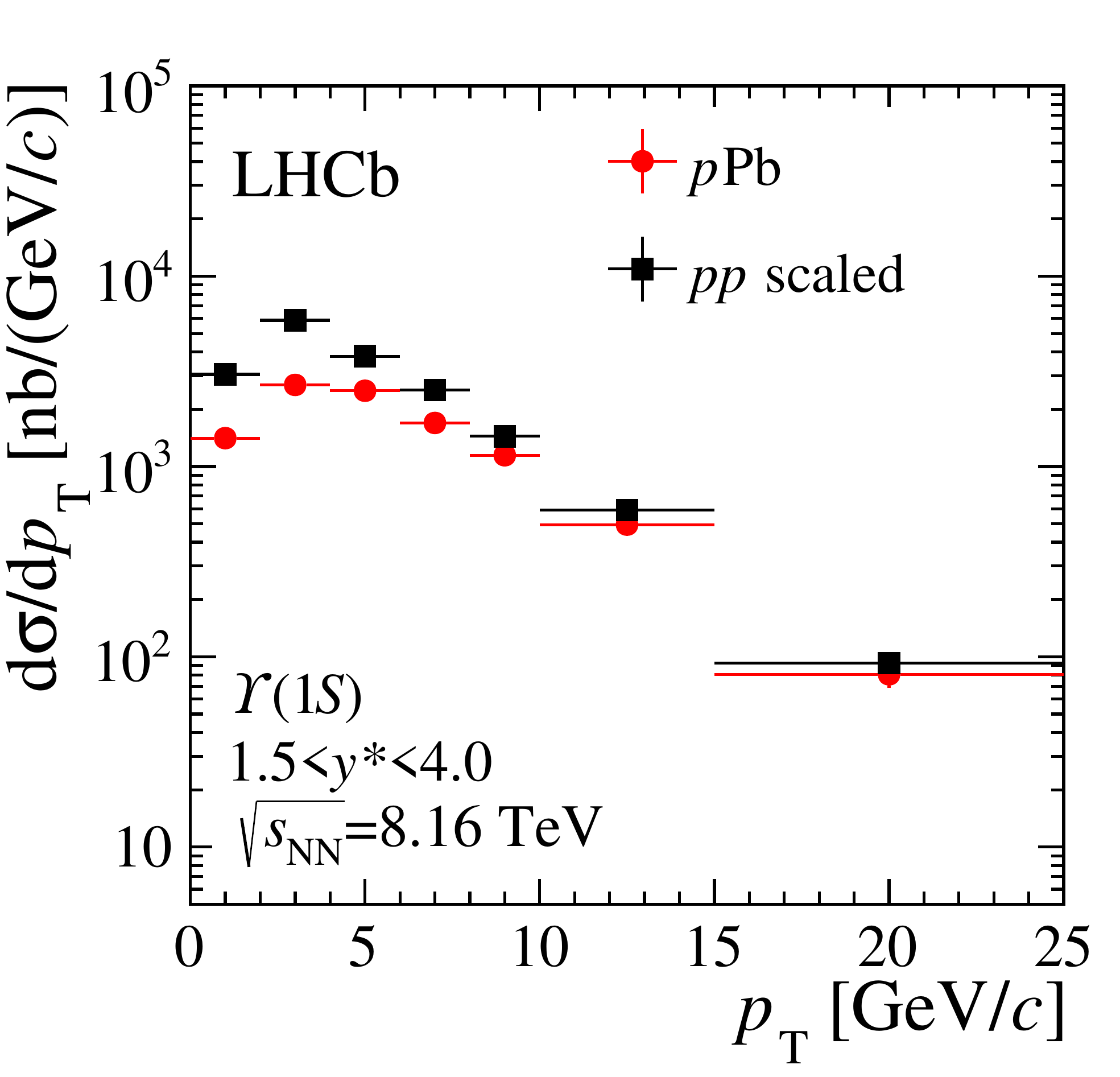}
     \includegraphics[width=0.49\linewidth]{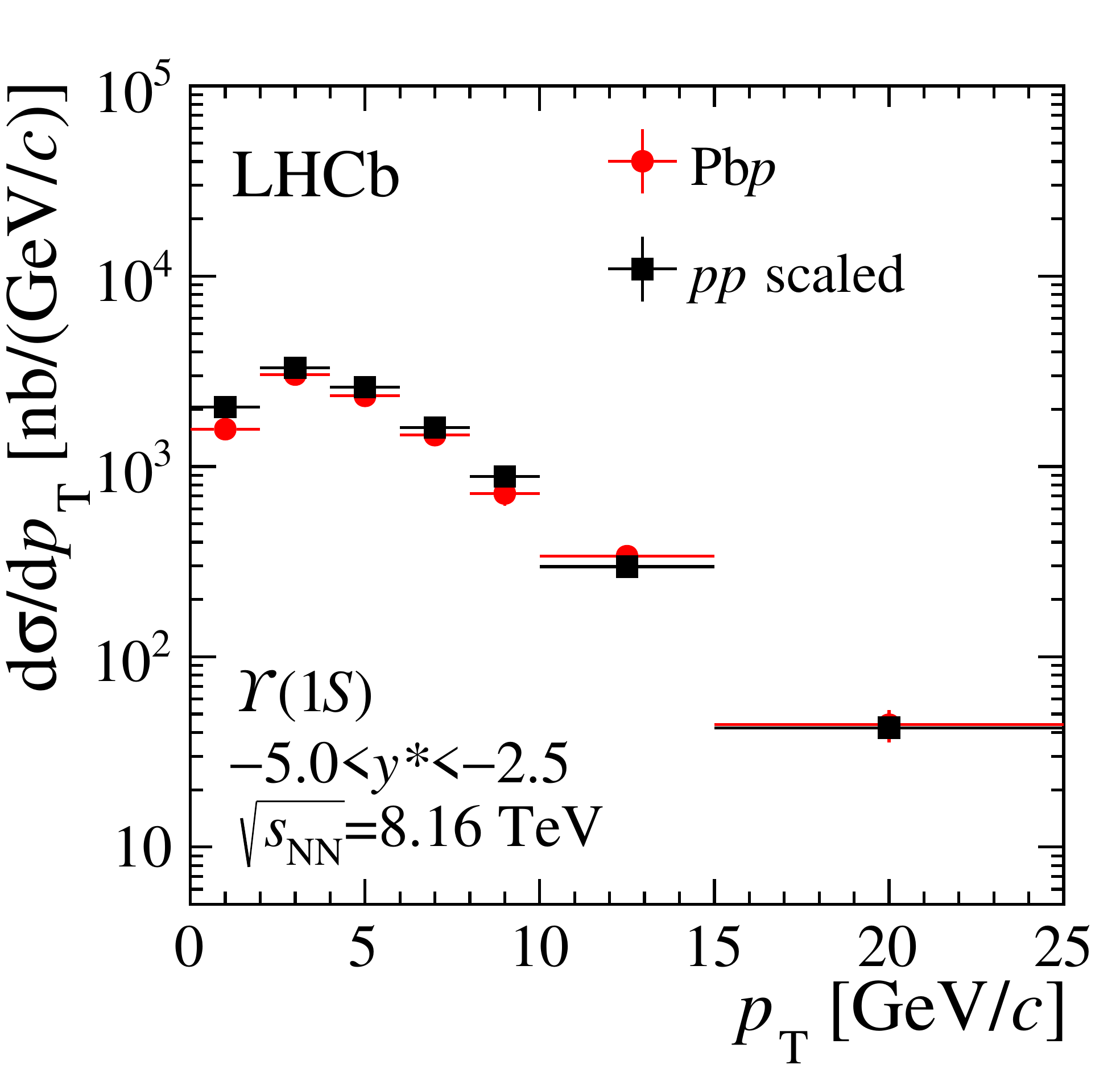}
     \includegraphics[width=0.49\linewidth]{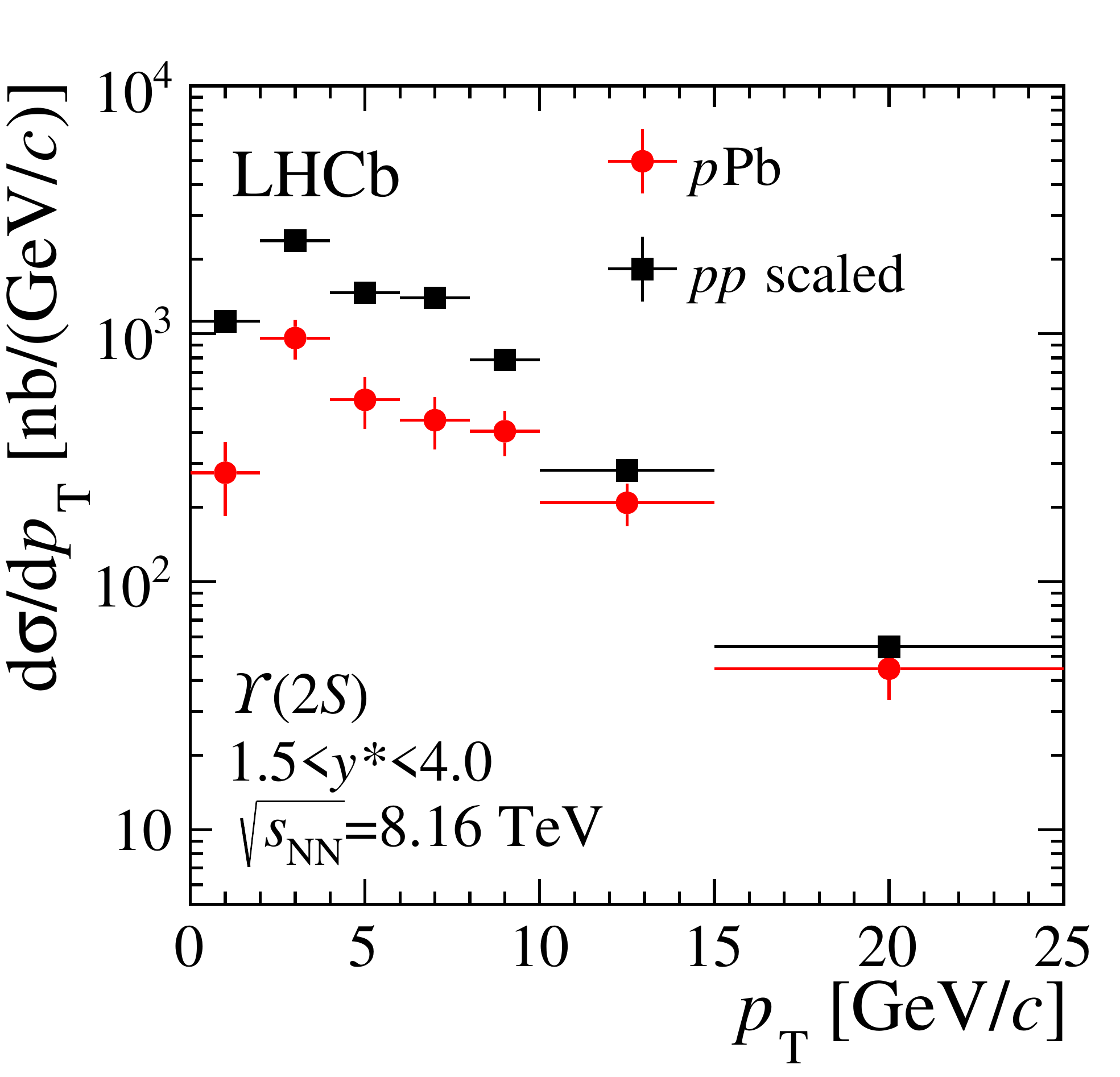}
     \includegraphics[width=0.49\linewidth]{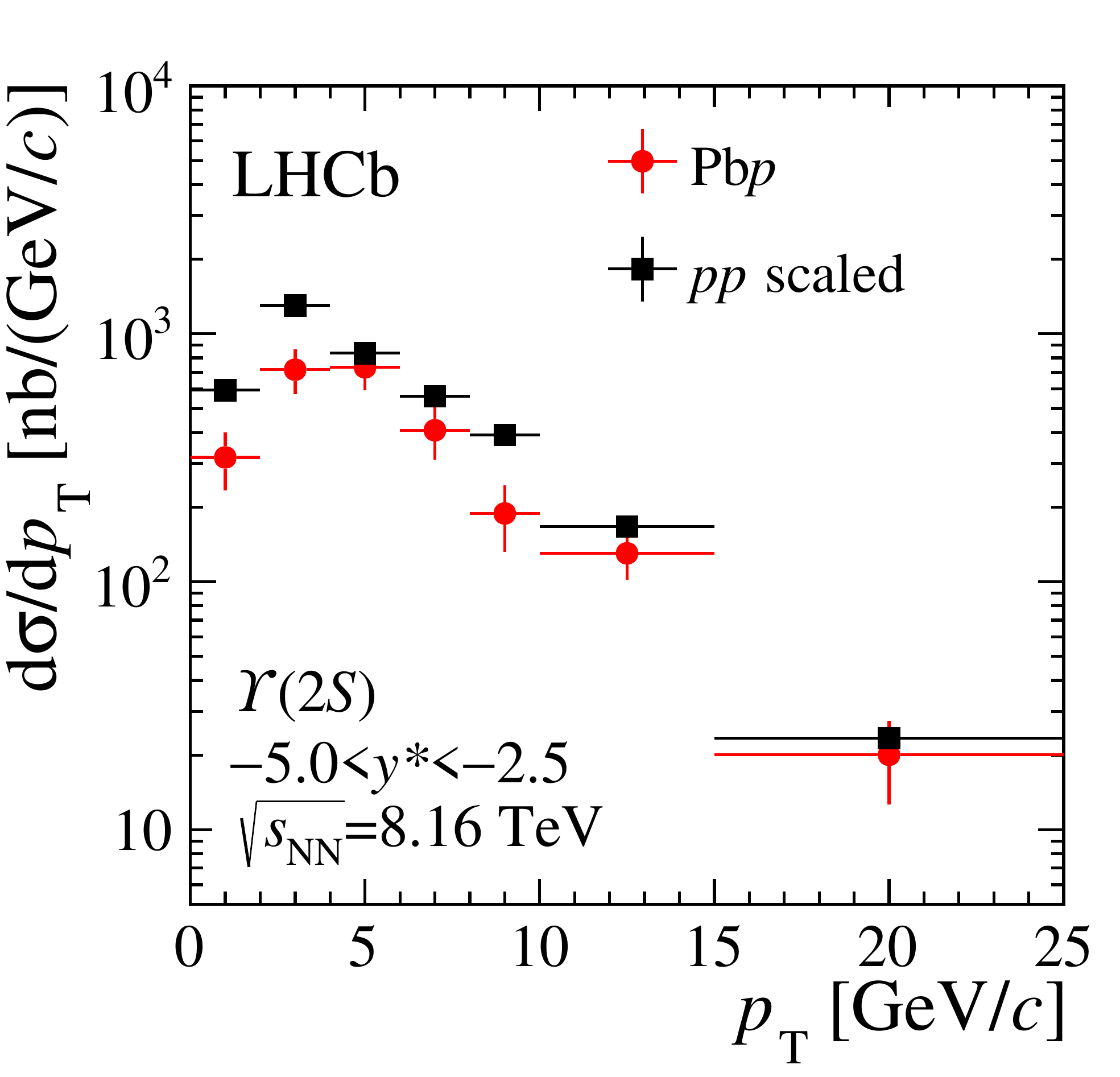}

   \end{center}
   \caption{
Cross-section of (top) \OneS\ and (bottom) \TwoS\ production as a function of \pt integrated over $y^*$ 
for the  (left)  forward and  (right)  backward samples compared to 
the cross-section measured in $pp$, interpolated to $\sqsnn=8.16\TeV$.
    }
       \label{fig:CS_pt_1S}
 \end{figure}

The total \nS\ cross-sections in the kinematic region $\pt$ $ <$ 25 GeV/c and $1.5 < y^* < 4.0$ $(-5.0< y^*<-2.5)$ for $p$Pb (Pb$p$) sample are measured to be
\begin{align*}
\sigma_{p\rm Pb}^{\PUpsilon(1S)}  &= 22.8 \pm 0.9\, \stat \pm 2.1\, \syst\, \mu{\rm b},\\
\sigma_{p\rm Pb}^{\PUpsilon(2S)}  &= \phantom{0}6.4 \pm 0.6\, \stat\pm 0.8\, \syst\,  \mu{\rm b},\\
\sigma_{p\rm Pb}^{\PUpsilon(3S)}  &= \phantom{0}2.5 \pm 0.4\, \stat\pm 0.3\, \syst\,  \mu{\rm b},\\
\sigma_{{\rm Pb}p}^{\PUpsilon(1S)}  &= 20.3 \pm 0.8\, \stat\pm 2.6\, \syst\,  \mu{\rm b},\\
\sigma_{{\rm Pb}p}^{\PUpsilon(2S)}  &= \phantom{0}6.0 \pm  0.5\, \stat\pm 0.9\, \syst\,  \mu{\rm b},\\
\sigma_{{\rm Pb}p}^{\PUpsilon(3S)}  &= \phantom{0}1.2 \pm 0.3\, \stat\pm 0.2\, \syst\,  \mu{\rm b}.
\end{align*}
The cross-sections are also  evaluated as a function 
of \pt and $y^*$ for the \OneS\ and \TwoS\ states.
The double-differential cross-section for the 
\OneS\ state is shown in Fig.~\ref{fig:CS_pt_ystr_1S}.
It is 
 integrated over \pt to form a differential cross-section as a function of $y^*$, as shown in Fig.~\ref{fig:CS_ystr_1S} (left), and integrated over $y^*$ to form a differential cross-section as a function of \pt, as shown
 in Fig.~\ref{fig:CS_pt_1S} (left).\footnote{
The numerical results of all cross-section measurements
shown in this section can be found in Appendix~\ref{sec:A1}. 
}
Similarly, for the \TwoS\ state the differential cross-section as a function of $y^*$ and \pt are shown in Fig.~\ref{fig:CS_ystr_1S} (right) and Fig.~\ref{fig:CS_pt_1S} (right), respectively.
For the $\ThreeS$ state, due to the limited sample size, only the cross-section 
integrated over \pt\ and $y^*$ is measured. 

\begin{figure}[tb]
   \begin{center}
     \includegraphics[width=0.49\linewidth]{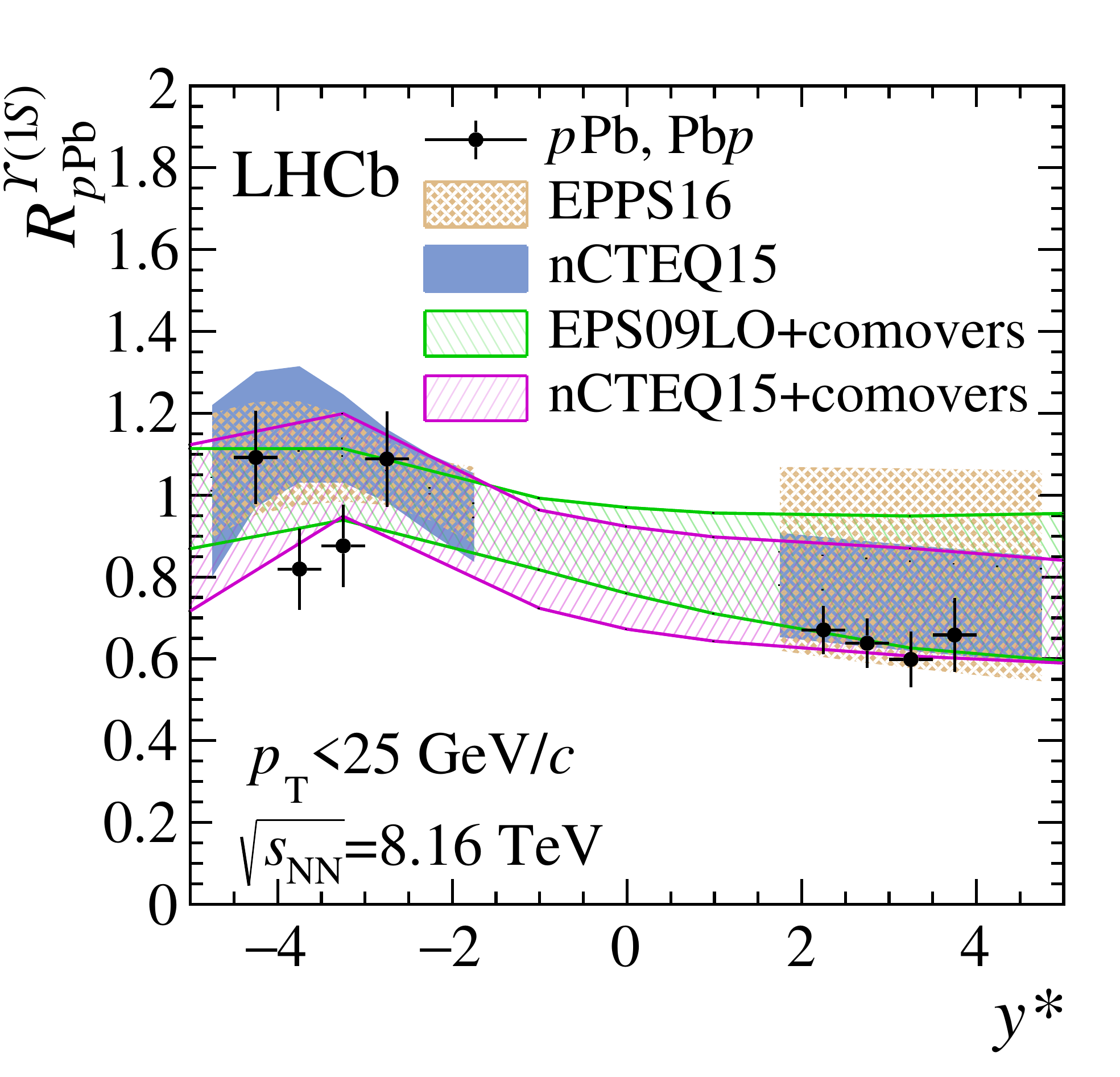}
     \includegraphics[width=0.49\linewidth]{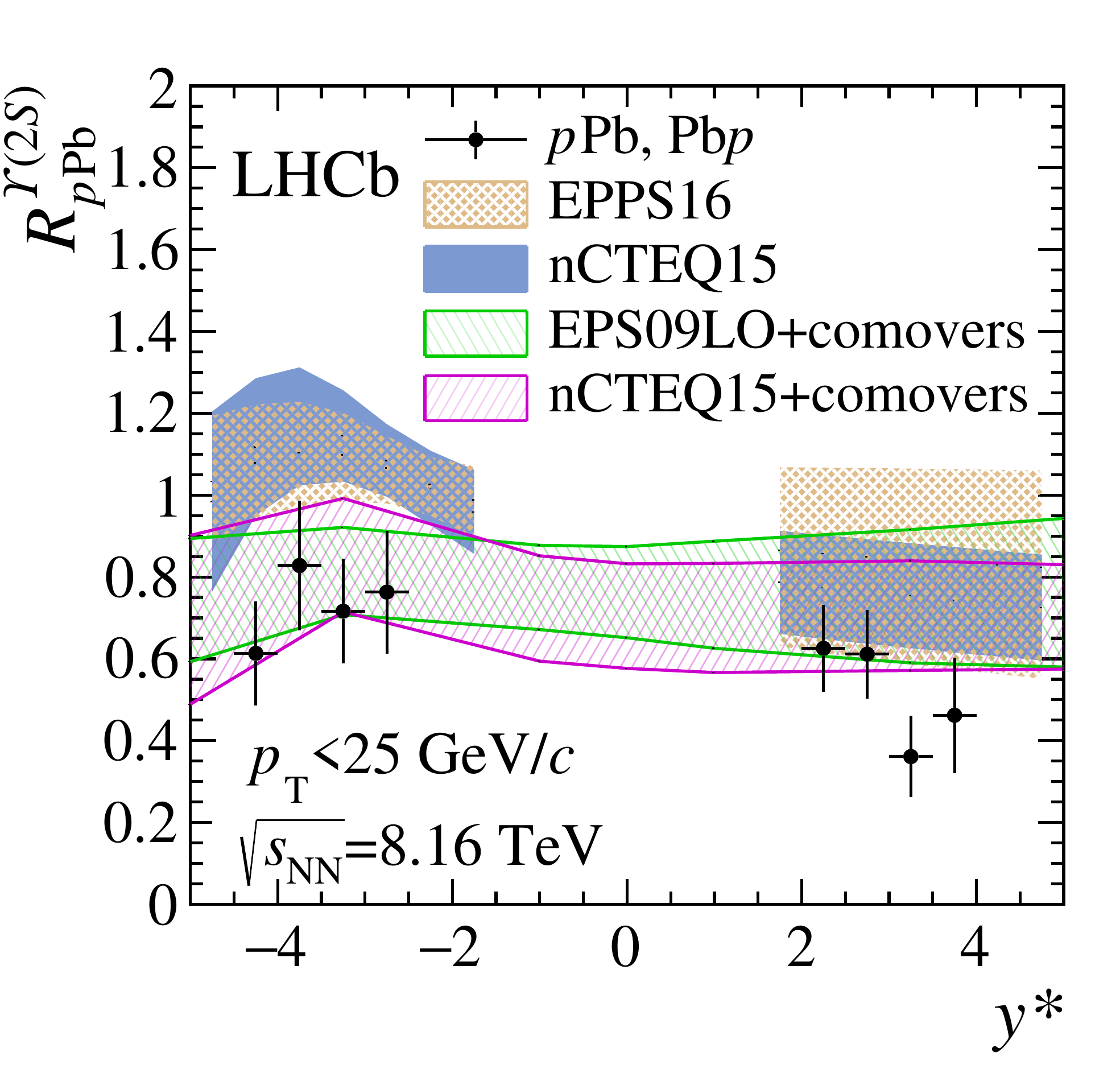}
   \end{center}
   \caption{
Nuclear modification factors of the (left) \OneS\ and (right) \TwoS\ mesons as a function of $y^*$ integrated over \pt 
for the forward and backward samples.
The bands correspond to the theoretical predictions for the 
nCTEQ15 and EPPS16 NNPDF sets, and the comovers model as reported in the text.
    }
       \label{fig:NMF_ystr_1S}
 \end{figure}

\begin{figure}[tb]
   \begin{center}
     \includegraphics[width=0.49\linewidth]{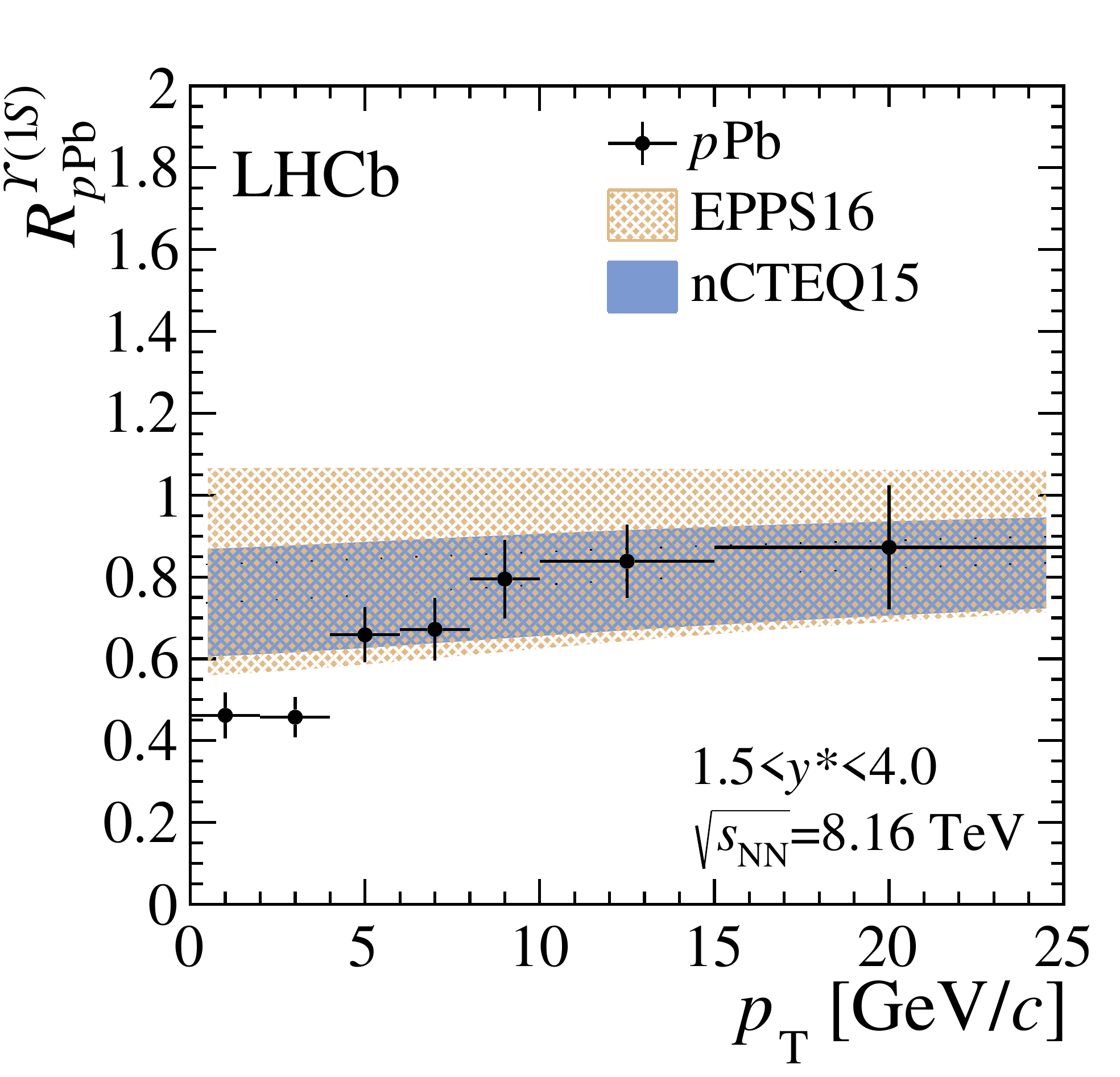}
     \includegraphics[width=0.49\linewidth]{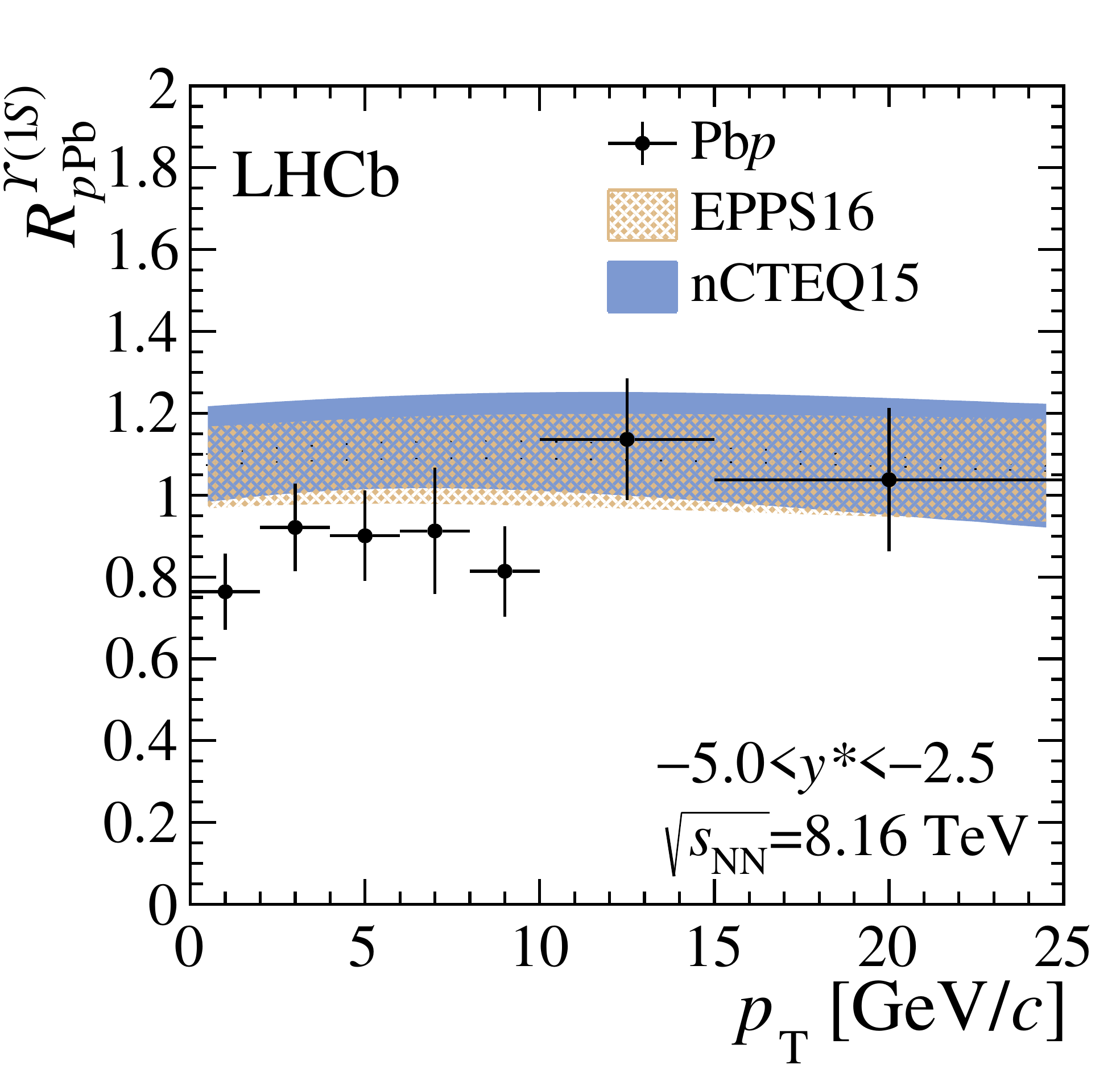}
     
     \includegraphics[width=0.49\linewidth]{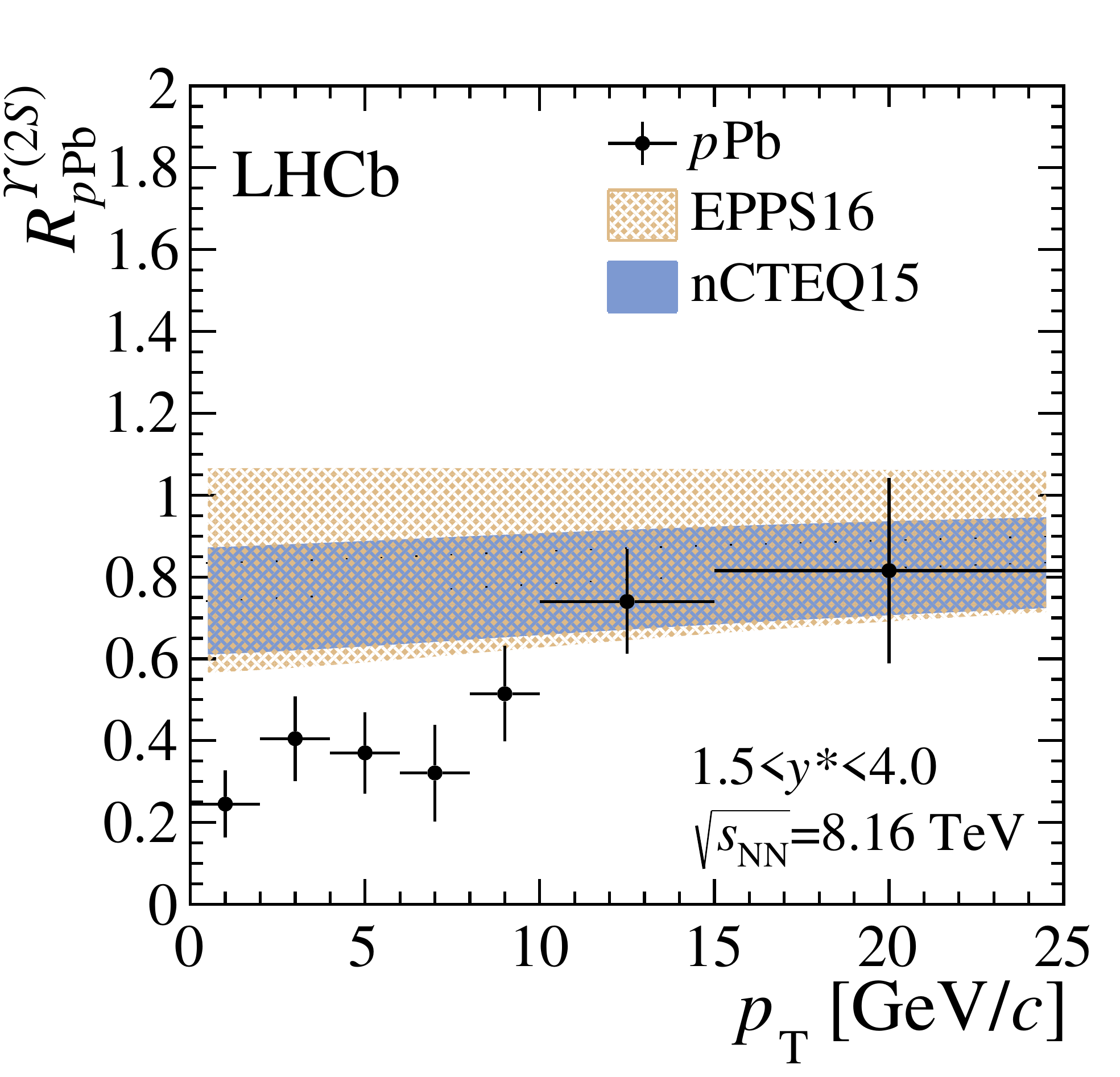}
     \includegraphics[width=0.49\linewidth]{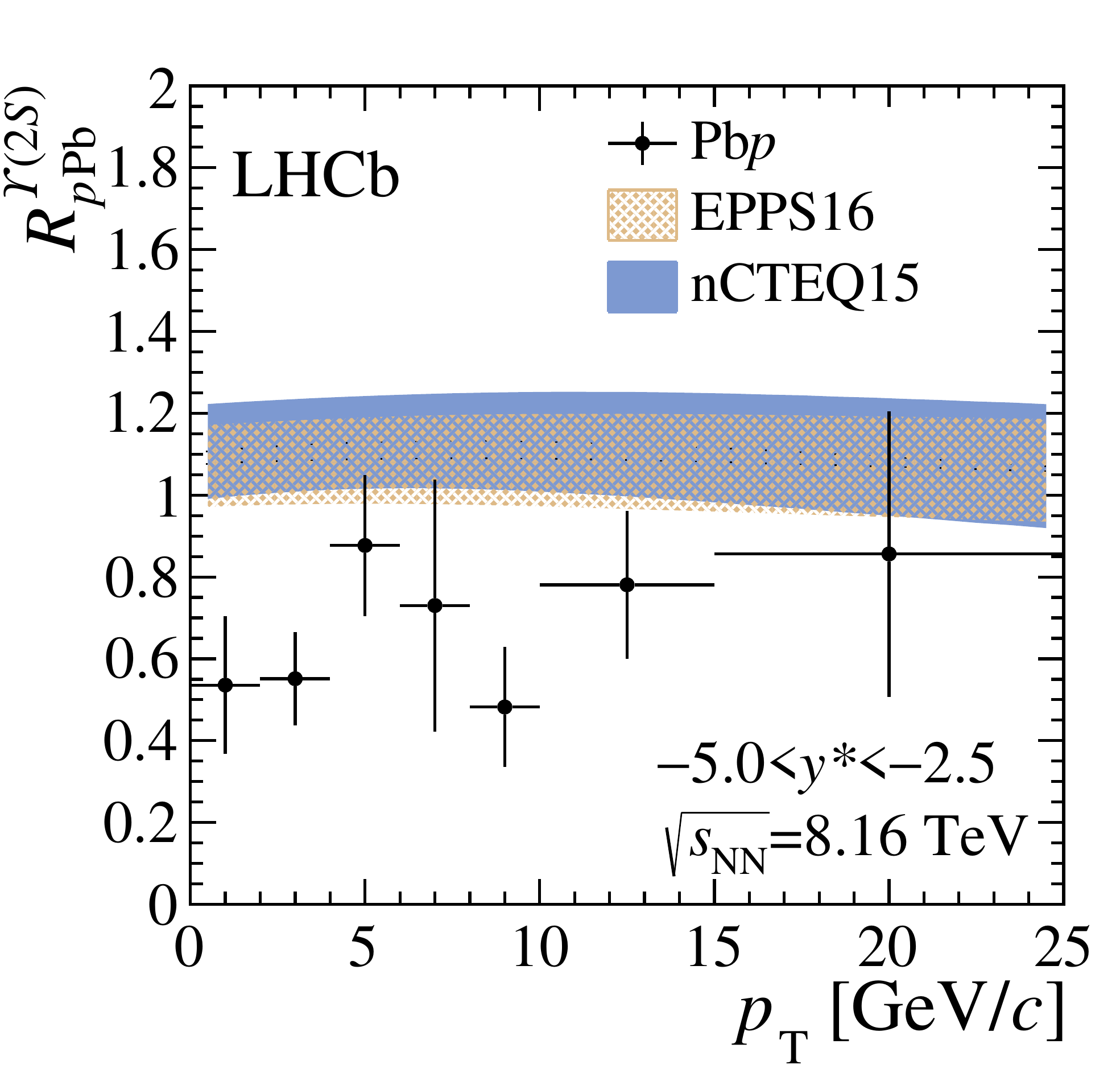}
   \end{center}
   \caption{
Nuclear modification factors of the (top) \OneS\ and (bottom) \TwoS\ mesons as a function of \pt integrated over $y^*$ 
for the (left) forward and  (right) backward samples.
The bands correspond to the theoretical predictions for the 
nCTEQ15 and EPPS16 NNPDF sets  as reported in the text. 
    }
       \label{fig:NMF_pt_1S}
 \end{figure}

To measure the nuclear modification 
factor, a measurement of the $pp$ cross-section at the 
same centre-of-mass energy is needed.
In the absence of a direct measurement, the value
of the $\nS$ cross-section in $pp$ collisions at $\sqrt{s}=8.16$~\tev\
is obtained by interpolating between the values measured in $pp$ collisions by 
\lhcb at 2.76, 7,
 8 and 13~\tev~\cite{LHCb-PAPER-2013-066,LHCb-PAPER-2015-045,LHCb-PAPER-2018-002} using a second-order polynomial function.
The differences between the scale factors obtained using the nominal second-order polynomial fits and alternative fits using exponential functions are assigned as systematic uncertainties on the interpolated cross-sections.
The values of the \OneS\ and \TwoS\  differential cross-sections in \pt ($y^*$) integrated over $y^*$ (\pt) 
 in $pp$ collisions at $\sqrt{s}=8.16$~TeV 
 are shown in Figs.~\ref{fig:CS_ystr_1S} to \ref{fig:CS_pt_1S}, and their numerical values are provided in Appendix~\ref{sec:App}.
The production of both \OneS\ and \TwoS\ is suppressed in the forward $p$Pb
region with respect to the scaled value from $pp$ collisions, as 
already observed in the prompt \jpsi\ measurement~\cite{LHCb-PAPER-2017-014},
while no significant suppression is visible in the backward Pb$p$ region.
The nuclear modification factors are evaluated as functions
of \pt and $y^*$ for the \OneS\ and \TwoS\ states,\footnote{
In the nuclear modification factors, the systematic uncertainty related to branching ratios cancels.} and 
compared with different theoretical calculations:
\begin{enumerate}
\item 
A calculation based on the ``HELAC-Onia" framework~\cite{Lansberg:2016deg,Shao:2015vga,Shao:2012iz}, where the modification 
of the parton flux due to CNM is treated within the 
collinear factorisation framework using two different 
nuclear parton distribution functions (nPDFs), the EPPS16 NNPDF~\cite{Eskola:2018ghi} and 
nCTEQ15 set~\cite{Kovarik:2015cma}.
\item Calculations based on the {\em comovers} model of \nS\ production~\cite{Ferreiro:2018wbd, Ferreiro:2014bia}, which implements final state interaction of the quarkonia states and nuclear parton distribution function modification via EPS09 at leading order~\cite{Eskola:2009uj},
and the nCTEQ15 set already described.
\end{enumerate}
The measurements and the calculations are shown in Figs.~\ref{fig:NMF_ystr_1S} and \ref{fig:NMF_pt_1S}. 
For the \OneS\ state the nuclear modification factor is about 0.5 (0.8) at low \pt in the forward (backward) region, and is consistent with unity for \pt larger than 10 \gevc, as predicted by the models. As a function of rapidity, $R_{p\rm{Pb}}$ is consistent with unity in the 
Pb$p$ region at negative $|y^*|$, while a suppression is observed in the 
$p$Pb region, where it averages around 0.7, consistent with the models analysed.
The nuclear modification factor for \TwoS\ is smaller than \OneS, which is consistent with the comovers models.
The corresponding numerical results can be found in Appendix~\ref{sec:A2}. 
The same trend as for the \OneS\ state is observed for the \TwoS state, 
although the suppression seems more pronounced for the \TwoS state,
as already observed by other experiments~\cite{Aaboud:2017cif},
especially in the backward region.

\begin{figure}[tb]
   \begin{center}
     \includegraphics[width=0.49\linewidth]{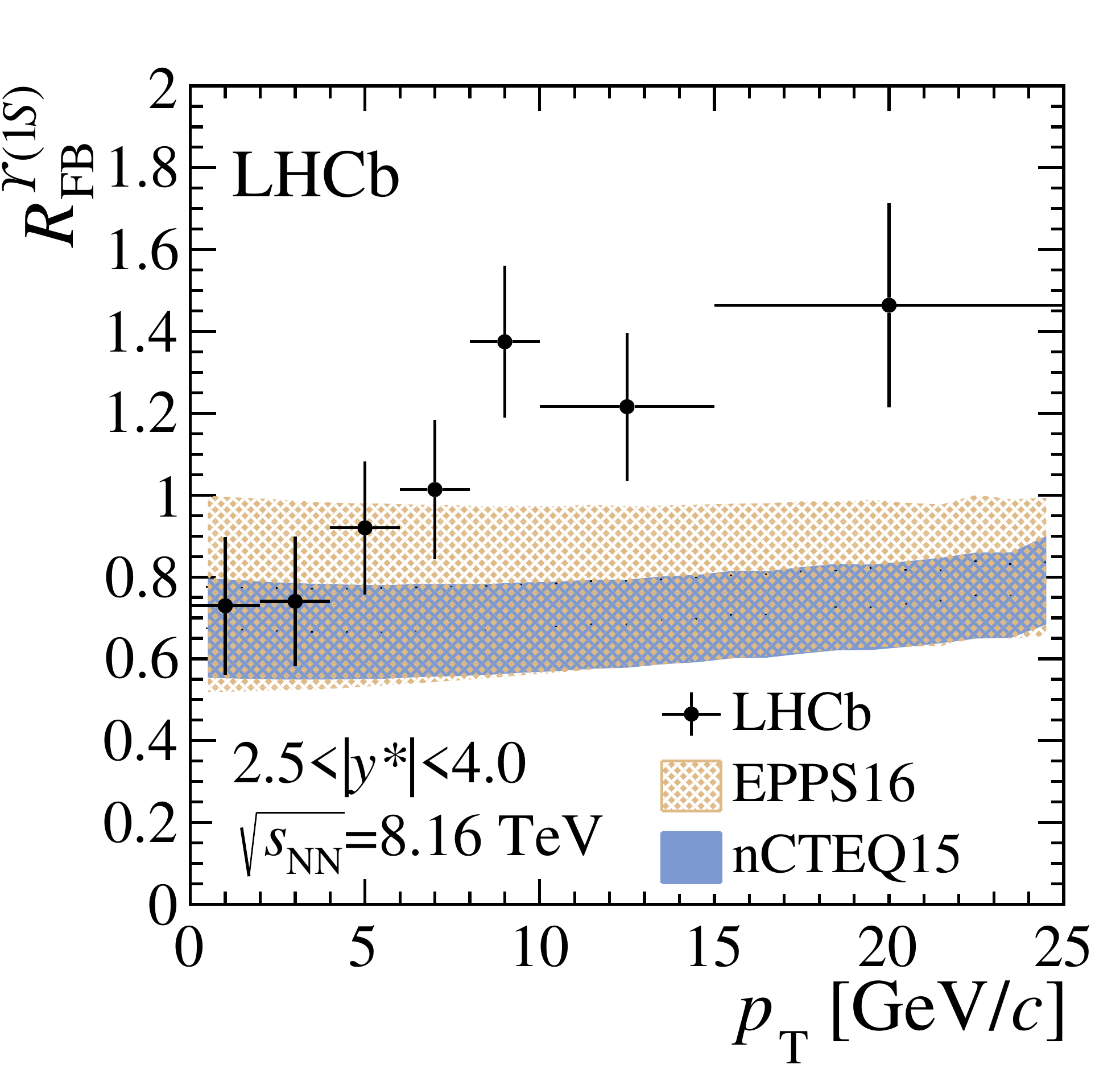}
     \includegraphics[width=0.49\linewidth]{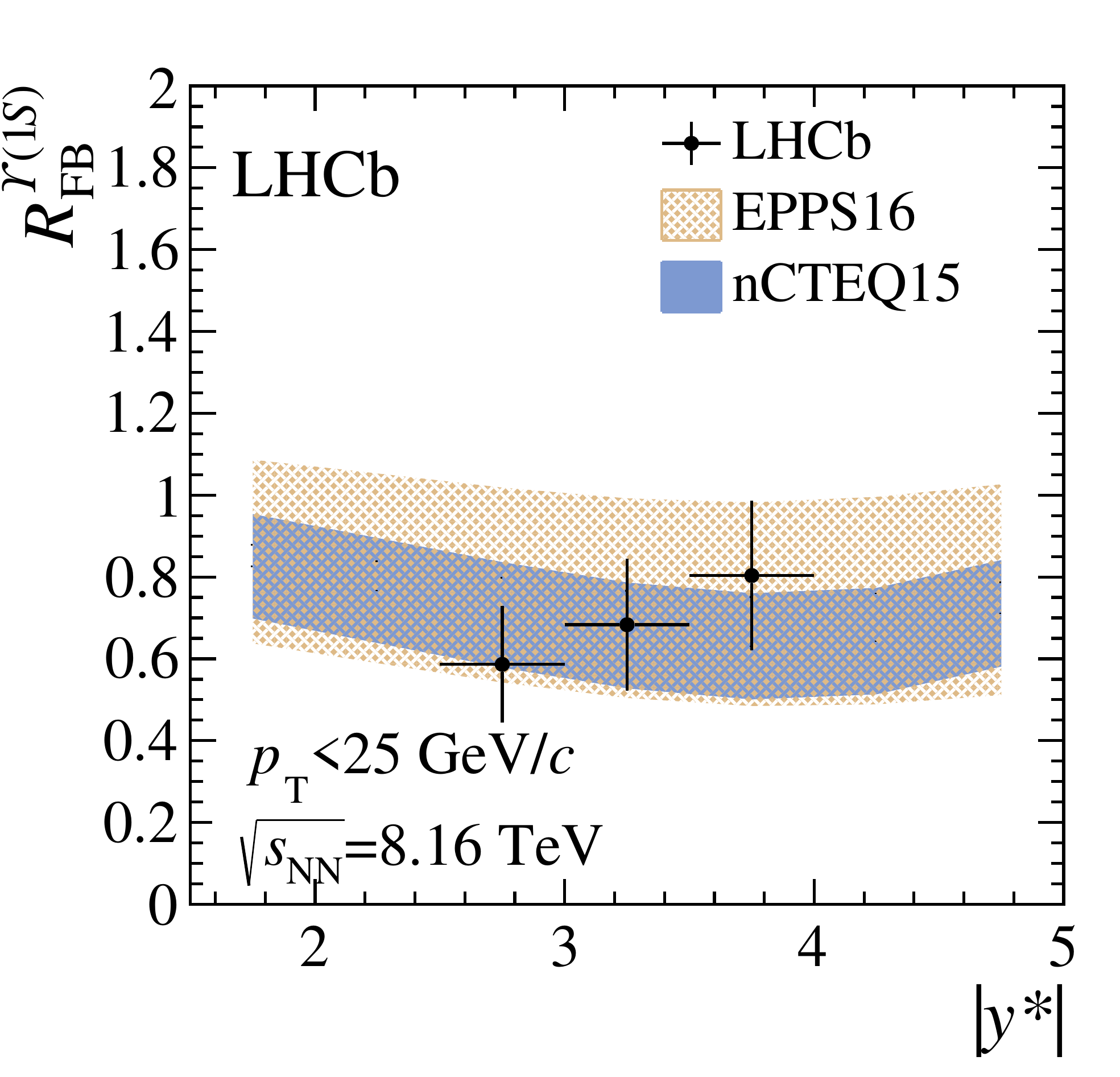}
   \end{center}
   \caption{
Forward-backward ratio for the \OneS\ as a function of  (left) \pt integrated over $y^*$ and  (right) as a function of $|y^*|$ integrated over \pt.
The bands correspond to the theoretical calculations for the 
nCTEQ15 and EPPS16 NNPDF sets  as reported in the text.
    }
       \label{fig:FB_pt_1S}
 \end{figure}

The forward-backward asymmetry is evaluated only for the 
\OneS\ meson as a function 
of \pt and $y^*$, see Fig.~\ref{fig:FB_pt_1S}, whereas for the \TwoS\ meson it is integrated over both $y^*$ and \pt as shown in Fig.~\ref{fig:FB_pt_2S}.
The corresponding numerical results can be found in Appendix~\ref{sec:A3}.\footnote{In the forward-backward ratio, the systematic uncertainty related to branching ratios cancels.}
  \begin{figure}[tb]
   \begin{center}
     \includegraphics[width=0.49\linewidth]{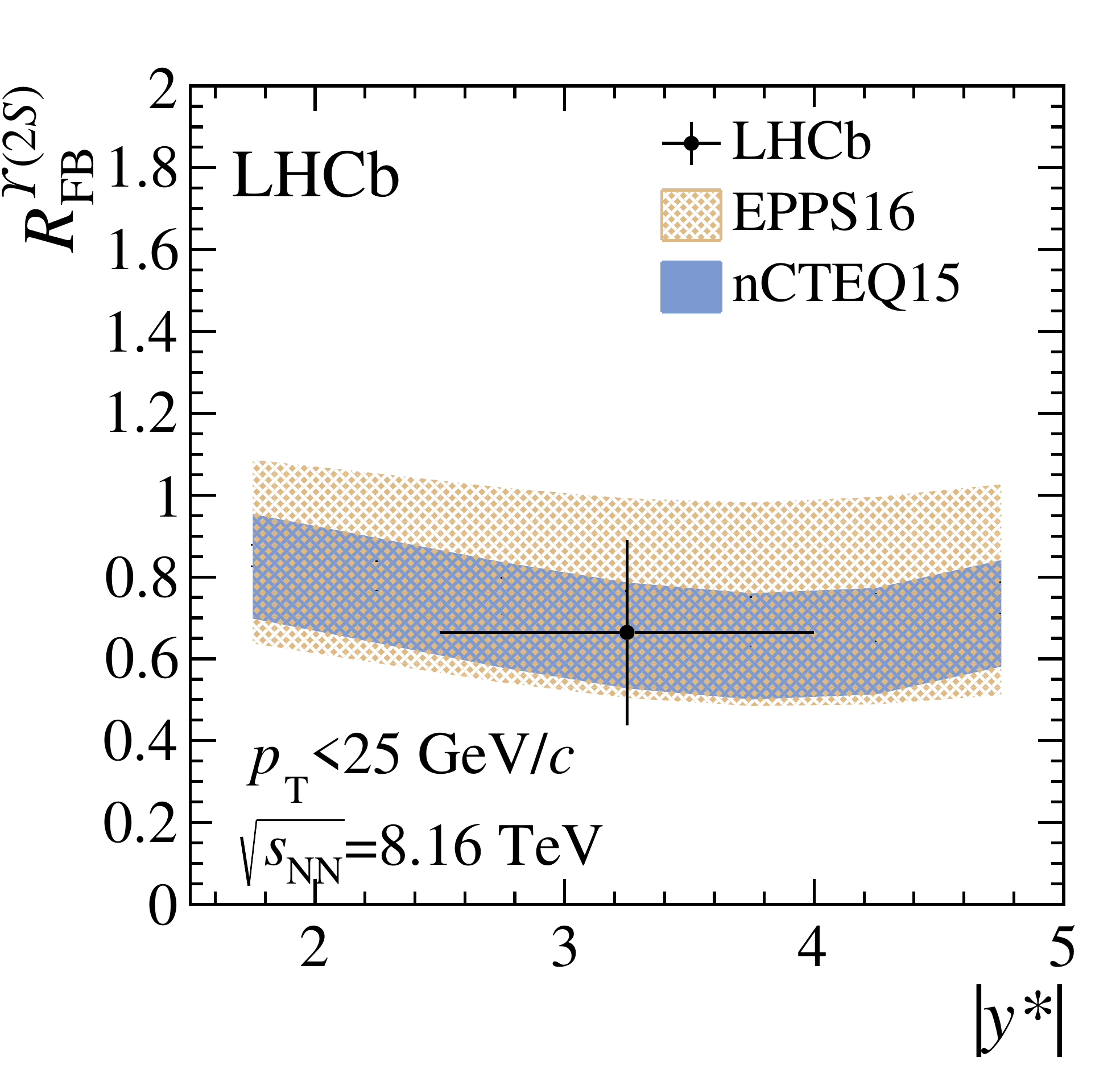}
   \end{center}
   \caption{
Forward-backward ratio for the \TwoS\ compared with 
theoretical calculations for the 
nCTEQ15 and EPPS16 NNPDF sets  as reported in the text.
    }
       \label{fig:FB_pt_2S}
 \end{figure}

The ratio of the cross-sections of $\TwoS$ and $\OneS$ mesons as a function of \pt, integrated over $y^*$, and as function of  $y^*$,  integrated over \pt, are shown in Fig.~\ref{fig:nS}. The corresponding numerical results can be found in Appendix~\ref{sec:A4}. The ratios confirmed a larger suppression for 
the excited states with respect to the ground state observed in proton-lead collisions 
compared to $pp$ collisions~\cite{LHCb-PAPER-2015-045}.
 \begin{figure}[tb]
   \begin{center}
     \includegraphics[width=0.49\linewidth]{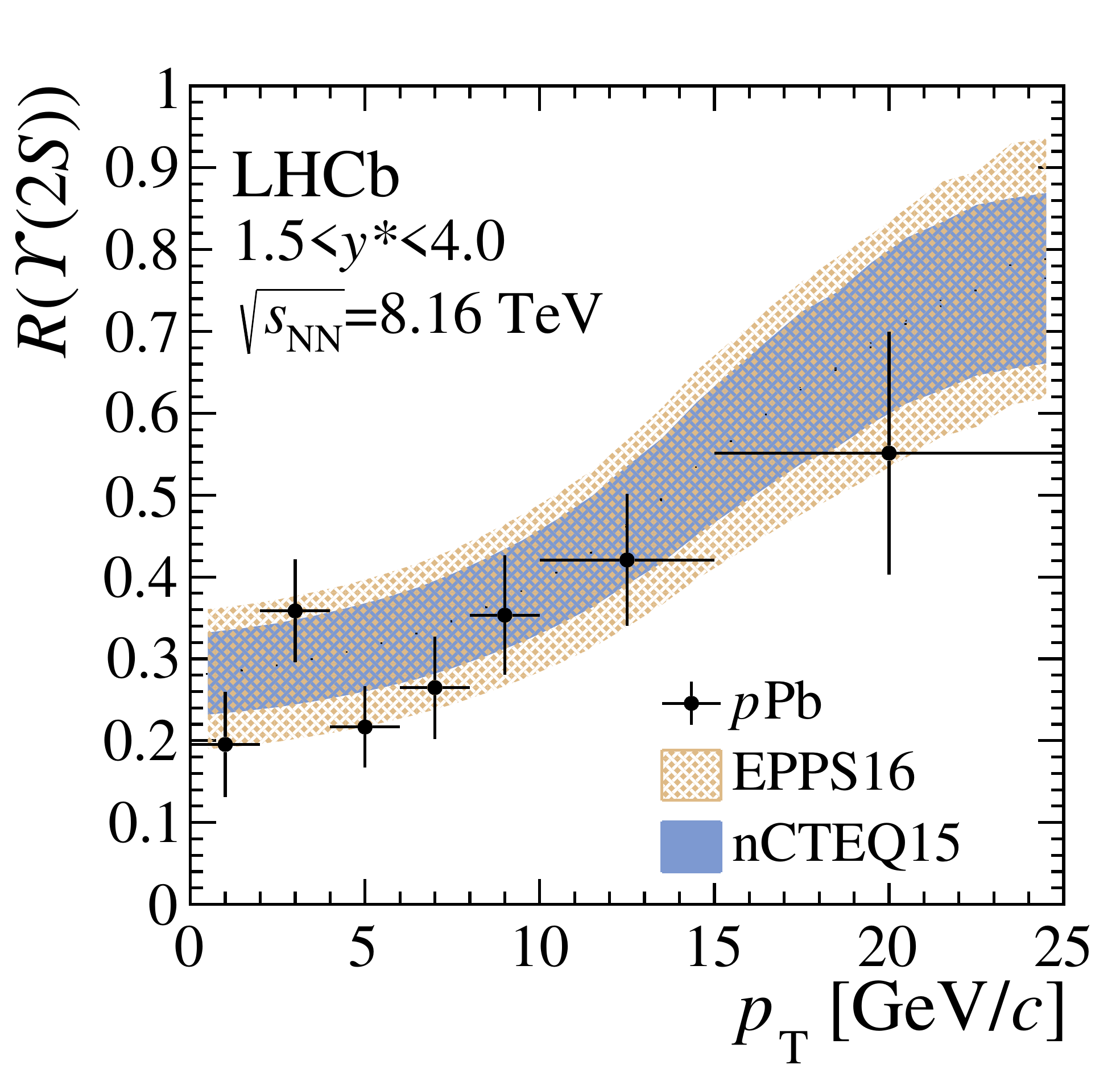}
     \includegraphics[width=0.49\linewidth]{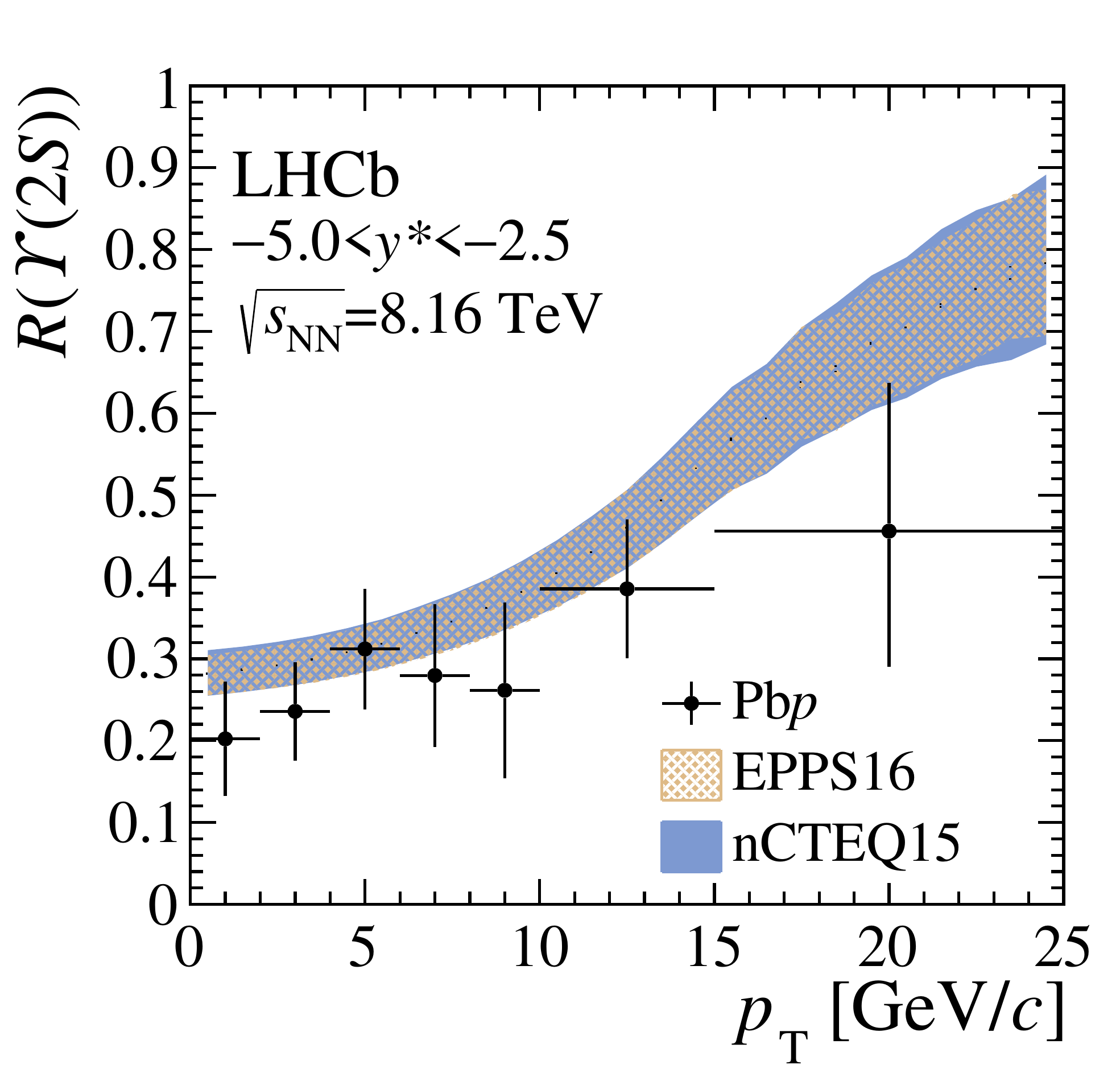}
     \includegraphics[width=0.49\linewidth]{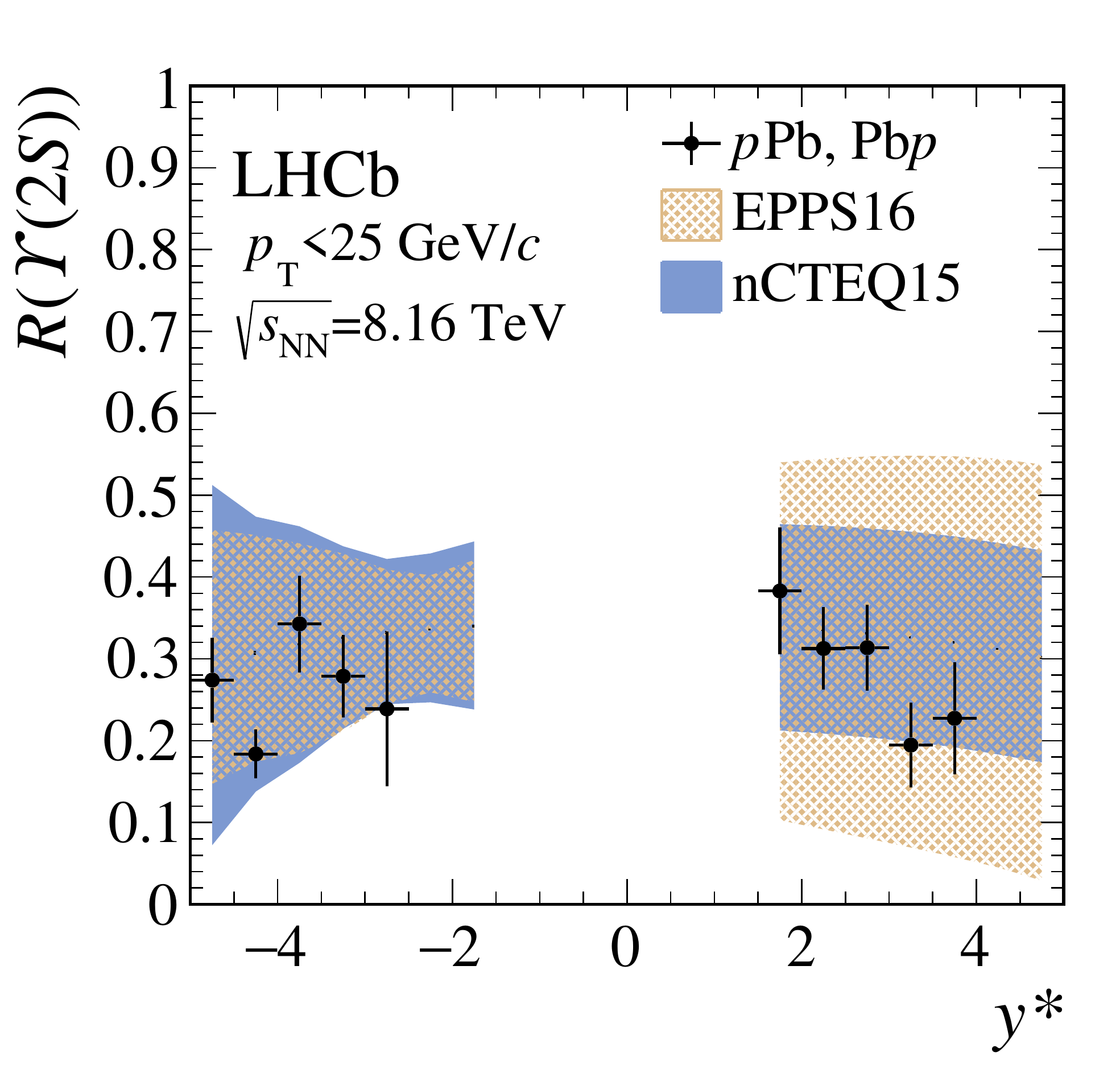}
   \end{center}
   \caption{
       Ratios between \TwoS\ and \OneS\ cross-sections as a function of (top) \pt integrated over $y^*$, 
and as function of (bottom) $y^*$  integrated over \pt, for $p$Pb and Pb$p$ collisions. 
The bands correspond to the theoretical predictions for the 
nCTEQ15 and EPPS16 NNPDF sets as reported in the text.
    }
       \label{fig:nS}
 \end{figure}
For the \ThreeS\ state, due to the limited size of the data sample, only an integral ratio is measured.  
In the determination of the ratio $R(\nS)$, most of the systematic uncertainties cancel, except that related to branching ratios. 

\begin{figure}[tb]
   \begin{center}
     \includegraphics[width=0.49\linewidth]{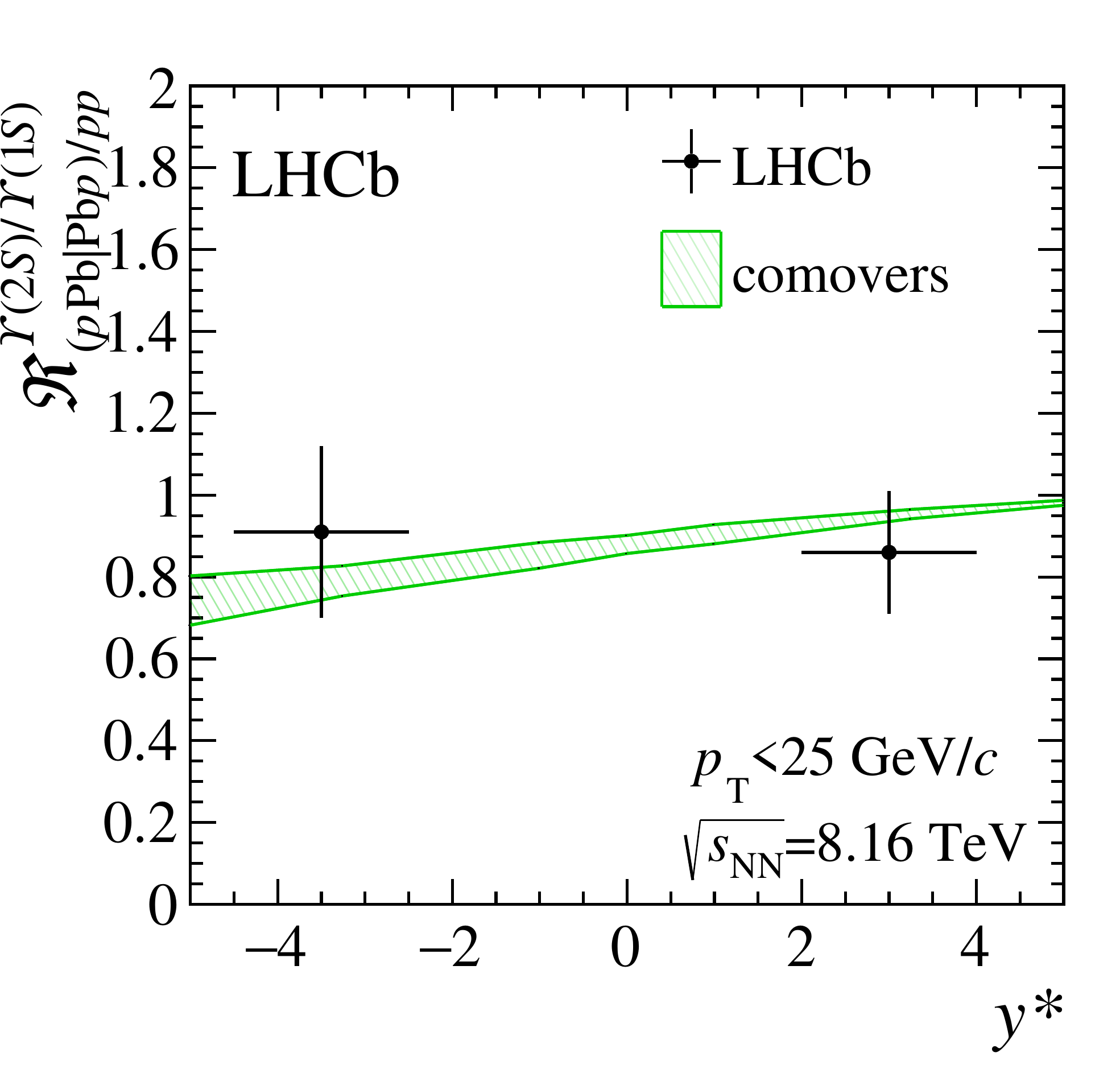}
     \includegraphics[width=0.49\linewidth]{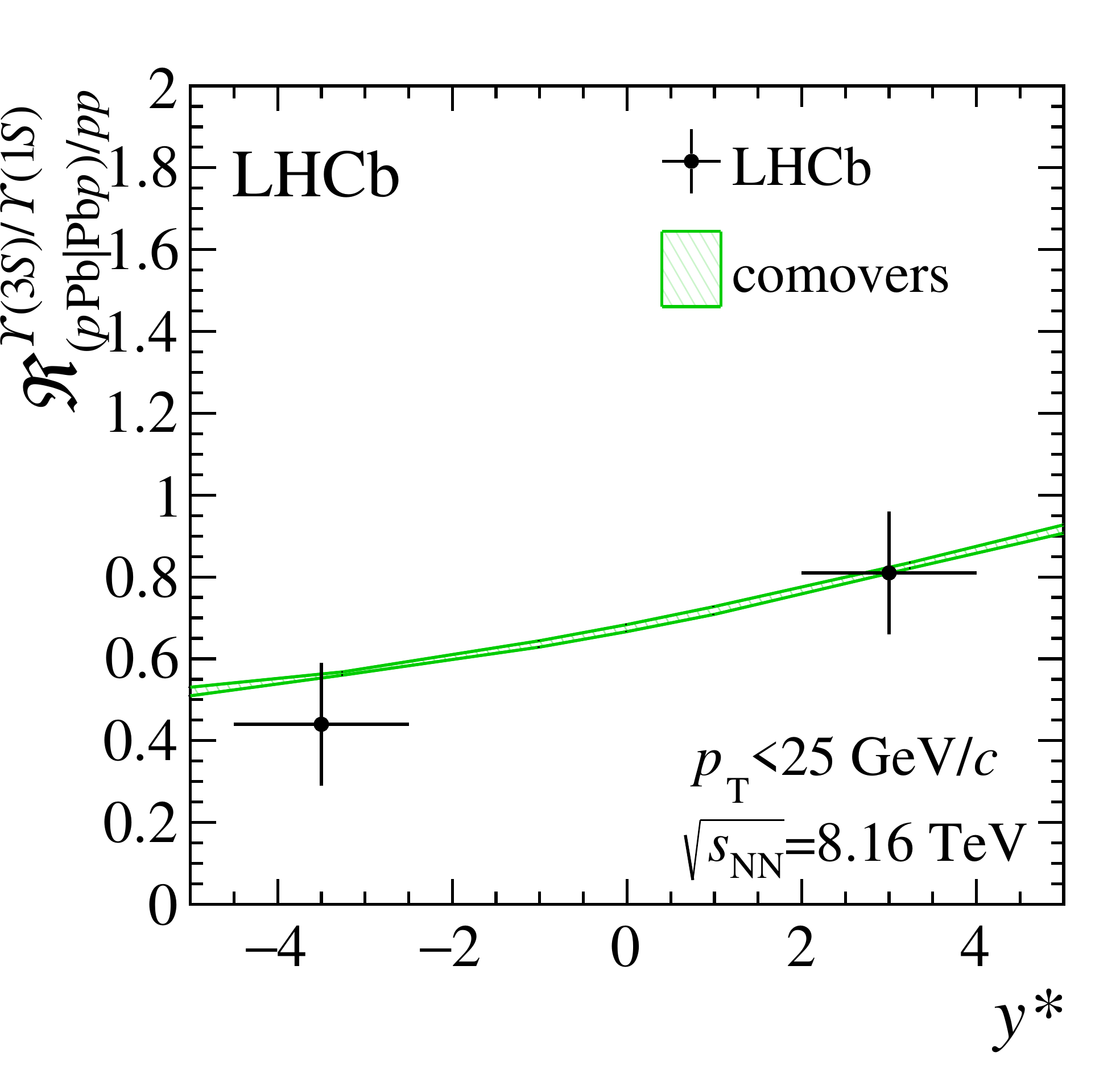}
   \end{center}
   \caption{
      Double ratios for (left) \TwoS\ and (right) \ThreeS. The bands correspond to the theoretical prediction for the 
       comovers model as reported in the text.
    }
       \label{fig:doubleR}
 \end{figure}

The integrated ratios are summarised in Table~\ref{tab:nSRatio}, where values are also reported for $pp$ collisions. The corresponding double-ratio results are shown in Fig.~\ref{fig:doubleR} (left), together with the comovers model calculations, and the numerical results are
 
\begin{figure}[ht]
   \begin{center}
     \includegraphics[width=0.49\linewidth]{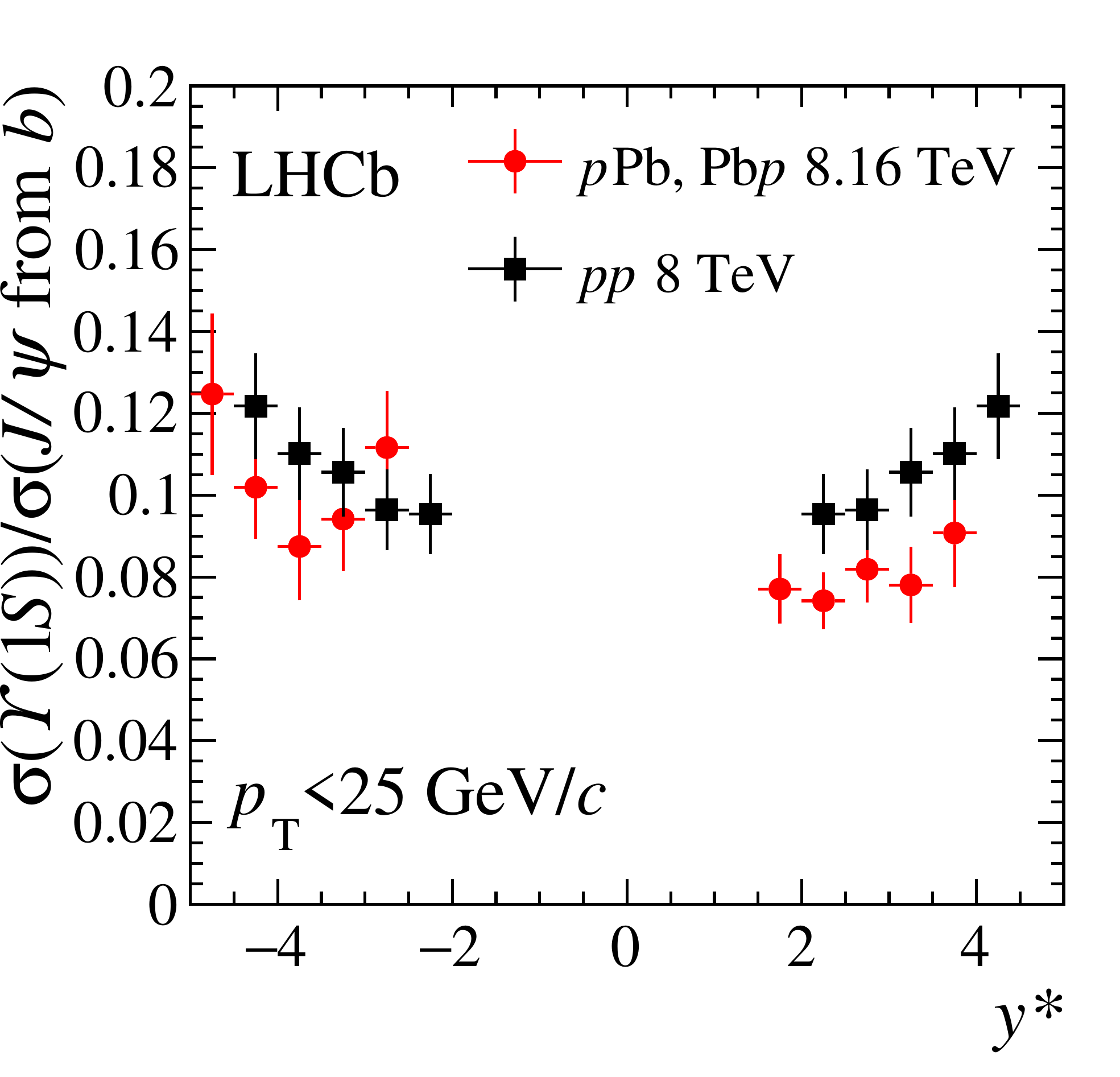}
   \end{center}
   \caption{
       Ratio of \OneS\ to nonprompt \jpsi\ cross-sections as a function of $y^*$  integrated over \pt, for $p$Pb and Pb$p$ collisions. 
    }
       \label{fig:OneSToJpsiRatio}
 \end{figure}

 \begin{table}
 \caption{
 Ratio $R(\nS)$ in $pp$, $p$Pb, and Pb$p$ samples. The uncertainties are combinations of statistical and systematical components. 
    }
\begin{center}\begin{tabular}{rr@{$\,<\,$}c@{$\,<\,$}lcc}  
     \hline        
    \multicolumn{4}{c}{Sample}          & $R(\TwoS)$& $R(\ThreeS)$      \\              
     \hline          
$pp$&$2.0$ & $y^*$& $\phantom{-}4.0$ & $0.328\pm 0.004$& $0.137\pm 0.002$  \\
$pp$&$-4.5$ & $y^*$& $-2.5$ & $0.325\pm 0.004$& $0.137\pm 0.002$  \\
$p$Pb&$2.0$ & $y^*$& $\phantom{-}4.0$ & $0.282\pm 0.049$ & $0.111\pm 0.021 $ \\
Pb$p$&$-4.5$ & $y^*$& $-2.5$ & $0.296\pm 0.070$ & $ 0.060\pm 0.016$\\
     \hline                                                                                                                               
   \end{tabular}\end{center}   
   \label{tab:nSRatio}
   \end{table}
\begin{align*}
\mathfrak{R}^{\TwoS/\OneS}_{p{\rm Pb}/pp} & =  0.86 \pm 0.15,\\
\mathfrak{R}^{\ThreeS/\OneS}_{p{\rm Pb}/pp}& = 0.81 \pm 0.15, \\
\mathfrak{R}^{\TwoS/\OneS}_{{\rm Pb}p/pp}& = 0.91 \pm 0.21, \\
\mathfrak{R}^{\ThreeS/\OneS}_{{\rm Pb}p/pp}& = 0.44 \pm 0.15.\\
\end{align*}
For the double ratio of the \ThreeS\ over \OneS\ in the backward 
a clear indication of stronger suppression is observed, in agreement with the 
comovers model as shown in Fig.~\ref{fig:doubleR} (right).
The ratio of the \OneS\ and nonprompt \jpsi\ cross-sections in $p$Pb and Pb$p$ collisions is also measured, where the 
the nonprompt \jpsi\ cross-section was measured previously by 
LHCb~\cite{LHCb-PAPER-2017-014} using the same data sample. 
The ratio is shown in Fig.~\ref{fig:OneSToJpsiRatio}  compared to the corresponding result observed in $pp$ collisions. The numerical results are 
reported in Appendix~\ref{sec:A5}. 
A small suppression is visible, which could be attributed to 
final-state CNM effects.
More data are needed in order to have a more definite 
indication of a different suppression mechanism for  
bottomonium and open beauty, such as \OneS\ and nonprompt $J/\psi$ states, as indicated by Refs.~\cite{PhysRevC.97.014908,Winn:2016amg}.

\section{Summary}

The production of \nS states is studied in 
proton-lead collisions at \mbox{$\sqsnn=8.16~\TeV$} using data collected
by the LHCb detector in 2016.
The cross-sections, nuclear modification factors 
and forward-backward ratios are measured double-differentially (\OneS) and single-differentially (\TwoS). 
The ratios of the production cross-sections of the different \nS\ 
states are also measured as functions of transverse momentum and 
rapidity in the nucleon-nucleon centre-of-mass frame. 
The results are  consistent with previous observations and with the theoretical model calculations, 
indicating some suppression of \nS\ production
in proton-lead collisions, more pronounced for the excited $\PUpsilon$ states.


\section*{Acknowledgements}
%
%
\noindent We thank the theorists who provided predictions for 
our measurements: J.-P.~Lansberg, H.-S.~Shao and E.~Gonzalez-Ferreiro. 
We express our gratitude to our colleagues in the CERN
accelerator departments for the excellent performance of the LHC. We
thank the technical and administrative staff at the LHCb
institutes.
We acknowledge support from CERN and from the national agencies:
CAPES, CNPq, FAPERJ and FINEP (Brazil); 
MOST and NSFC (China); 
CNRS/IN2P3 (France); 
BMBF, DFG and MPG (Germany); 
INFN (Italy); 
NWO (Netherlands); 
MNiSW and NCN (Poland); 
MEN/IFA (Romania); 
MSHE (Russia); 
MinECo (Spain); 
SNSF and SER (Switzerland); 
NASU (Ukraine); 
STFC (United Kingdom); 
NSF (USA).
We acknowledge the computing resources that are provided by CERN, IN2P3
(France), KIT and DESY (Germany), INFN (Italy), SURF (Netherlands),
PIC (Spain), GridPP (United Kingdom), RRCKI and Yandex
LLC (Russia), CSCS (Switzerland), IFIN-HH (Romania), CBPF (Brazil),
PL-GRID (Poland) and OSC (USA).
We are indebted to the communities behind the multiple open-source
software packages on which we depend.
Individual groups or members have received support from
AvH Foundation (Germany);
EPLANET, Marie Sk\l{}odowska-Curie Actions and ERC (European Union);
ANR, Labex P2IO and OCEVU, and R\'{e}gion Auvergne-Rh\^{o}ne-Alpes (France);
Key Research Program of Frontier Sciences of CAS, CAS PIFI, and the Thousand Talents Program (China);
RFBR, RSF and Yandex LLC (Russia);
GVA, XuntaGal and GENCAT (Spain);
the Royal Society
and the Leverhulme Trust (United Kingdom);
Laboratory Directed Research and Development program of LANL (USA).

\clearpage 

{\noindent\normalfont\bfseries\Large Appendices}   

\appendix  

\section{Cross-section}   
\label{sec:A1}

 Tables~\ref{tab:A1} and~\ref{tab:A2} list the double-differential cross-section for \OneS\ in $p$Pb forward and backward samples.  Tables~\ref{tab:A3} and~\ref{tab:A4} list the differential cross-section for \OneS\ in bins of transverse momentum and rapidity. 
The corresponding values for the \TwoS\ state are listed in Tables~\ref{tab:A5} and~\ref{tab:A6}.
In all tables, the quoted uncertainties are the sum in quadrature of the statistical and systematic components.
%
%
%
%
%
%
%

 \begin{table}[!ht]
   \caption{ \OneS\   production cross-section in $p$Pb, as a function of $\pt$ and $y^*$. }
%
%
 \begin{center}\begin{tabular}{r@{$\,<\,$}c@{$\,<\,$}lr@{$\,<\,$}c@{$\,<\,$}lr@{\,$\pm$\,}l}
      \hline
       \multicolumn{3}{c}{$\pt$  [\GeVc]} & \multicolumn{3}{c}{$y^*$ } & \multicolumn{2}{c}{$\dfrac{{\rm d}^2\sigma}{{\rm d}p_{\rm T}{\rm d}y^*}$ [nb/(\GeVc)]}  \\
      \hline
      0 & $\pt$ & 2 & 1.5 & $y^*$ & 2.0 & ~~~~~~~~644 & 142  \\
0 & $\pt$ & 2 & 2.0 & $y^*$ & 2.5 & 656 & 106  \\
0 & $\pt$ & 2 & 2.5 & $y^*$ & 3.0 & 641 & 119  \\
0 & $\pt$ & 2 & 3.0 & $y^*$ & 3.5 & 486 & 92  \\
0 & $\pt$ & 2 & 3.5 & $y^*$ & 4.0 & 345 & 50  \\
2 & $\pt$ & 4 & 1.5 & $y^*$ & 2.0 & 1134 & 227  \\
2 & $\pt$ & 4 & 2.0 & $y^*$ & 2.5 & 1312 & 163  \\
2 & $\pt$ & 4 & 2.5 & $y^*$ & 3.0 & 1226 & 171  \\
2 & $\pt$ & 4 & 3.0 & $y^*$ & 3.5 & 794 & 129  \\
2 & $\pt$ & 4 & 3.5 & $y^*$ & 4.0 & 765 & 147  \\
4 & $\pt$ & 6 & 1.5 & $y^*$ & 2.0 & 1162 & 184  \\
4 & $\pt$ & 6 & 2.0 & $y^*$ & 2.5 & 1130 & 128  \\
4 & $\pt$ & 6 & 2.5 & $y^*$ & 3.0 & 1121 & 135  \\
4 & $\pt$ & 6 & 3.0 & $y^*$ & 3.5 & 915 & 147  \\
4 & $\pt$ & 6 & 3.5 & $y^*$ & 4.0 & 586 & 132  \\
6 & $\pt$ & 8 & 1.5 & $y^*$ & 2.0 & 908 & 171  \\
6 & $\pt$ & 8 & 2.0 & $y^*$ & 2.5 & 851 & 135  \\
6 & $\pt$ & 8 & 2.5 & $y^*$ & 3.0 & 690 & 106  \\
6 & $\pt$ & 8 & 3.0 & $y^*$ & 3.5 & 625 & 111  \\
6 & $\pt$ & 8 & 3.5 & $y^*$ & 4.0 & 570 & 131  \\
8 & $\pt$ & 10 & 1.5 & $y^*$ & 2.0 & 651 & 145  \\
8 & $\pt$ & 10 & 2.0 & $y^*$ & 2.5 & 474 & 83  \\
8 & $\pt$ & 10 & 2.5 & $y^*$ & 3.0 & 525 & 79  \\
8 & $\pt$ & 10 & 3.0 & $y^*$ & 3.5 & 384 & 71  \\
8 & $\pt$ & 10 & 3.5 & $y^*$ & 4.0 & 285 & 79  \\
10 & $\pt$ & 15 & 1.5 & $y^*$ & 2.0 & 224 & 61  \\
10 & $\pt$ & 15 & 2.0 & $y^*$ & 2.5 & 237 & 36  \\
10 & $\pt$ & 15 & 2.5 & $y^*$ & 3.0 & 190 & 30  \\
10 & $\pt$ & 15 & 3.0 & $y^*$ & 3.5 & 140 & 28  \\
10 & $\pt$ & 25 & 3.5 & $y^*$ & 4.0 & 33 & 11  \\
15 & $\pt$ & 25 & 1.5 & $y^*$ & 2.0 & 62 & 20  \\
15 & $\pt$ & 25 & 2.0 & $y^*$ & 2.5 & 41 & 9  \\
15 & $\pt$ & 25 & 2.5 & $y^*$ & 3.0 & 29 & 8  \\
15 & $\pt$ & 25 & 3.0 & $y^*$ & 3.5 & 23 & 7  \\
       \hline
     \end{tabular}\end{center}
   \label{tab:A1}
   \end{table}      
%
%
%
 \clearpage

 \begin{table}[!ht]
   \caption{\OneS\  production cross-section in Pb$p$, as a function of $\pt$ and $y^*$. }
 \begin{center}\begin{tabular}{r@{$\,<\,$}c@{$\,<\,$}lr@{$\,<\,$}c@{$\,<\,$}lr@{\,$\pm$\,}l}
      \hline
       \multicolumn{3}{c}{$\pt$  [\GeVc]} & \multicolumn{3}{c}{$y^*$ } & \multicolumn{2}{c}{$\dfrac{{\rm d}^2\sigma}{{\rm d}p_{\rm T}{\rm d}y^*}$ [\nb/(\GeVc)]}  \\
      \hline
      0 & $\pt$ & 2 & $-$3.0 & $y^*$ & $-$2.5 & ~~~~~~~~839 & 130  \\
0 & $\pt$ & 2 & $-$3.5 & $y^*$ & $-$3.0 & 740 & 114  \\
0 & $\pt$ & 2 & $-$4.0 & $y^*$ & $-$3.5 & 627 & 129  \\
0 & $\pt$ & 2 & $-$4.5 & $y^*$ & $-$4.0 & 523 & 90  \\
0 & $\pt$ & 2 & $-$5.0 & $y^*$ & $-$4.5 & 318 & 77  \\
2 & $\pt$ & 4 & $-$3.0 & $y^*$ & $-$2.5 & 1661 & 228  \\
2 & $\pt$ & 4 & $-$3.5 & $y^*$ & $-$3.0 & 1478 & 225  \\
2 & $\pt$ & 4 & $-$4.0 & $y^*$ & $-$3.5 & 1366 & 216  \\
2 & $\pt$ & 4 & $-$4.5 & $y^*$ & $-$4.0 & 913 & 164  \\
2 & $\pt$ & 4 & $-$5.0 & $y^*$ & $-$4.5 & 503 & 99  \\
4 & $\pt$ & 6 & $-$3.0 & $y^*$ & $-$2.5 & 1538 & 243  \\
4 & $\pt$ & 6 & $-$3.5 & $y^*$ & $-$3.0 & 1199 & 204  \\
4 & $\pt$ & 6 & $-$4.0 & $y^*$ & $-$3.5 & 869 & 165  \\
4 & $\pt$ & 6 & $-$4.5 & $y^*$ & $-$4.0 & 895 & 152  \\
4 & $\pt$ & 6 & $-$5.0 & $y^*$ & $-$4.5 & 406 & 107  \\
6 & $\pt$ & 8 & $-$3.0 & $y^*$ & $-$2.5 & 1313 & 222  \\
6 & $\pt$ & 8 & $-$3.5 & $y^*$ & $-$3.0 & 859 & 149  \\
6 & $\pt$ & 8 & $-$4.0 & $y^*$ & $-$3.5 & 518 & 99  \\
6 & $\pt$ & 8 & $-$4.5 & $y^*$ & $-$4.0 & 242 & 69  \\
6 & $\pt$ & 8 & $-$5.0 & $y^*$ & $-$4.5 & 240 & 45  \\
8 & $\pt$ & 10 & $-$3.0 & $y^*$ & $-$2.5 & 608 & 156  \\
8 & $\pt$ & 10 & $-$3.5 & $y^*$ & $-$3.0 & 449 & 83  \\
8 & $\pt$ & 10 & $-$4.0 & $y^*$ & $-$3.5 & 263 & 53  \\
8 & $\pt$ & 10 & $-$4.5 & $y^*$ & $-$4.0 & 88 & 40  \\
8 & $\pt$ & 10 & $-$5.0 & $y^*$ & $-$4.5 & 82 & 47  \\
10 & $\pt$ & 15 & $-$3.0 & $y^*$ & $-$2.5 & 336 & 75  \\
10 & $\pt$ & 15 & $-$3.5 & $y^*$ & $-$3.0 & 181 & 33  \\
10 & $\pt$ & 25 & $-$4.0 & $y^*$ & $-$3.5 & 39 & 7  \\
10 & $\pt$ & 25 & $-$4.5 & $y^*$ & $-$4.0 & 24 & 5  \\
10 & $\pt$ & 25 & $-$5.0 & $y^*$ & $-$4.5 & 9 & 6  \\
15 & $\pt$ & 25 & $-$3.0 & $y^*$ & $-$2.5 & 43 & 15  \\
15 & $\pt$ & 25 & $-$3.5 & $y^*$ & $-$3.0 & 26 & 8  \\
       \hline
     \end{tabular}\end{center}
   \label{tab:A2}
   \end{table}  

\clearpage

 \begin{table}[!ht]
   \caption{ \OneS\  production cross-section in $p$Pb and Pb$p$, as a function of $\pt$. }
 \begin{center}\begin{tabular}{r@{\,$<$\,}c@{\,$<$\,}lr@{\,$\pm$\,}lr@{\,$\pm$\,}l}
      \hline
       \multicolumn{3}{c}{$\pt$  (\GeVc)} & \multicolumn{2}{c}{$\dfrac{{\rm d}\sigma}{{\rm d}p_{\rm T}}$ in $p$Pb [nb/(\GeVc)]} & \multicolumn{2}{c}{$\dfrac{{\rm d}\sigma}{{\rm d}p_{\rm T}}$ in Pb$p$ [nb/(\GeVc)]} \\    
      \hline
      0 & $\pt$ & 2 & ~~~~~~~~~~1409 & 164 & ~~~~~~~~~~1570 & 234  \\
2 & $\pt$ & 4 & 2683 & 287  & 3040 & 437 \\
4 & $\pt$ & 6 & 2500 & 268  & 2349 & 341  \\
6 & $\pt$ & 8 & 1693 & 197  & 1461 & 203  \\
8 & $\pt$ & 10 & 1145 & 142   & 721 & 107  \\
10 & $\pt$ & 15 & 495 & 61 & 338 & 48   \\
15 & $\pt$ & 25 & 81 & 13  & 44 & 9  \\
       \hline
     \end{tabular}\end{center}
   \label{tab:A3}
   \end{table}

 \begin{table}[!ht]
   \caption{ \OneS\  production cross-section in $p$Pb and Pb$p$, as a function of $y^*$. }
 \begin{center}\begin{tabular}{r@{\,$<$\,}c@{\,$<$\,}lr@{\,$\pm$\,}l}
      \hline
        \multicolumn{3}{c}{$y^*$ }  & \multicolumn{2}{c}{$\dfrac{{\rm d}\sigma}{{\rm d}y^*}$ [nb]}  \\    
      \hline     
  $-$5.0 & $y^*$ & $-$4.5 & 4050 & 646  \\
  $-$4.5 & $y^*$ & $-$4.0 & 5572 & 720  \\
  $-$4.0 & $y^*$ & $-$3.5 & 7333 & 1109  \\
  $-$3.5 & $y^*$ & $-$3.0 & 10300 & 1399  \\
              $-$3.0 & $y^*$ & $-$2.5 & 15531 & 1868  \\
        1.5 & $y^*$ & 2.0 & 11500 & 1266  \\
  2.0 & $y^*$ & 2.5 & 10175 & 955  \\
  2.5 & $y^*$ & 3.0 & 9107 & 908  \\
  3.0 & $y^*$ & 3.5 & 7038 & 843  \\
  3.5 & $y^*$ & 4.0 & 5891 & 862  \\
       \hline
     \end{tabular}\end{center}
   \label{tab:A4}
   \end{table}  

   \clearpage
   
 \begin{table}[!ht]
   \caption{\TwoS\  production cross-section in $p$Pb and Pb$p$, as a function of $\pt$. }
 \begin{center}\begin{tabular}{r@{\,$<$\,}c@{\,$<$\,}lr@{\,$\pm$\,}lr@{\,$\pm$\,}l}
      \hline
       \multicolumn{3}{c}{$\pt$  [\GeVc]} & \multicolumn{2}{c}{$\dfrac{{\rm d}\sigma}{{\rm d}p_{\rm T}}$ in $p$Pb [nb/(\GeVc)]}  & \multicolumn{2}{c}{$\dfrac{{\rm d}\sigma}{{\rm d}p_{\rm T}}$ in Pb$p$ [nb/(\GeVc)]} \\    
      \hline
      0 & $\pt$ & 2 & ~~~~~~~~~~275 & 91  & ~~~~~~~~~~317 & 83  \\
2 & $\pt$ & 4 & 962 & 179  & 717 & 148  \\
4 & $\pt$ & 6 & 542 & 129  & 733 & 142  \\
6 & $\pt$ & 8 & 448 & 109  & 409 & 97  \\
8 & $\pt$ & 10 & 405 & 86  & 189 & 57  \\
10 & $\pt$ & 15 & 208 & 42 & 130 & 28  \\
15 & $\pt$ & 25 & 45 & 11  & 20 & 7  \\
       \hline
     \end{tabular}\end{center}
   \label{tab:A5}
   \end{table}  
   
 \begin{table}[!ht]
   \caption{\TwoS\  production cross-section in $p$Pb, as a function of $y^*$. }
 \begin{center}\begin{tabular}{r@{\,$<$\,}c@{\,$<$\,}lr@{\,$\pm$\,}l}
      \hline
        \multicolumn{3}{c}{$y^*$ }  & \multicolumn{2}{c}{$\dfrac{{\rm d}\sigma}{{\rm d}y^*}$ [nb]}  \\    
      \hline
  $-5.0$ & $y^*$ & $-4.5$ & 1058 & 414  \\
  $-4.5$ & $y^*$ & $-4.0$ & 979 & 202  \\
  $-4.0$ & $y^*$ & $-3.5$ & 2400 & 458  \\
  $-3.5$ & $y^*$ & $-3.0$ & 2716 & 485  \\
        $-3.0$ & $y^*$ & $-2.5$ & 3565 & 702  \\
        1.5 & $y^*$ & 2.0 & 4402 & 898  \\
  2.0 & $y^*$ & 2.5 & 3180 & 551  \\
  2.5 & $y^*$ & 3.0 & 2856 & 515  \\
 3.0 & $y^*$ & 3.5 & 1369 & 381  \\
  3.5 & $y^*$ & 4.0 & 1339 & 416  \\
       \hline
     \end{tabular}\end{center}
   \label{tab:A6}
   \end{table}  

 \clearpage 
 \section{\boldmath Scaled $\PUpsilon(1S)$ and $\PUpsilon(2S)$ differential cross-sections in $pp$ collisions}
 \label{sec:App}
 
 Tables~\ref{tab:pprefCS} 
 and~\ref{tab:pprefCS1} show the $\OneS$ and $\TwoS$ differential cross-sections scaled to the cross-section in $pp$ collisions at $\sqsnn=8.16\TeV$ in \pt integrated over $y$ in region $2.0<y<4.5$ and in $y$ over \pt in region $\pt<25$\,\gevc.
%
%
 
  \begin{table}[!ht]
   \caption{
  Scaled $pp$ differential cross-section in \pt at $\sqsnn=8.16$\,TeV. The first uncertainty is 
statistical, the second  
is systematic,
     which includes the systematic uncertainty from the $pp$ measurement 
and that estimated by changing the interpolation function.  
    }
 \begin{center}\begin{tabular}{r@{\,$<$\,}c@{\,$<$\,}lr@{\,$\pm$\,}c@{\,$\pm$\,}lr@{\,$\pm$\,}c@{\,$\pm$\,}l}
      \hline
      \multicolumn{3}{c}{$\pt$  [\GeVc]} & \multicolumn{3}{c}{$\PUpsilon(1S)$ $\dfrac{{\rm d}\sigma}{{\rm d}p_{\rm T}}$  [nb/(\GeVc)]}  & \multicolumn{3}{c}{$\PUpsilon(2S)$ $\dfrac{{\rm d}\sigma}{{\rm d}p_{\rm T}}$  [nb/(\GeVc)]}  \\
      \hline
      0  & $\pt$ & 2 &~~~~~~~~1995 & 14 & 31 &~~~~~~~~555 & 9 & 11 \\
      2  & $\pt$ &  4 &3626 & 18 & 51 &1052 & 11 & 19 \\
      4  & $\pt$ &  6 &2898 & 16 & 40 &910 & 11 & 15 \\
      6  & $\pt$ &  8 &1786 & 12 & 28 &634 & 9 & 14 \\
      8  & $\pt$ &  10 &1009 & 9 & 15 &394 & 7 & 7 \\
      10  & $\pt$ & 15 &382 & 5 & 7 &169 & 4 & 4 \\
      15  & $\pt$ &  25 &54 & 2 & 1 &29 & 1 & 1 \\
       \hline
     \end{tabular}\end{center}
   \label{tab:pprefCS}
   \end{table}
   
 \begin{table}[!ht]
   \caption{
  Scaled $pp$ differential cross-section in $y$ at $\sqsnn=8.16$\,TeV. The first uncertainty is 
statistical, the second  
is systematic,
     which includes the systematic uncertainty from the $pp$ measurement 
and that estimated by changing the interpolation function.  
    }
 \begin{center}\begin{tabular}{r@{\,$<$\,}c@{\,$<$\,}lr@{\,$\pm$\,}c@{\,$\pm$\,}lr@{\,$\pm$\,}c@{\,$\pm$\,}l}
      \hline
      \multicolumn{3}{c}{$y$ } & \multicolumn{3}{c}{$\PUpsilon(1S)$ $\dfrac{{\rm d}\sigma}{{\rm d}y}$  [nb]}  & \multicolumn{3}{c}{$\PUpsilon(2S)$ $\dfrac{{\rm d}\sigma}{{\rm d}y}$  [nb]}  \\
      \hline
      2.0  & $y$ &  2.5 &15171 & 143 & 250 &5083 & 105 & 110 \\
      2.5  & $y$ &  3.0 &14273 & 82 & 193 &4672 & 60 & 79 \\
      3.0  & $y$ &  3.5 &11758 & 66 & 170 &3792 & 49 & 71 \\
      3.5  & $y$ &  4.0 &8950 & 65 & 137 &2898 & 46 & 61 \\
      4.0 & $y$ &  4.5 &5103 & 73 & 90 &1596 & 50 & 42 \\
       \hline
     \end{tabular}\end{center}
   \label{tab:pprefCS1}
   \end{table}
 
\section{Nuclear modification factor}   
\label{sec:A2}

Tables~\ref{tab:B1} and~\ref{tab:B2} list the nuclear modification factors  $R_{p{\rm Pb}}^{\PUpsilon(1S)}$  for \OneS\ in transverse momentum bins and in rapidity bins. 
Tables~\ref{tab:B3} and~\ref{tab:B4} listed the nuclear modification factors for \OneS\ $R_{p{\rm Pb}}^{\PUpsilon(2S)}$  for \TwoS\ in transverse momentum bins and in rapidity bins. In all tables, the quoted uncertainties are the sum in quadrature of the statistical and systematic components.

 \begin{table}[!ht]
   \caption{$\PUpsilon(1S)$ nuclear modification factor, $R_{p{\rm Pb}}^{\PUpsilon(1S)}$, in $p$Pb and Pb$p$ as a function of $\pt$ integrated over
    $y^*$ in the range $1.5 < y^* < 4.0$ for $p$Pb and $-5.0 < y^* < -2.5$ for Pb$p$.  }
 \begin{center}\begin{tabular}{r@{\,$<$\,}c@{\,$<$\,}lr@{\,$\pm$\,}lr@{\,$\pm$\,}l}
      \hline
       \multicolumn{3}{c}{$\pt$  [\GeVc]} & \multicolumn{2}{c}{$R_{p{\rm Pb}}^{\PUpsilon(1S)}$ in $p$Pb}   & \multicolumn{2}{c}{$R_{p{\rm Pb}}^{\PUpsilon(1S)}$ in Pb$p$} \\    
      \hline
0 & $\pt$ & 2 & ~~0.46 & 0.06   & ~~0.76 & 0.11\\
2 & $\pt$ & 4 & 0.46 & 0.05  & 0.92 & 0.13 \\
4 & $\pt$ & 6 & 0.66 & 0.07 & 0.90 & 0.13 \\
6 & $\pt$ & 8 & 0.67 & 0.08  & 0.91 & 0.17\\
8 & $\pt$ & 10 & 0.79 & 0.10 & 0.81 & 0.12 \\
10 & $\pt$ & 15 & 0.84 & 0.10 & 1.14 & 0.16 \\
15 & $\pt$ & 25 & 0.87 & 0.16  & 1.04 & 0.18\\
       \hline
     \end{tabular}\end{center}
   \label{tab:B1}
   \end{table}  

  \begin{table}[!ht]
   \caption{$\PUpsilon(1S)$ nuclear modification factor, $R_{p{\rm Pb}}^{\PUpsilon(1S)}$, in $p$Pb and Pb$p$ as a function of $y^*$ integrated over $\pt$ in the range $0 < \pt < 25$~\GeVc.}
 \begin{center}\begin{tabular}{r@{\,$<$\,}c@{\,$<$\,}lr@{\,$\pm$\,}l}
      \hline
        \multicolumn{3}{c}{$y^*$ }  & \multicolumn{2}{c}{$R_{p{\rm Pb}}^{\PUpsilon(1S)}$} \\    
      \hline
  $-$4.5 & $y^*$ & $-$4.0 & 1.09 & 0.14  \\
  $-$4.0 & $y^*$ & $-$3.5 & 0.82 & 0.12  \\
  $-$3.5 & $y^*$ & $-$3.0 & 0.88 & 0.12  \\
        $-$3.0 & $y^*$ & $-$2.5 & 1.09 & 0.13  \\
  2.0 & $y^*$ & 2.5 & 0.67 & 0.06  \\
  2.5 & $y^*$ & 3.0 & 0.64 & 0.06  \\
  3.0 & $y^*$ & 3.5 & 0.60 & 0.07  \\
  3.5 & $y^*$ & 4.0 & 0.66 & 0.10  \\
       \hline
     \end{tabular}\end{center}
   \label{tab:B2}
   \end{table}  
 
 \clearpage 

 \begin{table}[!ht]
   \caption{$\PUpsilon(2S)$ nuclear modification factor, $R_{p{\rm Pb}}^{\PUpsilon(2S)}$, in $p$Pb and Pb$p$ as a function of $\pt$ integrated over
    $y^*$ in the range $1.5 < y^* < 4.0$ for $p$Pb and $-5.0 < y^* < -2.5$ for Pb$p$.  }
 \begin{center}\begin{tabular}{r@{\,$<$\,}c@{\,$<$\,}lr@{\,$\pm$\,}lr@{\,$\pm$\,}l}
      \hline
       \multicolumn{3}{c}{$\pt$  [\GeVc]} & \multicolumn{2}{c}{$R_{p{\rm Pb}}^{\PUpsilon(2S)}$ in $p$Pb}   & \multicolumn{2}{c}{$R_{p{\rm Pb}}^{\PUpsilon(2S)}$ in Pb$p$}\\    
      \hline
0 & $\pt$ & 2 & ~~0.22 & 0.08    & ~~0.54 & 0.17\\
2 & $\pt$ & 4 & 0.38 & 0.10  & 0.55 & 0.11 \\
4 & $\pt$ & 6 & 0.35 & 0.09  & 0.88 & 0.17 \\
6 & $\pt$ & 8 & 0.30 & 0.11   & 0.73 & 0.31\\
8 & $\pt$ & 10 & 0.49 & 0.11 & 0.48 & 0.15 \\
10 & $\pt$ & 15 & 0.69 & 0.12 & 0.78 & 0.18 \\
15 & $\pt$ & 25 & 0.78 & 0.22  & 0.86 & 0.35\\
       \hline
     \end{tabular}\end{center}
   \label{tab:B3}
   \end{table}

  \begin{table}[!ht]
   \caption{$\PUpsilon(2S)$ nuclear modification factor, $R_{p{\rm Pb}}^{\PUpsilon(2S)}$, in $p$Pb and Pb$p$ as a function of $y^*$ integrated over $\pt$ in the range $0 < \pt < 25$ \GeVc. }
 \begin{center}\begin{tabular}{r@{\,$<$\,}c@{\,$<$\,}lr@{\,$\pm$\,}l}
      \hline
        \multicolumn{3}{c}{$y^*$ }  & \multicolumn{2}{c}{$R_{p{\rm Pb}}^{\PUpsilon(2S)}$}  \\    
      \hline
  $-$4.5 & $y^*$ & $-$4.0 & 0.61 & 0.13  \\
  $-$4.0 & $y^*$ & $-$3.5 & 0.83 & 0.16  \\
  $-$3.5 & $y^*$ & $-$3.0 & 0.72 & 0.13  \\
        $-$3.0 & $y^*$ & $-$2.5 & 0.76 & 0.15  \\
  2.0 & $y^*$ & 2.5 & 0.63 & 0.11  \\
  2.5 & $y^*$ & 3.0 & 0.61 & 0.11  \\
  3.0 & $y^*$ & 3.5 & 0.36 & 0.10  \\
  3.5 & $y^*$ & 4.0 & 0.46 & 0.14  \\
       \hline
     \end{tabular}\end{center}
   \label{tab:B4}
   \end{table}  
 
 \clearpage 
\section{Forward-to-backward ratios}   
\label{sec:A3}

Tables~\ref{tab:C1} and~\ref{tab:C2} list the forward-to-backward ratios  $R_{{\rm FB}}^{\PUpsilon(1S)}$  for \OneS\ in transverse momentum bins and in rapidity bins.
In all tables, the quoted uncertainties are the sum in quadrature of the statistical and systematic components.
  The ratio $R_{{\rm FB}}^{\PUpsilon(2S)}$ integrated over $\vert y^* \vert$ in the range $2.5 < \vert y^* \vert < 4.0$, and over $\pt$ in the range $0 < \pt < 25$ $\GeVc$ is $0.66\pm0.23$.

  \begin{table}[!ht]
   \caption{$\PUpsilon(1S)$ forward-to-backward ratio, $R_{{\rm FB}}^{\PUpsilon(1S)}$, as a function of $\pt$ integrated over $\vert y^* \vert$ in the range $2.5 < \vert y^* \vert < 4.0$. }
 \begin{center}\begin{tabular}{r@{\,$<$\,}c@{\,$<$\,}lr@{\,$\pm$\,}l}
      \hline
         \multicolumn{3}{c}{$\pt$  [\GeVc]} & \multicolumn{2}{c}{$R_{{\rm FB}}^{\PUpsilon(1S)}$} \\    
      \hline
0 & $\pt$ & 2 & 0.73 & 0.19  \\
2 & $\pt$ & 4 & 0.74 & 0.18  \\
4 & $\pt$ & 6 & 0.92 & 0.19  \\
6 & $\pt$ & 8 & 1.01 & 0.19  \\
8 & $\pt$ & 10 & 1.37 & 0.20  \\
10 & $\pt$ & 15 & 1.22 & 0.20  \\
15 & $\pt$ & 25 & 1.46 & 0.26  \\
       \hline
     \end{tabular}\end{center}
   \label{tab:C1}
   \end{table}  
   
  \begin{table}[!ht]
   \caption{$\PUpsilon(1S)$ forward-to-backward ratio, $R_{{\rm FB}}^{\PUpsilon(1S)}$, as a function of $\vert y^* \vert$ integrated over $\pt$ in the range $0 < \pt < 25$ \GeVc. }
 \begin{center}\begin{tabular}{r@{\,$<$\,}c@{\,$<$\,}lr@{\,$\pm$\,}l}
      \hline
         \multicolumn{3}{c}{$\vert y^* \vert$ }  & \multicolumn{2}{c}{$R_{{\rm FB}}^{\PUpsilon(1S)}$} \\    
      \hline
  2.5 & $|y^*|$ & 3.0 & 0.59 & 0.16  \\
  3.0 & $|y^*|$ & 3.5 & 0.68 & 0.18  \\
  3.5 & $|y^*|$ & 4.0 & 0.80 & 0.21  \\
       \hline
     \end{tabular}\end{center}
   \label{tab:C2}
   \end{table}

\clearpage
\section{Ratios between excited states}   
\label{sec:A4}

Tables~\ref{tab:D1} and~\ref{tab:D2} list the $\PUpsilon(2S)$ to $\PUpsilon(1S)$ ratios in bins of transverse momentum bins and rapidity. In all tables, the quoted uncertainties are the sum in quadrature of the statistical and systematic components.

  \begin{table}[!ht]
   \caption{$\PUpsilon(2S)$ to $\PUpsilon(1S)$ ratio, $R(\TwoS)$,  in $p$Pb and Pb$p$ as a function of $\pt$ integrated over $y^*$ in the range  $1.5 < y^* < 4.0$ for $p$Pb and $-5.0 < y^* < -2.5$ for Pb$p$. }
 \begin{center}\begin{tabular}{r@{\,$<$\,}c@{\,$<$\,}lr@{\,$\pm$\,}lr@{\,$\pm$\,}l}
      \hline
         \multicolumn{3}{c}{$\pt$  [\GeVc]} & \multicolumn{2}{c}{$R(\TwoS)$ in $p$Pb} &  \multicolumn{2}{c}{$R(\TwoS)$ in Pb$p$}   \\    
      \hline
0 & $\pt$ & 2 & ~~~~0.20 & 0.06   & ~~~~0.21 & 0.07\\
2 & $\pt$ & 4 & 0.36 & 0.06 & 0.25 & 0.06   \\
4 & $\pt$ & 6 & 0.22 & 0.05 & 0.33 & 0.08   \\
6 & $\pt$ & 8 & 0.26 & 0.06  & 0.29 & 0.09  \\
8 & $\pt$ & 10 & 0.35 & 0.07  & 0.28 & 0.11  \\
10 & $\pt$ & 15 & 0.42 & 0.08  & 0.41 & 0.09 \\
15 & $\pt$ & 25 & 0.55 & 0.15  & 0.49 & 0.19  \\
       \hline
     \end{tabular}\end{center}
   \label{tab:D1}
   \end{table}  

  \begin{table}[!ht]
   \caption{$\PUpsilon(2S)$ to $\PUpsilon(1S)$ ratio, $R(\TwoS)$,  in $p$Pb and Pb$p$ as a function of $y^*$ integrated over $\pt$ in the range $0 < \pt < 25$ \GeVc.}
 \begin{center}\begin{tabular}{r@{\,$<$\,}c@{\,$<$\,}lr@{\,$\pm$\,}l}
      \hline
        \multicolumn{3}{c}{ $y^*$}   & \multicolumn{2}{c}{$R(2S)$} \\    
      \hline
  $-$5.0 & $y^*$ & $-$4.5 & 0.27 & 0.05  \\
  $-$4.5 & $y^*$ & $-$4.0 & 0.18 & 0.03  \\
  $-$4.0 & $y^*$ & $-$3.5 & 0.34 & 0.06  \\
  $-$3.5 & $y^*$ & $-$3.0 & 0.28 & 0.05  \\
  $-$3.0 & $y^*$ & $-$2.5 & 0.24 & 0.09  \\
    1.5 & $y^*$ & 2.0 & 0.38 & 0.08  \\
  2.0 & $y^*$ & 2.5 & 0.31 & 0.05  \\
  2.5 & $y^*$ & 3.0 & 0.31 & 0.05  \\
  3.0 & $y^*$ & 3.5 & 0.19 & 0.05  \\
  3.5 & $y^*$ & 4.0 & 0.23 & 0.07  \\
       \hline
     \end{tabular}\end{center}
   \label{tab:D2}
   \end{table}  

\clearpage
\section{ \boldmath $\PUpsilon(1S)$ to  nonprompt $J/\psi$ ratios}   
\label{sec:A5}

Table~\ref{tab:E1} lists the $\PUpsilon(1S)$ to  nonprompt $J/\psi$ ratios in rapidity bins. 

  \begin{table}[!ht]
   \caption{$\PUpsilon(1S)$ to nonprompt $J/\psi$, in $p$Pb and Pb$p$ as a function of $y^*$ integrated over $\pt$ in the range $0 < \pt < 25$ \GeVc. The quoted uncertainties are the sum in quadrature of the statistical and systematic components.}
 \begin{center}\begin{tabular}{r@{\,$<$\,}c@{\,$<$\,}lr@{\,$\pm$\,}l}
      \hline
        \multicolumn{3}{c}{ $y^*$ }  & \multicolumn{2}{c}{$\PUpsilon(1S)$ to  $J/\psi$-from-b} \\    
      \hline
  $-$5.0 & $y^*$ & $-$4.5 & ~~~~~~0.125 & 0.020  \\
  $-$4.5 & $y^*$ & $-$4.0 & 0.102 & 0.013  \\
  $-$4.0 & $y^*$ & $-$3.5 & 0.087 & 0.013  \\
  $-$3.5 & $y^*$ & $-$3.0 & 0.094 & 0.013  \\
  $-$3.0 & $y^*$ & $-$2.5 & 0.112 & 0.014  \\
  1.5 & $y^*$ & 2.0 & 0.077 & 0.008  \\
  2.0 & $y^*$ & 2.5 & 0.074 & 0.007  \\
  2.5 & $y^*$ & 3.0 & 0.082 & 0.008  \\
  3.0 & $y^*$ & 3.5 & 0.078 & 0.009  \\
  3.5 & $y^*$ & 4.0 & 0.091 & 0.013  \\
       \hline
     \end{tabular}\end{center}
   \label{tab:E1}
   \end{table}  
 \clearpage 


\addcontentsline{toc}{section}{References}
%
%
\setboolean{inbibliography}{true}
\bibliographystyle{LHCb}
\bibliography{main,LHCb-PAPER,LHCb-CONF,LHCb-DP,LHCb-TDR,giulia}

\ifx\mcitethebibliography\mciteundefinedmacro
\PackageError{LHCb.bst}{mciteplus.sty has not been loaded}
{This bibstyle requires the use of the mciteplus package.}\fi
\providecommand{\href}[2]{#2}
\begin{mcitethebibliography}{10}
\mciteSetBstSublistMode{n}
\mciteSetBstMaxWidthForm{subitem}{\alph{mcitesubitemcount})}
\mciteSetBstSublistLabelBeginEnd{\mcitemaxwidthsubitemform\space}
{\relax}{\relax}

\bibitem{Adams:2005dq}
STAR collaboration, J.~Adams {\em et~al.},
  \ifthenelse{\boolean{articletitles}}{\emph{{Experimental and theoretical
  challenges in the search for the quark-gluon plasma: The STAR Collaboration's
  critical assessment of the evidence from RHIC collisions}},
  }{}\href{http://dx.doi.org/10.1016/j.nuclphysa.2005.03.085}{Nucl.\ Phys.\
  \textbf{A757} (2005) 102},
  \href{http://arxiv.org/abs/nucl-ex/0501009}{{\normalfont\ttfamily
  arXiv:nucl-ex/0501009}}\relax
\mciteBstWouldAddEndPuncttrue
\mciteSetBstMidEndSepPunct{\mcitedefaultmidpunct}
{\mcitedefaultendpunct}{\mcitedefaultseppunct}\relax
\EndOfBibitem
\bibitem{Maciula:2013wg}
R.~Maciula and A.~Szczurek, \ifthenelse{\boolean{articletitles}}{\emph{{Open
  charm production at the LHC: $k_{t}$-factorization approach}},
  }{}\href{http://dx.doi.org/10.1103/PhysRevD.87.094022}{Phys.\ Rev.\
  \textbf{D87} (2013) 094022},
  \href{http://arxiv.org/abs/1301.3033}{{\normalfont\ttfamily
  arXiv:1301.3033}}\relax
\mciteBstWouldAddEndPuncttrue
\mciteSetBstMidEndSepPunct{\mcitedefaultmidpunct}
{\mcitedefaultendpunct}{\mcitedefaultseppunct}\relax
\EndOfBibitem
\bibitem{Matsui:1986dk}
T.~Matsui and H.~Satz, \ifthenelse{\boolean{articletitles}}{\emph{{$J/\psi$
  suppression by quark-gluon plasma formation}},
  }{}\href{http://dx.doi.org/10.1016/0370-2693(86)91404-8}{Phys.\ Lett.\
  \textbf{B178} (1986) 416}\relax
\mciteBstWouldAddEndPuncttrue
\mciteSetBstMidEndSepPunct{\mcitedefaultmidpunct}
{\mcitedefaultendpunct}{\mcitedefaultseppunct}\relax
\EndOfBibitem
\bibitem{deFlorian:2011fp}
D.~de~Florian, R.~Sassot, P.~Zurita, and M.~Stratmann,
  \ifthenelse{\boolean{articletitles}}{\emph{{Global Analysis of Nuclear Parton
  Distributions}},
  }{}\href{http://dx.doi.org/10.1103/PhysRevD.85.074028}{Phys.\ Rev.\
  \textbf{D85} (2012) 074028},
  \href{http://arxiv.org/abs/1112.6324}{{\normalfont\ttfamily
  arXiv:1112.6324}}\relax
\mciteBstWouldAddEndPuncttrue
\mciteSetBstMidEndSepPunct{\mcitedefaultmidpunct}
{\mcitedefaultendpunct}{\mcitedefaultseppunct}\relax
\EndOfBibitem
\bibitem{Owens:2012bv}
J.~F. Owens, A.~Accardi, and W.~Melnitchouk,
  \ifthenelse{\boolean{articletitles}}{\emph{{Global parton distributions with
  nuclear and finite-$Q^2$ corrections}},
  }{}\href{http://dx.doi.org/10.1103/PhysRevD.87.094012}{Phys.\ Rev.\
  \textbf{D87} (2013) 094012},
  \href{http://arxiv.org/abs/1212.1702}{{\normalfont\ttfamily
  arXiv:1212.1702}}\relax
\mciteBstWouldAddEndPuncttrue
\mciteSetBstMidEndSepPunct{\mcitedefaultmidpunct}
{\mcitedefaultendpunct}{\mcitedefaultseppunct}\relax
\EndOfBibitem
\bibitem{Eskola:2009uj}
K.~J. Eskola, H.~Paukkunen, and C.~A. Salgado,
  \ifthenelse{\boolean{articletitles}}{\emph{{EPS09: A New Generation of NLO
  and LO Nuclear Parton Distribution Functions}},
  }{}\href{http://dx.doi.org/10.1088/1126-6708/2009/04/065}{JHEP \textbf{04}
  (2009) 065}, \href{http://arxiv.org/abs/0902.4154}{{\normalfont\ttfamily
  arXiv:0902.4154}}\relax
\mciteBstWouldAddEndPuncttrue
\mciteSetBstMidEndSepPunct{\mcitedefaultmidpunct}
{\mcitedefaultendpunct}{\mcitedefaultseppunct}\relax
\EndOfBibitem
\bibitem{Kovarik:2015cma}
K.~Kovarik {\em et~al.}, \ifthenelse{\boolean{articletitles}}{\emph{{nCTEQ15 -
  Global analysis of nuclear parton distributions with uncertainties in the
  CTEQ framework}},
  }{}\href{http://dx.doi.org/10.1103/PhysRevD.93.085037}{Phys.\ Rev.\
  \textbf{D93} (2016) 085037},
  \href{http://arxiv.org/abs/1509.00792}{{\normalfont\ttfamily
  arXiv:1509.00792}}\relax
\mciteBstWouldAddEndPuncttrue
\mciteSetBstMidEndSepPunct{\mcitedefaultmidpunct}
{\mcitedefaultendpunct}{\mcitedefaultseppunct}\relax
\EndOfBibitem
\bibitem{Arleo:2014oha}
F.~Arleo and S.~Peigné, \ifthenelse{\boolean{articletitles}}{\emph{{Quarkonium
  suppression in heavy-ion collisions from coherent energy loss in cold nuclear
  matter}}, }{}\href{http://dx.doi.org/10.1007/JHEP10(2014)073}{JHEP
  \textbf{10} (2014) 073},
  \href{http://arxiv.org/abs/1407.5054}{{\normalfont\ttfamily
  arXiv:1407.5054}}\relax
\mciteBstWouldAddEndPuncttrue
\mciteSetBstMidEndSepPunct{\mcitedefaultmidpunct}
{\mcitedefaultendpunct}{\mcitedefaultseppunct}\relax
\EndOfBibitem
\bibitem{Arleo:2013zua}
F.~Arleo, R.~Kolevatov, S.~Peigné, and M.~Rustamova,
  \ifthenelse{\boolean{articletitles}}{\emph{{Centrality and $p_{\perp}$
  dependence of $\jpsi$ suppression in proton-nucleus collisions from parton
  energy loss}}, }{}\href{http://dx.doi.org/10.1007/JHEP05(2013)155}{JHEP
  \textbf{05} (2013) 155},
  \href{http://arxiv.org/abs/1304.0901}{{\normalfont\ttfamily
  arXiv:1304.0901}}\relax
\mciteBstWouldAddEndPuncttrue
\mciteSetBstMidEndSepPunct{\mcitedefaultmidpunct}
{\mcitedefaultendpunct}{\mcitedefaultseppunct}\relax
\EndOfBibitem
\bibitem{Arleo:2012rs}
F.~Arleo and S.~Peign\'e,
  \ifthenelse{\boolean{articletitles}}{\emph{{Heavy-quarkonium suppression in
  p-A collisions from parton energy loss in cold QCD matter}},
  }{}\href{http://dx.doi.org/10.1007/JHEP03(2013)122}{JHEP \textbf{03} (2013)
  122}, \href{http://arxiv.org/abs/1212.0434}{{\normalfont\ttfamily
  arXiv:1212.0434}}\relax
\mciteBstWouldAddEndPuncttrue
\mciteSetBstMidEndSepPunct{\mcitedefaultmidpunct}
{\mcitedefaultendpunct}{\mcitedefaultseppunct}\relax
\EndOfBibitem
\bibitem{GERSCHEL1988253}
C.~Gerschel and J.~H{\"u}fner, \ifthenelse{\boolean{articletitles}}{\emph{A
  contribution to the suppression of the \jpsi meson produced in high-energy
  nucleus-nucleus collisions},
  }{}\href{http://dx.doi.org/10.1016/0370-2693(88)90570-9}{Phys.\ Lett.\
  \textbf{B207} (1988) 253}\relax
\mciteBstWouldAddEndPuncttrue
\mciteSetBstMidEndSepPunct{\mcitedefaultmidpunct}
{\mcitedefaultendpunct}{\mcitedefaultseppunct}\relax
\EndOfBibitem
\bibitem{Albacete:2013ei}
J.~L. Albacete {\em et~al.},
  \ifthenelse{\boolean{articletitles}}{\emph{{Predictions for $p+$Pb collisions
  at \sqsnnref = 5 TeV}},
  }{}\href{http://dx.doi.org/10.1142/S0218301313300075}{Int.\ J.\ Mod.\ Phys.\
  \textbf{E22} (2013) 1330007},
  \href{http://arxiv.org/abs/1301.3395}{{\normalfont\ttfamily
  arXiv:1301.3395}}\relax
\mciteBstWouldAddEndPuncttrue
\mciteSetBstMidEndSepPunct{\mcitedefaultmidpunct}
{\mcitedefaultendpunct}{\mcitedefaultseppunct}\relax
\EndOfBibitem
\bibitem{Adeluyi:2013tuu}
A.~Adeluyi and T.~Nguyen, \ifthenelse{\boolean{articletitles}}{\emph{{Coherent
  photoproduction of $\psi$ and $\PUpsilon$ mesons in ultraperipheral pPb and
  PbPb collisions at the CERN Large Hadron Collider at \mbox{$\sqsnnref =
  5\tev$} and \sqsnnref = 2.76 \TeV}},
  }{}\href{http://dx.doi.org/10.1103/PhysRevC.87.027901}{Phys.\ Rev.\
  \textbf{C87} (2013) 027901},
  \href{http://arxiv.org/abs/1302.4288}{{\normalfont\ttfamily
  arXiv:1302.4288}}\relax
\mciteBstWouldAddEndPuncttrue
\mciteSetBstMidEndSepPunct{\mcitedefaultmidpunct}
{\mcitedefaultendpunct}{\mcitedefaultseppunct}\relax
\EndOfBibitem
\bibitem{Chirilli:2012jd}
G.~A. Chirilli, B.-W. Xiao, and F.~Yuan,
  \ifthenelse{\boolean{articletitles}}{\emph{{Inclusive hadron productions in
  pA collisions}},
  }{}\href{http://dx.doi.org/10.1103/PhysRevD.86.054005}{Phys.\ Rev.\
  \textbf{D86} (2012) 054005},
  \href{http://arxiv.org/abs/1203.6139}{{\normalfont\ttfamily
  arXiv:1203.6139}}\relax
\mciteBstWouldAddEndPuncttrue
\mciteSetBstMidEndSepPunct{\mcitedefaultmidpunct}
{\mcitedefaultendpunct}{\mcitedefaultseppunct}\relax
\EndOfBibitem
\bibitem{Chirilli:2012sk}
G.~A. Chirilli, \ifthenelse{\boolean{articletitles}}{\emph{{High-Energy QCD
  factorization from DIS to pA collisions}},
  }{}\href{http://dx.doi.org/10.1142/S2010194512009245}{Int.\ J.\ Mod.\ Phys.\
  Conf.\ Ser.\  \textbf{20} (2012) 200},
  \href{http://arxiv.org/abs/1209.1614}{{\normalfont\ttfamily
  arXiv:1209.1614}}\relax
\mciteBstWouldAddEndPuncttrue
\mciteSetBstMidEndSepPunct{\mcitedefaultmidpunct}
{\mcitedefaultendpunct}{\mcitedefaultseppunct}\relax
\EndOfBibitem
\bibitem{Ferreiro:2013pua}
E.~G. Ferreiro, F.~Fleuret, J.~P. Lansberg, and A.~Rakotozafindrabe,
  \ifthenelse{\boolean{articletitles}}{\emph{{Impact of the nuclear
  modification of the gluon densities on $J/\psi$ production in $p$Pb
  collisions at \sqsnnref = 5 TeV}},
  }{}\href{http://dx.doi.org/10.1103/PhysRevC.88.047901}{Phys.\ Rev.\
  \textbf{C88} (2013) 047901},
  \href{http://arxiv.org/abs/1305.4569}{{\normalfont\ttfamily
  arXiv:1305.4569}}\relax
\mciteBstWouldAddEndPuncttrue
\mciteSetBstMidEndSepPunct{\mcitedefaultmidpunct}
{\mcitedefaultendpunct}{\mcitedefaultseppunct}\relax
\EndOfBibitem
\bibitem{Ferreiro:2018wbd}
E.~G. Ferreiro and J.-P. Lansberg,
  \ifthenelse{\boolean{articletitles}}{\emph{{Is bottomonium suppression in
  proton-nucleus and nucleus-nucleus collisions at LHC energies due to the same
  effects?}}, }{}\href{http://arxiv.org/abs/1804.04474}{{\normalfont\ttfamily
  arXiv:1804.04474}}\relax
\mciteBstWouldAddEndPuncttrue
\mciteSetBstMidEndSepPunct{\mcitedefaultmidpunct}
{\mcitedefaultendpunct}{\mcitedefaultseppunct}\relax
\EndOfBibitem
\bibitem{Ferreiro:2014bia}
E.~G. Ferreiro, \ifthenelse{\boolean{articletitles}}{\emph{{Excited charmonium
  suppression in proton–nucleus collisions as a consequence of comovers}},
  }{}\href{http://dx.doi.org/10.1016/j.physletb.2015.07.066}{Phys.\ Lett.\
  \textbf{B749} (2015) 98},
  \href{http://arxiv.org/abs/1411.0549}{{\normalfont\ttfamily
  arXiv:1411.0549}}\relax
\mciteBstWouldAddEndPuncttrue
\mciteSetBstMidEndSepPunct{\mcitedefaultmidpunct}
{\mcitedefaultendpunct}{\mcitedefaultseppunct}\relax
\EndOfBibitem
\bibitem{Du:2015wha}
X.~Du and R.~Rapp, \ifthenelse{\boolean{articletitles}}{\emph{{Sequential
  Regeneration of Charmonia in Heavy-Ion Collisions}},
  }{}\href{http://dx.doi.org/10.1016/j.nuclphysa.2015.09.006}{Nucl.\ Phys.\
  \textbf{A943} (2015) 147},
  \href{http://arxiv.org/abs/1504.00670}{{\normalfont\ttfamily
  arXiv:1504.00670}}\relax
\mciteBstWouldAddEndPuncttrue
\mciteSetBstMidEndSepPunct{\mcitedefaultmidpunct}
{\mcitedefaultendpunct}{\mcitedefaultseppunct}\relax
\EndOfBibitem
\bibitem{Ma:2017rsu}
Y.-Q. Ma, R.~Venugopalan, K.~Watanabe, and H.-F. Zhang,
  \ifthenelse{\boolean{articletitles}}{\emph{{$\psi(2S)$ versus $J/\psi$
  suppression in proton-nucleus collisions from factorization violating soft
  color exchanges}},
  }{}\href{http://dx.doi.org/10.1103/PhysRevC.97.014909}{Phys.\ Rev.\
  \textbf{C97} (2018) 014909},
  \href{http://arxiv.org/abs/1707.07266}{{\normalfont\ttfamily
  arXiv:1707.07266}}\relax
\mciteBstWouldAddEndPuncttrue
\mciteSetBstMidEndSepPunct{\mcitedefaultmidpunct}
{\mcitedefaultendpunct}{\mcitedefaultseppunct}\relax
\EndOfBibitem
\bibitem{Chatrchyan:2011pe}
CMS collaboration, S.~Chatrchyan {\em et~al.},
  \ifthenelse{\boolean{articletitles}}{\emph{{Indications of suppression of
  excited $\PUpsilon$ states in PbPb collisions at $\sqrt{S_{NN}}$ = 2.76
  TeV}}, }{}\href{http://dx.doi.org/10.1103/PhysRevLett.107.052302}{Phys.\
  Rev.\ Lett.\  \textbf{107} (2011) 052302},
  \href{http://arxiv.org/abs/1105.4894}{{\normalfont\ttfamily
  arXiv:1105.4894}}\relax
\mciteBstWouldAddEndPuncttrue
\mciteSetBstMidEndSepPunct{\mcitedefaultmidpunct}
{\mcitedefaultendpunct}{\mcitedefaultseppunct}\relax
\EndOfBibitem
\bibitem{Aaboud:2017cif}
ATLAS, M.~Aaboud {\em et~al.},
  \ifthenelse{\boolean{articletitles}}{\emph{{Measurement of quarkonium
  production in proton–lead and proton–proton collisions at $5.02\tev$ with
  the ATLAS detector}},
  }{}\href{http://dx.doi.org/10.1140/epjc/s10052-018-5624-4}{Eur.\ Phys.\ J.\
  \textbf{C78} (2018) 171},
  \href{http://arxiv.org/abs/1709.03089}{{\normalfont\ttfamily
  arXiv:1709.03089}}\relax
\mciteBstWouldAddEndPuncttrue
\mciteSetBstMidEndSepPunct{\mcitedefaultmidpunct}
{\mcitedefaultendpunct}{\mcitedefaultseppunct}\relax
\EndOfBibitem
\bibitem{Ye:2017fwv}
Z.~Ye, \ifthenelse{\boolean{articletitles}}{\emph{{$\PUpsilon$ measurements in
  p+p, p+Au and Au+Au collisions at \sqsnnref = 200GeV with the STAR
  experiment}},
  }{}\href{http://dx.doi.org/10.1016/j.nuclphysa.2017.06.040}{Nucl.\ Phys.\
  \textbf{A967} (2017) 600}\relax
\mciteBstWouldAddEndPuncttrue
\mciteSetBstMidEndSepPunct{\mcitedefaultmidpunct}
{\mcitedefaultendpunct}{\mcitedefaultseppunct}\relax
\EndOfBibitem
\bibitem{Khachatryan:2016xxp}
CMS collaboration, V.~Khachatryan {\em et~al.},
  \ifthenelse{\boolean{articletitles}}{\emph{{Suppression of $\OneS, \TwoS$ and
  $\ThreeS$ quarkonium states in PbPb collisions at \sqsnnref = 2.76 TeV}},
  }{}\href{http://dx.doi.org/10.1016/j.physletb.2017.04.031}{Phys.\ Lett.\
  \textbf{B770} (2017) 357},
  \href{http://arxiv.org/abs/1611.01510}{{\normalfont\ttfamily
  arXiv:1611.01510}}\relax
\mciteBstWouldAddEndPuncttrue
\mciteSetBstMidEndSepPunct{\mcitedefaultmidpunct}
{\mcitedefaultendpunct}{\mcitedefaultseppunct}\relax
\EndOfBibitem
\bibitem{Chatrchyan:2013nza}
CMS collaboration, S.~Chatrchyan {\em et~al.},
  \ifthenelse{\boolean{articletitles}}{\emph{{Event activity dependence of
  $\PUpsilon(nS)$ production in $\sqsnn=5.02$ TeV pPb and $\sqrt{s}$=2.76 TeV
  pp collisions}}, }{}\href{http://dx.doi.org/10.1007/JHEP04(2014)103}{JHEP
  \textbf{04} (2014) 103},
  \href{http://arxiv.org/abs/1312.6300}{{\normalfont\ttfamily
  arXiv:1312.6300}}\relax
\mciteBstWouldAddEndPuncttrue
\mciteSetBstMidEndSepPunct{\mcitedefaultmidpunct}
{\mcitedefaultendpunct}{\mcitedefaultseppunct}\relax
\EndOfBibitem
\bibitem{Sirunyan:2017lzi}
CMS collaboration, A.~M. Sirunyan {\em et~al.},
  \ifthenelse{\boolean{articletitles}}{\emph{{Suppression of excited
  $\PUpsilon$ states relative to the ground state in Pb-Pb collisions at
  \sqsnnref = 5.02 \TeV}},
  }{}\href{http://dx.doi.org/10.1103/PhysRevLett.120.142301}{Phys.\ Rev.\
  Lett.\  \textbf{120} (2018) 142301},
  \href{http://arxiv.org/abs/1706.05984}{{\normalfont\ttfamily
  arXiv:1706.05984}}\relax
\mciteBstWouldAddEndPuncttrue
\mciteSetBstMidEndSepPunct{\mcitedefaultmidpunct}
{\mcitedefaultendpunct}{\mcitedefaultseppunct}\relax
\EndOfBibitem
\bibitem{Abelev:2014oea}
ALICE collaboration, B.~B. Abelev {\em et~al.},
  \ifthenelse{\boolean{articletitles}}{\emph{{Production of inclusive
  $\PUpsilon$(1S) and $\PUpsilon$(2S) in p-Pb collisions at
  $\mathbf{\sqrt{s_{{\rm NN}}} = 5.02}$ TeV}},
  }{}\href{http://dx.doi.org/10.1016/j.physletb.2014.11.041}{Phys.\ Lett.\
  \textbf{B740} (2015) 105},
  \href{http://arxiv.org/abs/1410.2234}{{\normalfont\ttfamily
  arXiv:1410.2234}}\relax
\mciteBstWouldAddEndPuncttrue
\mciteSetBstMidEndSepPunct{\mcitedefaultmidpunct}
{\mcitedefaultendpunct}{\mcitedefaultseppunct}\relax
\EndOfBibitem
\bibitem{Aaij:2014mza}
LHCb collaboration, R.~Aaij {\em et~al.},
  \ifthenelse{\boolean{articletitles}}{\emph{{Study of $\PUpsilon$ production
  and cold nuclear matter effects in $p$Pb collisions at \sqsnnref = 5\TeV}},
  }{}\href{http://dx.doi.org/10.1007/JHEP07(2014)094}{JHEP \textbf{07} (2014)
  094}, \href{http://arxiv.org/abs/1405.5152}{{\normalfont\ttfamily
  arXiv:1405.5152}}\relax
\mciteBstWouldAddEndPuncttrue
\mciteSetBstMidEndSepPunct{\mcitedefaultmidpunct}
{\mcitedefaultendpunct}{\mcitedefaultseppunct}\relax
\EndOfBibitem
\bibitem{LHCb-PAPER-2017-014}
LHCb collaboration, R.~Aaij {\em et~al.},
  \ifthenelse{\boolean{articletitles}}{\emph{{Prompt and nonprompt \jpsi\
  production and nuclear modification in $p$Pb collisions at $\sqsnn =
  8.16$\,TeV}},
  }{}\href{http://dx.doi.org/10.1016/j.physletb.2017.09.058}{Phys.\ Lett.\
  \textbf{B774} (2017) 159},
  \href{http://arxiv.org/abs/1706.07122}{{\normalfont\ttfamily
  arXiv:1706.07122}}\relax
\mciteBstWouldAddEndPuncttrue
\mciteSetBstMidEndSepPunct{\mcitedefaultmidpunct}
{\mcitedefaultendpunct}{\mcitedefaultseppunct}\relax
\EndOfBibitem
\bibitem{Alves:2008zz}
LHCb collaboration, A.~A. Alves~Jr.\ {\em et~al.},
  \ifthenelse{\boolean{articletitles}}{\emph{{The \lhcb detector at the LHC}},
  }{}\href{http://dx.doi.org/10.1088/1748-0221/3/08/S08005}{JINST \textbf{3}
  (2008) S08005}\relax
\mciteBstWouldAddEndPuncttrue
\mciteSetBstMidEndSepPunct{\mcitedefaultmidpunct}
{\mcitedefaultendpunct}{\mcitedefaultseppunct}\relax
\EndOfBibitem
\bibitem{LHCb-DP-2014-002}
LHCb collaboration, R.~Aaij {\em et~al.},
  \ifthenelse{\boolean{articletitles}}{\emph{{LHCb detector performance}},
  }{}\href{http://dx.doi.org/10.1142/S0217751X15300227}{Int.\ J.\ Mod.\ Phys.\
  \textbf{A30} (2015) 1530022},
  \href{http://arxiv.org/abs/1412.6352}{{\normalfont\ttfamily
  arXiv:1412.6352}}\relax
\mciteBstWouldAddEndPuncttrue
\mciteSetBstMidEndSepPunct{\mcitedefaultmidpunct}
{\mcitedefaultendpunct}{\mcitedefaultseppunct}\relax
\EndOfBibitem
\bibitem{LHCb-DP-2014-001}
R.~Aaij {\em et~al.}, \ifthenelse{\boolean{articletitles}}{\emph{{Performance
  of the LHCb Vertex Locator}},
  }{}\href{http://dx.doi.org/10.1088/1748-0221/9/09/P09007}{JINST \textbf{9}
  (2014) P09007}, \href{http://arxiv.org/abs/1405.7808}{{\normalfont\ttfamily
  arXiv:1405.7808}}\relax
\mciteBstWouldAddEndPuncttrue
\mciteSetBstMidEndSepPunct{\mcitedefaultmidpunct}
{\mcitedefaultendpunct}{\mcitedefaultseppunct}\relax
\EndOfBibitem
\bibitem{LHCb-DP-2013-003}
R.~Arink {\em et~al.}, \ifthenelse{\boolean{articletitles}}{\emph{{Performance
  of the LHCb Outer Tracker}},
  }{}\href{http://dx.doi.org/10.1088/1748-0221/9/01/P01002}{JINST \textbf{9}
  (2014) P01002}, \href{http://arxiv.org/abs/1311.3893}{{\normalfont\ttfamily
  arXiv:1311.3893}}\relax
\mciteBstWouldAddEndPuncttrue
\mciteSetBstMidEndSepPunct{\mcitedefaultmidpunct}
{\mcitedefaultendpunct}{\mcitedefaultseppunct}\relax
\EndOfBibitem
\bibitem{LHCb-DP-2012-003}
M.~Adinolfi {\em et~al.},
  \ifthenelse{\boolean{articletitles}}{\emph{{Performance of the \lhcb RICH
  detector at the LHC}},
  }{}\href{http://dx.doi.org/10.1140/epjc/s10052-013-2431-9}{Eur.\ Phys.\ J.\
  \textbf{C73} (2013) 2431},
  \href{http://arxiv.org/abs/1211.6759}{{\normalfont\ttfamily
  arXiv:1211.6759}}\relax
\mciteBstWouldAddEndPuncttrue
\mciteSetBstMidEndSepPunct{\mcitedefaultmidpunct}
{\mcitedefaultendpunct}{\mcitedefaultseppunct}\relax
\EndOfBibitem
\bibitem{LHCb-DP-2012-002}
A.~A. Alves~Jr.\ {\em et~al.},
  \ifthenelse{\boolean{articletitles}}{\emph{{Performance of the LHCb muon
  system}}, }{}\href{http://dx.doi.org/10.1088/1748-0221/8/02/P02022}{JINST
  \textbf{8} (2013) P02022},
  \href{http://arxiv.org/abs/1211.1346}{{\normalfont\ttfamily
  arXiv:1211.1346}}\relax
\mciteBstWouldAddEndPuncttrue
\mciteSetBstMidEndSepPunct{\mcitedefaultmidpunct}
{\mcitedefaultendpunct}{\mcitedefaultseppunct}\relax
\EndOfBibitem
\bibitem{LHCb-DP-2012-004}
R.~Aaij {\em et~al.}, \ifthenelse{\boolean{articletitles}}{\emph{{The \lhcb
  trigger and its performance in 2011}},
  }{}\href{http://dx.doi.org/10.1088/1748-0221/8/04/P04022}{JINST \textbf{8}
  (2013) P04022}, \href{http://arxiv.org/abs/1211.3055}{{\normalfont\ttfamily
  arXiv:1211.3055}}\relax
\mciteBstWouldAddEndPuncttrue
\mciteSetBstMidEndSepPunct{\mcitedefaultmidpunct}
{\mcitedefaultendpunct}{\mcitedefaultseppunct}\relax
\EndOfBibitem
\bibitem{LHCb-PROC-2015-011}
G.~Dujany and B.~Storaci, \ifthenelse{\boolean{articletitles}}{\emph{{Real-time
  alignment and calibration of the LHCb Detector in Run II}},
  }{}\href{http://dx.doi.org/10.1088/1742-6596/664/8/082010}{{J.\ Phys.\ Conf.\
  Ser.\ } \textbf{664} (2015) 082010}\relax
\mciteBstWouldAddEndPuncttrue
\mciteSetBstMidEndSepPunct{\mcitedefaultmidpunct}
{\mcitedefaultendpunct}{\mcitedefaultseppunct}\relax
\EndOfBibitem
\bibitem{LHCb-DP-2016-001}
R.~Aaij {\em et~al.}, \ifthenelse{\boolean{articletitles}}{\emph{{Tesla: an
  application for real-time data analysis in High Energy Physics}},
  }{}\href{http://dx.doi.org/10.1016/j.cpc.2016.07.022}{Comput.\ Phys.\
  Commun.\  \textbf{208} (2016) 35},
  \href{http://arxiv.org/abs/1604.05596}{{\normalfont\ttfamily
  arXiv:1604.05596}}\relax
\mciteBstWouldAddEndPuncttrue
\mciteSetBstMidEndSepPunct{\mcitedefaultmidpunct}
{\mcitedefaultendpunct}{\mcitedefaultseppunct}\relax
\EndOfBibitem
\bibitem{LHCB-PAPER-2016-064}
LHCb collaboration, R.~Aaij {\em et~al.},
  \ifthenelse{\boolean{articletitles}}{\emph{{Study of \jpsi production in
  jets}}, }{}\href{http://dx.doi.org/10.1103/PhysRevLett.118.192001}{Phys.\
  Rev.\ Lett.\  \textbf{118} (2017) 192001},
  \href{http://arxiv.org/abs/1701.05116}{{\normalfont\ttfamily
  arXiv:1701.05116}}\relax
\mciteBstWouldAddEndPuncttrue
\mciteSetBstMidEndSepPunct{\mcitedefaultmidpunct}
{\mcitedefaultendpunct}{\mcitedefaultseppunct}\relax
\EndOfBibitem
\bibitem{EPOS}
T.~Pierog {\em et~al.}, \ifthenelse{\boolean{articletitles}}{\emph{{EPOS LHC:
  Test of collective hadronization with data measured at the CERN Large Hadron
  Collider}}, }{}\href{http://dx.doi.org/10.1103/PhysRevC.92.034906}{Phys.\
  Rev.\  \textbf{C92} (2015) 034906},
  \href{http://arxiv.org/abs/1306.0121}{{\normalfont\ttfamily
  arXiv:1306.0121}}\relax
\mciteBstWouldAddEndPuncttrue
\mciteSetBstMidEndSepPunct{\mcitedefaultmidpunct}
{\mcitedefaultendpunct}{\mcitedefaultseppunct}\relax
\EndOfBibitem
\bibitem{Sjostrand:2006za}
T.~Sj\"{o}strand, S.~Mrenna, and P.~Skands,
  \ifthenelse{\boolean{articletitles}}{\emph{{PYTHIA 6.4 physics and manual}},
  }{}\href{http://dx.doi.org/10.1088/1126-6708/2006/05/026}{JHEP \textbf{05}
  (2006) 026}, \href{http://arxiv.org/abs/hep-ph/0603175}{{\normalfont\ttfamily
  arXiv:hep-ph/0603175}}\relax
\mciteBstWouldAddEndPuncttrue
\mciteSetBstMidEndSepPunct{\mcitedefaultmidpunct}
{\mcitedefaultendpunct}{\mcitedefaultseppunct}\relax
\EndOfBibitem
\bibitem{Sjostrand:2007gs}
T.~Sj\"{o}strand, S.~Mrenna, and P.~Skands,
  \ifthenelse{\boolean{articletitles}}{\emph{{A brief introduction to PYTHIA
  8.1}}, }{}\href{http://dx.doi.org/10.1016/j.cpc.2008.01.036}{Comput.\ Phys.\
  Commun.\  \textbf{178} (2008) 852},
  \href{http://arxiv.org/abs/0710.3820}{{\normalfont\ttfamily
  arXiv:0710.3820}}\relax
\mciteBstWouldAddEndPuncttrue
\mciteSetBstMidEndSepPunct{\mcitedefaultmidpunct}
{\mcitedefaultendpunct}{\mcitedefaultseppunct}\relax
\EndOfBibitem
\bibitem{Allison:2006ve}
Geant4 collaboration, J.~Allison {\em et~al.},
  \ifthenelse{\boolean{articletitles}}{\emph{{Geant4 developments and
  applications}}, }{}\href{http://dx.doi.org/10.1109/TNS.2006.869826}{IEEE
  Trans.\ Nucl.\ Sci.\  \textbf{53} (2006) 270}\relax
\mciteBstWouldAddEndPuncttrue
\mciteSetBstMidEndSepPunct{\mcitedefaultmidpunct}
{\mcitedefaultendpunct}{\mcitedefaultseppunct}\relax
\EndOfBibitem
\bibitem{Agostinelli:2002hh}
Geant4 collaboration, S.~Agostinelli {\em et~al.},
  \ifthenelse{\boolean{articletitles}}{\emph{{Geant4: A simulation toolkit}},
  }{}\href{http://dx.doi.org/10.1016/S0168-9002(03)01368-8}{Nucl.\ Instrum.\
  Meth.\  \textbf{A506} (2003) 250}\relax
\mciteBstWouldAddEndPuncttrue
\mciteSetBstMidEndSepPunct{\mcitedefaultmidpunct}
{\mcitedefaultendpunct}{\mcitedefaultseppunct}\relax
\EndOfBibitem
\bibitem{LHCb-PROC-2011-006}
M.~Clemencic {\em et~al.}, \ifthenelse{\boolean{articletitles}}{\emph{{The
  \lhcb simulation application, Gauss: Design, evolution and experience}},
  }{}\href{http://dx.doi.org/10.1088/1742-6596/331/3/032023}{{J.\ Phys.\ Conf.\
  Ser.\ } \textbf{331} (2011) 032023}\relax
\mciteBstWouldAddEndPuncttrue
\mciteSetBstMidEndSepPunct{\mcitedefaultmidpunct}
{\mcitedefaultendpunct}{\mcitedefaultseppunct}\relax
\EndOfBibitem
\bibitem{Aaij:2017egv}
LHCb collaboration, R.~Aaij {\em et~al.},
  \ifthenelse{\boolean{articletitles}}{\emph{{Measurement of the $\PUpsilon$
  polarizations in $pp$ collisions at \sqsnnref = 7 and 8 \TeV}},
  }{}\href{http://dx.doi.org/10.1007/JHEP12(2017)110}{JHEP \textbf{12} (2017)
  110}, \href{http://arxiv.org/abs/1709.01301}{{\normalfont\ttfamily
  arXiv:1709.01301}}\relax
\mciteBstWouldAddEndPuncttrue
\mciteSetBstMidEndSepPunct{\mcitedefaultmidpunct}
{\mcitedefaultendpunct}{\mcitedefaultseppunct}\relax
\EndOfBibitem
\bibitem{LHCb-PAPER-2014-047}
LHCb collaboration, R.~Aaij {\em et~al.},
  \ifthenelse{\boolean{articletitles}}{\emph{{Precision luminosity measurements
  at LHCb}}, }{}\href{http://dx.doi.org/10.1088/1748-0221/9/12/P12005}{JINST
  \textbf{9} (2014) P12005},
  \href{http://arxiv.org/abs/1410.0149}{{\normalfont\ttfamily
  arXiv:1410.0149}}\relax
\mciteBstWouldAddEndPuncttrue
\mciteSetBstMidEndSepPunct{\mcitedefaultmidpunct}
{\mcitedefaultendpunct}{\mcitedefaultseppunct}\relax
\EndOfBibitem
\bibitem{PDG2018}
Particle Data Group, M.~Tanabashi {\em et~al.},
  \ifthenelse{\boolean{articletitles}}{\emph{{\href{http://pdg.lbl.gov/}{Review
  of particle physics}}},
  }{}\href{http://dx.doi.org/10.1103/PhysRevD.98.030001}{Phys.\ Rev.\ D
  \textbf{98} (2018) 030001}\relax
\mciteBstWouldAddEndPuncttrue
\mciteSetBstMidEndSepPunct{\mcitedefaultmidpunct}
{\mcitedefaultendpunct}{\mcitedefaultseppunct}\relax
\EndOfBibitem
\bibitem{Skwarnicki:1986xj}
T.~Skwarnicki, {\em {A study of the radiative cascade transitions between the
  Upsilon-prime and Upsilon resonances}}, PhD thesis, Institute of Nuclear
  Physics, Krakow, 1986,
  {\href{http://inspirehep.net/record/230779/}{DESY-F31-86-02}}\relax
\mciteBstWouldAddEndPuncttrue
\mciteSetBstMidEndSepPunct{\mcitedefaultmidpunct}
{\mcitedefaultendpunct}{\mcitedefaultseppunct}\relax
\EndOfBibitem
\bibitem{LHCb-PAPER-2013-066}
LHCb collaboration, R.~Aaij {\em et~al.},
  \ifthenelse{\boolean{articletitles}}{\emph{{Measurement of $\PUpsilon$
  production in $\proton\proton$ collisions at $\sqrt{s} = 2.76$\tev}},
  }{}\href{http://dx.doi.org/10.1140/epjc/s10052-014-2835-1}{Eur.\ Phys.\ J.\
  \textbf{C74} (2014) 2835},
  \href{http://arxiv.org/abs/1402.2539}{{\normalfont\ttfamily
  arXiv:1402.2539}}\relax
\mciteBstWouldAddEndPuncttrue
\mciteSetBstMidEndSepPunct{\mcitedefaultmidpunct}
{\mcitedefaultendpunct}{\mcitedefaultseppunct}\relax
\EndOfBibitem
\bibitem{LHCb-PAPER-2015-045}
LHCb collaboration, R.~Aaij {\em et~al.},
  \ifthenelse{\boolean{articletitles}}{\emph{{Forward production of $\PUpsilon$
  mesons in $\proton\proton$ collisions at $\sqrt{s}=7$ and $8$\,TeV}},
  }{}\href{http://dx.doi.org/10.1007/JHEP11(2015)103}{JHEP \textbf{11} (2015)
  103}, \href{http://arxiv.org/abs/1509.02372}{{\normalfont\ttfamily
  arXiv:1509.02372}}\relax
\mciteBstWouldAddEndPuncttrue
\mciteSetBstMidEndSepPunct{\mcitedefaultmidpunct}
{\mcitedefaultendpunct}{\mcitedefaultseppunct}\relax
\EndOfBibitem
\bibitem{LHCb-PAPER-2018-002}
LHCb collaboration, R.~Aaij {\em et~al.},
  \ifthenelse{\boolean{articletitles}}{\emph{{Measurement of $\PUpsilon$
  production cross-section in $pp$ collisions at $\sqrt{s} = 13$ TeV}},
  }{}\href{http://dx.doi.org/10.1007/JHEP07(2018)134}{JHEP \textbf{07} (2018)
  134}, \href{http://arxiv.org/abs/1804.09214}{{\normalfont\ttfamily
  arXiv:1804.09214}}\relax
\mciteBstWouldAddEndPuncttrue
\mciteSetBstMidEndSepPunct{\mcitedefaultmidpunct}
{\mcitedefaultendpunct}{\mcitedefaultseppunct}\relax
\EndOfBibitem
\bibitem{Lansberg:2016deg}
J.-P. Lansberg and H.-S. Shao,
  \ifthenelse{\boolean{articletitles}}{\emph{{Towards an automated tool to
  evaluate the impact of the nuclear modification of the gluon density on
  quarkonium, D and B meson production in proton–nucleus collisions}},
  }{}\href{http://dx.doi.org/10.1140/epjc/s10052-016-4575-x}{Eur.\ Phys.\ J.\
  \textbf{C77} (2017) 1},
  \href{http://arxiv.org/abs/1610.05382}{{\normalfont\ttfamily
  arXiv:1610.05382}}\relax
\mciteBstWouldAddEndPuncttrue
\mciteSetBstMidEndSepPunct{\mcitedefaultmidpunct}
{\mcitedefaultendpunct}{\mcitedefaultseppunct}\relax
\EndOfBibitem
\bibitem{Shao:2015vga}
H.-S. Shao, \ifthenelse{\boolean{articletitles}}{\emph{{HELAC-Onia 2.0: An
  upgraded matrix-element and event generator for heavy quarkonium physics}},
  }{}\href{http://dx.doi.org/10.1016/j.cpc.2015.09.011}{Comput.\ Phys.\
  Commun.\  \textbf{198} (2016) 238},
  \href{http://arxiv.org/abs/1507.03435}{{\normalfont\ttfamily
  arXiv:1507.03435}}\relax
\mciteBstWouldAddEndPuncttrue
\mciteSetBstMidEndSepPunct{\mcitedefaultmidpunct}
{\mcitedefaultendpunct}{\mcitedefaultseppunct}\relax
\EndOfBibitem
\bibitem{Shao:2012iz}
H.-S. Shao, \ifthenelse{\boolean{articletitles}}{\emph{{HELAC-Onia: An
  automatic matrix element generator for heavy quarkonium physics}},
  }{}\href{http://dx.doi.org/10.1016/j.cpc.2013.05.023}{Comput.\ Phys.\
  Commun.\  \textbf{184} (2013) 2562},
  \href{http://arxiv.org/abs/1212.5293}{{\normalfont\ttfamily
  arXiv:1212.5293}}\relax
\mciteBstWouldAddEndPuncttrue
\mciteSetBstMidEndSepPunct{\mcitedefaultmidpunct}
{\mcitedefaultendpunct}{\mcitedefaultseppunct}\relax
\EndOfBibitem
\bibitem{Eskola:2018ghi}
K.~J. Eskola, P.~Paakkinen, H.~Paukkunen, and C.~A. Salgado,
  \ifthenelse{\boolean{articletitles}}{\emph{{EPPS16 - Bringing nuclear PDFs to
  the LHC era}}, }{} in {\em {12th International Workshop on High-pT Physics in
  the RHIC/LHC Era (HPT 2017) Bergen, Norway, October 2-5, 2017}}, 2018.
\newblock \href{http://arxiv.org/abs/1802.00713}{{\normalfont\ttfamily
  arXiv:1802.00713}}\relax
\mciteBstWouldAddEndPuncttrue
\mciteSetBstMidEndSepPunct{\mcitedefaultmidpunct}
{\mcitedefaultendpunct}{\mcitedefaultseppunct}\relax
\EndOfBibitem
\bibitem{PhysRevC.97.014908}
X.~Yao and B.~M\"uller, \ifthenelse{\boolean{articletitles}}{\emph{Approach to
  equilibrium of quarkonium in quark-gluon plasma},
  }{}\href{http://dx.doi.org/10.1103/PhysRevC.97.014908}{Phys.\ Rev.\ C
  \textbf{97} (2018) 014908},
  \href{http://arxiv.org/abs/1709.03539}{{\normalfont\ttfamily
  arXiv:1709.03539}}\relax
\mciteBstWouldAddEndPuncttrue
\mciteSetBstMidEndSepPunct{\mcitedefaultmidpunct}
{\mcitedefaultendpunct}{\mcitedefaultseppunct}\relax
\EndOfBibitem
\bibitem{Winn:2016amg}
M.~Winn, \ifthenelse{\boolean{articletitles}}{\emph{{Prospects for quarkonium
  measurements in p-A and A-A collisions at the LHC}},
  }{}\href{http://dx.doi.org/10.1007/s00601-016-1189-7}{Few Body Syst.\
  \textbf{58} (2017) 53},
  \href{http://arxiv.org/abs/1609.01135}{{\normalfont\ttfamily
  arXiv:1609.01135}}\relax
\mciteBstWouldAddEndPuncttrue
\mciteSetBstMidEndSepPunct{\mcitedefaultmidpunct}
{\mcitedefaultendpunct}{\mcitedefaultseppunct}\relax
\EndOfBibitem
\end{mcitethebibliography}

\newpage


 
\newpage
\centerline{\large\bf LHCb collaboration}
\begin{flushleft}
\small
R.~Aaij$^{28}$,
C.~Abell{\'a}n~Beteta$^{45}$,
B.~Adeva$^{42}$,
M.~Adinolfi$^{49}$,
C.A.~Aidala$^{76}$,
Z.~Ajaltouni$^{6}$,
S.~Akar$^{60}$,
P.~Albicocco$^{19}$,
J.~Albrecht$^{11}$,
F.~Alessio$^{43}$
% ...

{\footnotesize \it
% ...
$ ^{6}$Clermont Universit{\'e}, Universit{\'e} Blaise Pascal, CNRS/IN2P3, LPC, Clermont-Ferrand, France\\
% ...
$ ^{11}$Fakult{\"a}t Physik, Technische Universit{\"a}t Dortmund, Dortmund, Germany\\
% ...
$ ^{19}$INFN Laboratori Nazionali di Frascati, Frascati, Italy\\
% ...
$ ^{28}$Nikhef National Institute for Subatomic Physics, Amsterdam, Netherlands\\
% ...
$ ^{42}$Instituto Galego de F{\'\i}sica de Altas Enerx{\'\i}as (IGFAE), Universidade de Santiago de Compostela, Santiago de Compostela, Spain\\
$ ^{43}$European Organization for Nuclear Research (CERN), Geneva, Switzerland\\
% ...
$ ^{45}$Physik-Institut, Universit{\"a}t Z{\"u}rich, Z{\"u}rich, Switzerland\\
% ...
$ ^{49}$H.H. Wills Physics Laboratory, University of Bristol, Bristol, United Kingdom\\
% ...
$ ^{60}$University of Cincinnati, Cincinnati, OH, United States\\
% ...
$ ^{62}$Syracuse University, Syracuse, NY, United States\\
% ...
$ ^{76}$University of Michigan, Ann Arbor, United States, associated to $^{62}$\\
% ...
}
\end{flushleft}
\end{document}